\documentclass[12pt]{article}
\usepackage{amsmath,amssymb,epsfig,amsfonts}
\usepackage{graphicx,subfigure}
\usepackage{cite}
\usepackage[usenames,dvipsnames]{xcolor}
\usepackage{setspace}
\usepackage{lscape}
\usepackage{hyperref}
\usepackage{tikz}
\hypersetup{urlcolor=blue}
\usetikzlibrary{decorations.markings}
\tikzstyle{Vertex}=[circle, draw, inner sep=0pt, minimum size=6pt]

\usepackage[top=1.25in,bottom=1.25in,right=0.95in,left=0.95in]{geometry}

\makeatletter

\newcommand{\pqvec}[2]{\scriptsize \left( \begin{array}{c}
#1 \\
#2
\end{array} \right)}

\newcommand{\bZ}{\mathbb{Z}}
\newcommand{\bC}{\mathbb{C}}
\newcommand{\bP}{\mathbb{P}}

\newcommand{\cN}{\mathcal{N}}

\newcommand{\cO}{\mathcal{O}}

\newcommand{\g}{\mathfrak{g}}

\newcommand{\ep}{\varepsilon}

\def\ov#1{{\overline{#1}}}

\def\unit{{1\kern-.65ex {\rm l}}}
\def\1{{1\kern-.65ex {\rm l}}}

\newcommand{\ja}[1]{{\color{ForestGreen} \bf #1}}
\newcommand{\jb}[1]{{\color{Red} \bf #1}}
\newcommand{\jc}[1]{{\color{Cerulean} \bf #1}}
\newcommand{\jd}[1]{{\color{Black} \bf #1}}
\newcommand{\bface}[1]{{$\bf #1$}}
\newcommand{\obface}[1]{{$\overline{{\bf #1}}$}}

\definecolor{xdotcolor}{named}{ForestGreen}
\definecolor{tmotioncolor}{named}{Cerulean}

\newcount\hour \newcount\minute
\hour=\time \divide \hour by 60
\minute=\time
\def\now{%
\ifnum \hour<13
  \ifnum \hour=0 \advance \hour by 12 \number\hour:\else \number\hour:\fi%
     \ifnum \minute<10 0\fi%
     \number\minute%
\ A.M.%
\else \advance \hour by -12 \number\hour:%
  \ifnum \minute<10 0\fi%
  \number\minute%
  \ P.M.%
\fi%
}

\makeatother

\begin{document}

\baselineskip=18pt  
\numberwithin{equation}{section}  
\allowdisplaybreaks  



%
%


\thispagestyle{empty}

\vspace*{-2cm}
\begin{flushright}
{\tt NSF-KITP-13-074}
\end{flushright}

\vspace*{3.8cm}
\begin{center}
 {\LARGE Matter From Geometry Without Resolution}
 \vspace*{1.0cm}

Antonella Grassi$^1$, James Halverson$^2$, and Julius L. Shaneson$^1$

\vspace{1.0cm}

$^1$ Department of Mathematics, University of Pennsylvania\\
Philadelphia, PA 19104-6395 USA \\
$^2$ Kavli Institute for Theoretical Physics, University of California\\
Santa Barbara, CA 93106-4030 USA \\

\end{center}
\vspace*{1cm}

\abstract{ \vspace{.5cm} We utilize the deformation theory of
  algebraic singularities to study charged matter in compactifications
  of M-theory, F-theory, and type IIa string theory on elliptically
  fibered Calabi-Yau manifolds.  In F-theory, this description is more
  physical than that of resolution.  We describe how two-cycles can be
  identified and systematically studied after deformation.  For ADE
  singularities, we realize non-trivial ADE representations as
  sublattices of $\bZ^N$, where $N$ is the multiplicity of the
  codimension one singularity before deformation.  We give a method
  for the determination of Picard-Lefschetz vanishing cycles in this
  context and utilize this method for one-parameter smooth
  deformations of ADE singularities.  We give a general map from
  junctions to weights and demonstrate that Freudenthal's recursion
  formula applied to junctions correctly reproduces the structure of
  high-dimensional ADE representations, including the {\bf 126} of
  $SO(10)$ and the {\bf 43,758} of $E_6$. We identify the Weyl group
  action in some examples, and verify its order in others.  We
  describe the codimension two localization of matter in F-theory in
  the case of heterotic duality or simple normal crossing and
  demonstrate the branching of adjoint representations. Finally, we
  demonstrate geometrically that deformations correctly reproduce the
  appearance of non-simply-laced algebras induced by monodromy around
  codimension two singularities, showing the reduction of $D_4$ to
  $G_2$ in an example. A companion mathematical paper will follow.  }

%




\clearpage
\tableofcontents


\clearpage
\section{Introduction}
\onehalfspacing

The spectrum of particles which exist in Nature is detailed and rich.
The quarks and leptons of the standard model of particle physics fill
out non-trivial representations of the Lie algebra $SU(3)\times
SU(2)\times U(1)$, and these can be embedded into representations of higher
rank groups, such as the $\bf{10}+\bf{\ov 5}+{1}$ of $SU(5)$ or the
$\bf{16}$ of $SO(10)$.  Exotic particle representations are often
introduced in phenomenologically motivated extensions of the standard
model, sometimes of high dimension in grand unified theories.

An important physical question is whether there exist theoretical
constraints on the allowed particle representations.  Though anomaly
cancellation provides constraints on sets of fields in gauge theories,
no individual representation is ruled out on theoretical grounds.  By
contrast, the possibilities\footnote{See
  \cite{Dienes:1996yh,Dienes:1996wx} for a systematic study of
  constraints on matter representations in the free field
  heterotic string.} are more limited in
four-dimensional compactifications of string theory, F-theory, and
M-theory. For example, in the heterotic string matter representations
typically arise from branching the adjoint of $E_8$; in weakly coupled
type II string theory $U(N)$ gauge theories admit bifundamental,
symmetric tensor, and antisymmetric tensor representations, together
with their conjugates; in weakly coupled orientifold compactifications
it is not possible to realize the phenomenologically relevant
$\bf{16}$ of $SO(10)$.

Interestingly, more possibilities can  be realized outside of the weakly coupled
regime in M-theory \cite{Witten:1995ex} or F-theory
\cite{Vafa:1996xn}.  In these theories massless matter representations
are encoded in the structure of a singular compactification geometry,
and the possibilities are broader than in the weakly coupled
superstring theories. Studying matter in $d=4$ compactifications of
M-theory requires a detailed knowledge of codimension seven
singularities in compact singular $G_2$ manifolds, which, despite
much progress in the construction of compact smooth $G_2$ manifolds
\cite{Kovalev,KovalevLee,CortiHaskins1,CortiHaskins2}, is a difficult
mathematical problem.
Much more is known about singular elliptically fibered Calabi-Yau
varieties, and therefore compactifications of F-theory\footnote{Though
  we may use the language of F-theory throughout this paper, our
  results also apply in other contexts, as they are ultimately results
  about the homology of elliptic fibrations.  } on them may currently
be the broadest framework for studying matter representations in the
string landscape.

A great deal is already known
\cite{Katz:1996xe,Bershadsky:1996nh,Grassi:2000we,Aspinwall:2000kf,Morrison:2011mb,Grassi:2011hq}
about the structure of matter in compactifications on singular
elliptically fibered Calabi-Yau varieties with a section, namely their
Weierstrass model.  The resolution of generic
singularities\footnote{That is, a smooth point of the codimension
  one locus of the discriminant.} produces a new singular elliptic fiber above the general
point of the discriminant locus: it is a tree of holomorphic curves
whose dual graph structure coincides with the extended Dynkin diagram
of an ADE algebra $\g$ of rank $r$. All massless W-bosons of $\g$
appear by wrapping branes on these curves and taking a singular
limit. In codimension two, singularity enhancement gives rise to
matter representations of $\g$. If the enhancement is of a simple type
to another ADE algebra $\g'$ of rank $r+1$, the representation of the
localized matter\footnote{Ideas from both singularity resolution and deformation
were used in \cite{Katz:1996xe}.} can be determined \cite{Katz:1996xe} by the branching
rules of the adjoint of $\g'$; there exist more exotic possibilities
\cite{Grassi:2000we,Grassi:2011hq,Morrison:2011mb} at higher
codimension in moduli space. In addition, monodromy around codimension
two loci can induce an outer automorphism on $\g$ which reduces it
\cite{Bershadsky:1996nh} to a non-simply laced gauge algebra $\g'$.
We refer to this as \emph{outer monodromy} or O-monodromy.

Resolution has been an important tool for understanding other aspects
of physics in many recent works. This is particularly true in
F-theory, which has received a great deal of attention in the last
years, initiated by work \cite{Donagi:2008ca,Beasley:2008dc} on grand
unified models. Since then, there has been significant progress in
understanding F-theory compactifications to four dimensions, including
globally consistent models \cite{Andreas:2009uf, Marsano:2009ym,
  Collinucci:2009uh, Blumenhagen:2009up, Marsano:2009gv,
  Blumenhagen:2009yv, Marsano:2009wr, Grimm:2009yu, Cvetic:2010rq,
  Chen:2010ts, Chen:2010tp, Chung:2010bn,
  Chen:2010tg,Knapp:2011wk,Knapp:2011ip,Marsano:2012yc}, $U(1)$
symmetries \cite{Grimm:2010ez, Dolan:2011iu, Marsano:2011nn,
  Grimm:2011tb, Morrison:2012ei,Borchmann:2013jwa,Cvetic:2013nia},
instanton corrections \cite{Blumenhagen:2010ja,Cvetic:2010rq,
  Donagi:2010pd,
  Grimm:2011dj,Marsano:2011nn,Cvetic:2011gp,Bianchi:2011qh,Kerstan:2012cy,Cvetic:2012ts,Bianchi:2012kt},
and chirality inducing $G_4$-flux \cite{Marsano:2010ix,
  Collinucci:2010gz,Marsano:2011nn, Braun:2011zm, Marsano:2011hv,
  Krause:2011xj, Grimm:2011fx, Braun:2012nk, Kuntzler:2012bu,Lawrie:2012gg,
  Krause:2012he, Collinucci:2012as,Bianchi:2012kt,Marsano:2012bf};
there has also been progress in understanding the landscape of
six-dimensional F-theory compactifications
\cite{Kumar:2009ac,Kumar:2010ru,Kumar:2010am,Morrison:2012np,Morrison:2012ei}.  Singularity
resolution was utilized heavily in many of these works. While the
resolution of generic singularities is well understood and goes back
to Kodaira's work, the resolutions in higher codimensions have to be
worked out case by case and can be rather complicated, see for example
\cite{Esole:2011sm}.

Since four-dimensional $\cN=1$ compactifications of F-theory provide
an enormous class of vacua, continued progress in this direction is
critical for our understanding of the landscape.  Studying the physics
of these compactifications via resolution amounts to studying the
Coulomb branch of the defining three-dimensional $\cN=2$ M-theory
compactification. While this approach is useful in many cases --- for
example Chern-Simons terms in three dimensions beautifully encode the
structure of four-dimensional gauge anomalies \cite{Cvetic:2012xn} ---
it is rather indirect: the Coulomb branch does not exist in the
F-theory limit. It would be better, when possible, to study F-theory
by methods which exist in both the defining M-theory compactification
and also in the F-theory limit. This work is a step in that
direction.

We study the appearance of vector and matter multiplets in non-trivial
Lie algebra representations via the deformation theory of algebraic
singularities in elliptically fibered varieties $X$; see
\cite{Grassi:2000fk,Grassi:2001xu} for a study in heterotic M-theory.
This involves movement in the complex structure moduli space of $X$,
rather than the K\" ahler moduli space of resolution.  Though for
M-theory on $X$ deformation and resolution are complimentary
approaches for understanding gauge theoretic structure, only the
deformation picture is physical in the F-theory limit. For example,
ADE states which obtain mass upon Higgsing the gauge theory arise from
two-cycles whose non-zero volume is obtained from deformation, not
resolution. This picture is also necessitated by heterotic duality
\cite{Vafa:1996xn,Morrison:1996na,Morrison:1996pp}: at a generic point
in the moduli space of an $E_8$ gauge bundle the theory is completely
Higgsed and the $240$ W-bosons of $E_8$ are massive; in the dual
F-theory picture the heterotic bundle moduli map to complex structure
moduli which deform the $E_8$ singularity and give rise to the
necessary two-cycles.
 Klein showed that  resolutions and deformation of ADE surface singularities (also known as \emph{kleinan} singularities) are diffeomorphic. We use  deformation and junctions to naturally associate weights and other representations, in addition to the adjoint
  to these singularities.

It is simple to see how deformation differs from resolution, and why
it is physically relevant.  Consider an elliptic
fibration\footnote{Note that $X$ could be, but is not required to be,
  a Calabi-Yau variety.  Sometimes we utilize subscripts to denote the
  complex dimension of these spaces as $X_d$ and $B_{d-1}$.  We will
  often consider $X$ in Weierstrass form.  } $X \xrightarrow{\pi} B$
with discriminant $\Delta$.  If one of the components of the
discriminant is smooth, or locally smooth, it can be described by
$z=0$ and the discriminant takes the form $\Delta = z^N \tilde
\Delta$, where $\tilde \Delta$ is a residual piece which can be
computed in examples when $X$ is a Weierstrass model. If $N=1$, then
the Weierstrass model $X$ is generically smooth along $z=0$ and the
theory there  is completely Higgsed; if $N \geq 2$
generically $X$ has ADE singularities with algebra $\g$ along $z=0$, and
performing a small deformation of the elliptic fibration such that the
theory is completely Higgsed gives
\begin{equation}
\Delta \sim z^N  \qquad \xrightarrow{\text{deform}} \qquad \Delta \sim \prod_{i=1}^N (z-\ep_i)
\end{equation}
in a general neighborhood of $z=0$. We emphasize that $N$ is \emph{not}\footnote{The
cases $G=A_r$ or $G=D_r$ satisfy $N=r+1$ and $N=r+2$, respectively;
$E_6$ $E_7$ and $E_8$ satisfy $N=r+2$.} the rank of $\g$. In the
language of F-theory it is the number of seven-branes, and therefore $N$, and not just $r$,
must play a role in determining the structure of ADE states which end on seven-branes.

We demonstrate that the Lie-algebraic structure of these states is
determined by two-cycles arising from the deformation of ADE
singularities, which can be specified as vectors in $\bZ^N$.  Together
with an appropriate inner product, the two-cycles which naturally
arise from deformation give rise to root lattices which span
$r$-dimensional subspaces of $\bZ^N$.  We emphasize that these
deformation techniques and string junctions can be applied to
codimension two singularities in Weierstrass models to detect matter
and non-simply laced algebras; in section
\ref{eqn:D4G2WeierstrassSingular} we study the algebra $\g_2$ via deformation and string
junctions.  That is, we study matter without
    resolution.

We  reproduce the IIb
string junction formalism of \cite{DeWolfe:1998zf, Gaberdiel:1997ud,
  Mikhailov:1998bx}; as such, we will refer to these two-cycles as
junctions. There is naturally some overlap between the group theoretic
results we obtain and the results of \cite{DeWolfe:1998zf}, though we
generalize the group theory and emphasize the relationship to
deformation theory and geometry.  We will attempt to distinguish new results from
old, when possible. We also emphasize that the formalism of junctions
is computationally much simpler than that of singularity resolution;
in appendix \ref{sec:code} we provide a publicly available software
package for performing these computations. In the companion paper \cite{GrassiHalversonShaneson:Math} we analyze the mathematical underpinning of {\emph{matter without resolution}}.

\vspace{.3cm}

Let us summarize our results and the outline of this paper.

\vspace{.3cm}

In section \ref{sec:deformations and junctions} we lay out the basic
formalism for understanding two-cycles and deformations. The input
data for analysis is an ordered set $Z$ of vanishing one-cycles in the
elliptic fiber, obtained from geometry after singularity
deformation. Two-cycles arise naturally from this data, and for
particular sets $Z$ the $A_r$ algebras arise intuitively. More
generically, topological self-intersections induce a symmetric
bilinear form  $I$ which becomes the product on the root systems in the
case of deformations of ADE singularities. For a given set $Z$
this product can be represented as an $N\times N$ matrix.
The general setup admits a description in terms of Picard-Lefschetz monodromy and the related
theory for elliptic surfaces. The choice of $Z$ amounts to the choice
of a strong basis \cite{GuseinZade}, and the symmetric bilinear form on the algebra is
independent of this choice. We give a method for the explicit determination
of $Z$, which we exemplify later in the paper.

\vspace{.3cm} In section \ref{sec:deformations and ADE} we specialize
to the case of ADE deformations and representations. We review the
so-called canonical presentation of \cite{DeWolfe:1998zf}, which which
gives a set $Z$ for each ADE group. Our general formula for the
symmetric bilinear form reproduces the known results for ADE
deformations in the canonical presentation.  For each ADE group, we
specify a surface deformation which completely Higgses the gauge
group. We determine ordered sets $Z$
associated to these deformations, which differ from the results of
\cite{DeWolfe:1998zf}; nevertheless a junction analysis can be performed
and the representation theoretic results are consistent.

We discuss the appearance of non-trivial representations as junctions
with non-zero asymptotic charge. These are real two-manifolds
emanating from a complex codimension one locus with associated algebra
$\g$; studying the $\g$-representation theory only requires knowledge
of codimension one data. Utilizing the symmetric bilinear form $I$, we
construct a generic map from junctions to weights in the Dynkin basis
and demonstrate that Freudenthal's recursion formula also applies to
junctions, using $I$ as the product. This is essential since it allows
for the study of junctions in representations $\rho$ of $\g$ with
non-trivial weight multiplicities and lengths; moreover, it is
necessary for isomorphism. As a non-trivial check, we have verified
that the junction representations of the {\bf 126} of $SO(10)$ and the
{\bf 43,758} of $E_6$ appear in the expected way.

\clearpage
\vspace{.3cm} In section \ref{sec:codimension two} we discuss the
appearance of massless matter in codimension two at $z=t=0$.  As
emphasized, the representation theory is determined explicitly by
codimension one data, but the codimension two data is critical for
determining which representations become massless. We focus on the
cases where $X$ is $K3$-fibered or the codimension two singularity
arises from simple normal crossing; in these cases the localization of
matter in codimension two can be understood by studying families of
deformed elliptic surfaces parameterized by $t$. In some cases the
matter localized in codimension two follows from branching
junctions in the adjoint.  We argue that in the deformation picture
O-monodromy around codimension two loci can induce an outer
automorphism on junctions which reduces $\g$ to a non-simply laced
algebra $\g'$.

\vspace{.3cm} In section \ref{sec:examples} we provide a number of
illustrative examples with non-trivial behavior in codimension two.
Via explicit deformation we determine matter from the branching of
adjoints for $A_r\rightarrow A_{r+1}$ enhancement; in the canonical
basis we demonstrate the branching of the adjoints of $E_6$ and
$SO(8)$ to the {\bf 16} of $SO(10)$ and the ${\bf 6}$ of $SU(4)$,
describing weak coupling interpretations when possible. We explicitly
demonstrate that O-monodromy can be studied and understood in the
deformation picture. Specifically, we deform a $D_4$ singularity with
appropriate codimension two structure and demonstrate that O-monodromy
induces an outer automorphism on the $D_4$ roots, represented as
junctions, that reduces $D_4$ to $G_2$.  The basis we determine for
these deformations differs from the canonical basis and makes the
$\bZ_3$ automorphism more transparent; a similar computation was
performed in the canonical basis in \cite{Bonora:2010bu}, though the
results were not derived directly from a deformed geometry. We also
discuss junction realizations of ADE representations which play
important roles in particle physics models.

\section{From Deformations to Junctions: Basic Formalism}
\label{sec:deformations and junctions}
In this section we will lay out
formalism
in a general manner which includes, but is not limited, the case of
deformations of ADE singularities.  We begin by discussing an elliptic
fibration $X$ and an associated ordered set of vanishing one-cycles
$Z$.  The fiber above a component $\Delta_i$ of the discriminant
$\Delta$ of $X$ has a vanishing one-cycle $\pi_i$.  Two components of
the discriminant can be connected by strings; more generically
$n$-components can be connected by an $n$-pronged ``string junction.''
Such an object gives rise to a two-manifold in the total space $X$ and
can be represented as a vector in $\bZ^N$.  In a neighborhood of
nearby components $\Delta_i$ we define a product on these vectors
which amounts to the topological self-intersection of two-cycles.
This induces a symmetric bilinear form on $\bZ^N$ which becomes the
product on the ADE root system for ADE deformations. This formalism
admits a description in terms of Picard-Lefschetz theory. We give a
method for determining $Z$.

\clearpage
\subsection{Singularities, Deformations, and Seven-Branes}
\label{sec:singdefsevenbranes}

Consider the elliptic fibration $X$ and consider also the case where
there are $k\ge 2$ components of the discriminant.  Fix a base point
$P$ on $B\setminus \Delta$ and a basis for $H_1(\text{fiber})$. Above
each component $\Delta_i$ of the discriminant is an elliptic fiber
with a vanishing one-cycle $\pi_i$; after choosing a path in $B
\setminus \Delta$ which approaches $\Delta_i$, together with
generators of one-cycles in the fiber above $P$, we may write $\pi_i =
\pqvec{p_i}{q_i}\in\bZ^2$.  In an F-theory
compactification, such a geometry gives a $(p_i,q_i)$ seven-brane
along $\Delta_i=0$.  For any geometry it is critical to identify the
components of the discriminant, their vanishing cycles, and a choice
of path around them; this determines an ordered set $Z$ of vanishing
cycles which serves as input data for studying two-cycles.  We will see that
this is sufficient to begin uncovering Lie algebraic structure and
leave a detailed discussion of monodromy and the determination of $Z$
until section \ref{sec:picard lefschetz}.

As a brief aside, it is useful to review how well-known physical
scenarios can be described in terms of $(p,q)$ seven-branes, and also
to set conventions\footnote{When possible, we follow the conventions
  of \cite{DeWolfe:1998zf}.} which will be useful in discussing
deformations of ADE singularities and their associated algebras.  The
only seven-branes which exist in the weakly coupled type IIb limit
\cite{Sen:1996vd,Sen:1997gv} are D7-branes and O7-planes, and we
follow the convention in the literature where a D7-brane is called an
$A$-brane and has $\pi_A\equiv\pqvec{1}{0}$. In F-theory the string
coupling $g_s$ is finite and D(-1)-instanton effects split an O7-plane
into a $B$-brane and a $C$-brane, which have $\pi_B = \pqvec{1}{-1}$
and $\pi_C = \pqvec{1}{1}$, respectively. This splitting is precisely
equivalent to the quantum splitting due to instantons in
Seiberg-Witten theory \cite{Seiberg:1994rs}, where the $\cN=2$ $d=4$
gauge theory with $G=SU(2)$ is the worldvolume gauge theory of a
D3-brane probing the local $BC$-geometry \cite{Banks:1996nj}; the
gauge instantons of Seiberg-Witten theory are the D(-1)-instantons
inside the D3 probe\footnote{These effects play an important role in
  understanding certain globally consistent F-theory compactifications
  \cite{Cvetic:2010rq}.}. Singularities at codimension one in the base
are classified by Kodaira and are ADE singularities for minimal
Weierstrass models; after deformation these admit descriptions in
terms of particular $A$-branes, $B$-branes, and $C$-branes. See
section \ref{sec:deformations and ADE}.

\begin{figure}[t]
   \centering
   \includegraphics[scale=1.0]{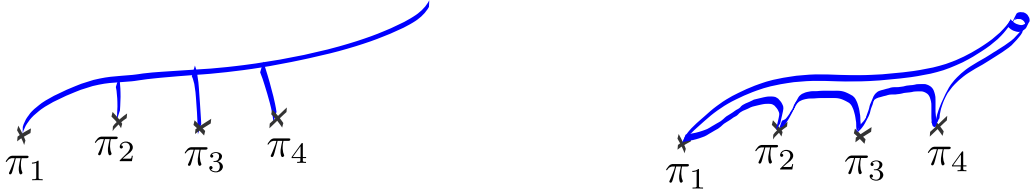}
\label{fig:fourjunc}
\caption{\small On the left is a four-pronged junction in $B$. The picture on the right is
of the corresponding real two-manifold in the total space $X$. The loose end wraps the one-cycle $a(J)$
in the elliptic fiber; i.e. the asymptotic charge.}
\end{figure}

The physical importance of deformation theory enters in a simple way.
Let $N_i \equiv deg(\Delta_i)$ be the degree of the component
$\Delta_i$, so that we write $\Delta = \prod_i \Delta_i^{N_i}$.  In
any given example, only a certain set of deformations are consistent
with the elliptic fibration $X$. In performing the deformation of the
geometry, the discriminant is changed such that the component
$\Delta_i$ will deform into some number of other components.  In the
case of ADE singularities the relevant deformations are known and the
degree $N$ component $\Delta_z$ splits into $N$ components, each of
which has an elliptic fiber above it with a particular vanishing cycle.

\clearpage
\subsection{Matter Fields from Junctions and Two-Spheres}
\label{sec:matter}
Armed with the knowledge of the components $\Delta_i$ and their vanishing
cycles $\pi_i$, it is possible to study two-cycles which arise
in the geometry. Physical objects, such as M2-branes, can
wrap these two-cycles and give rise to representations $\rho$ of the
gauge algebra $\g$. In F-theory these states must be described as
$n$-pronged string junctions in $B$, since only the dimensions of $B$ in the
total space $X$ are physical.

Consider first two separated components $\Delta_1$ and $\Delta_2$ of
the discriminant with vanishing one-cycles $\pi_1 = \pi_2$. An
oriented interval  $J$ exists from $\Delta_1$ to $\Delta_2$; in the
fiber above $\Delta_1$ the cycle $\pi_1$ vanishes, but upon moving
along $J$ towards $\Delta_2$ it grows and then vanishes again at
$\Delta_2$, giving a two-sphere with two marked points. Though $J$ is
an interval in $B$ represented by the ``brane charge'' vector
$(1,-1)$, denoting the interval orientation from $\Delta_1$ to
$\Delta_2$, we see that it is also two-sphere in $X$, since $1\times
\pi_1 + (-1)\times \pi_2 = \pqvec{0}{0}$. More generically, one could
consider $N$ components $\Delta_i$ connected in $B$ by an $n$-pronged
object represented by the vector $J=(J_1,\dots,J_N)\in \bZ^{N}$ with
$n$ of the $J_i\ne 0$. Such a $J$ is a \emph{junction} or \emph{string junction}.
 If $\sum_i J_i \pi_i = \pqvec{0}{0}$, this
gives a two-sphere in the total space $X$ with $n$ marked points.

Thus far we considered a general set of components $\Delta_i$ and
the associated two-cycles. It is possible, of course, that there is a
natural grouping of components into multiple sets according to
proximity to certain codimension one loci. For example, if a geometry
realizes an $A_{r_1}$ and a $D_{r_2}$ singularity along $\Delta_A$ and
$\Delta_D$, a generic but small deformation of the geometry deforms
$\Delta_A$ into $r_1+1$  nearby components, and similarly
deforms $\Delta_D$ into $r_2+2$ nearby components. As one might
expect, the structure of appropriate two-spheres constructed within
one set of components encodes the data of massive W-bosons, which
become massless in codimension one upon undoing the deformation.

More generically, consider two sets $\Delta_i$ and $\Delta_j$ with
$N_1$ and $N_2$ components and two-spheres which end on some number of
components in each set. These could be represented by a vector in
$J_{tot} \in \bZ^{N_1+N_2}$. However, this isn't necessary for some
purposes, since any representations arising from the $N_1$ components
should be determined only by how a junction ends on those
components. Truncating appropriately, we represent the junction prongs
ending on the $N_1$ components by a vector $J =(J_1,\cdots , J_{N_1})
\in \bZ^{N_1}$ with $\sum_i J_i \pi_i\ne \pqvec{0}{0}$, since the
two-sphere determined by $J_{tot}$ ends on some number of components
in $\Delta_j$. For determining which set of components in $\Delta_j$
the junction $J$ emanating from $\Delta_i$ can end on, it is important
to keep track of the asymptotic charge $a(J) \equiv \sum_i J_i \pi_i
\equiv \pqvec{p_J}{q_J}$. In type IIb language, the corresponding
state has $p_J$ units of $B$-field charge and $q_J$ units of
Ramond-Ramond $C_2$ charge. For example, the junctions $J$ giving
fundamentals of $SU(N)$ all have $a(J)=\pqvec{1}{0}$, which signifies
that they can end on $D7$-branes in codimension two. Antisymmetric
tensors, on the other hand, have $a(J)=\pqvec{2}{0}$, recovering the
fact that they become massless at intersections with
$O7$-planes.

In summary, we will represent junctions $J$ which emanate from a set
of $N$ components $\Delta_i$ near a common codimension one locus by a
vector $J\in \bZ^{N}$, recognizing that most $J$ will have asymptotic
charge.  The ordered set of vanishing cycles $Z$ naturally determines
a basis for $\bZ^N$, the ``junction basis.''
Whether a particular state $J$
becomes massless at a given codimension two locus depends on the
asymptotic charge $a(J)$ and the vanishing cycles $\pi_j$ of the
components $\Delta_j$; if the conditions on these quantities are
appropriate, there will be vanishing two-cycles in codimension
two. This will become clear in examples.

We refer the reader to appendix \ref{app:junction} for a detailed depiction
of a non-trivial junction.

\subsection{An Interlude on $A_r$ Algebras}
\label{sec:Ar interlude}
Though more generic cases will require further thought, the discussion
thus far is sufficient for showing that the structure of $A_r$
algebras arises from deformation theory and junctions. It is our hope
that this builds geometric and physical intuition before moving to the
more complicated $D_r$ and $E_{6,7,8}$ cases.

Suppose the the local geometry of $X$ has  an $A_r$ singularity along $\Delta_z$.
In this case $N=r+1$ and we have $\Delta = z^{r+1} \tilde \Delta$,
where $\tilde \Delta$ is a residual piece which would exist if this
were a local description in a globally defined geometry. After
performing the one-parameter\footnote{Deformations with more
  parameters exist; we study one-parameter deformations which
  completely Higgs the gauge group.} deformation (\ref{eq:Ar
  deformation}), one arrives at $\Delta = \tilde \Delta \,\,
\prod_{i=1}^{r+1} (z_i-\ep_i) \equiv \tilde \Delta \prod_{i=1}^{r+1} \Delta_i$
where $\ep_i$ are appropriate roots of the
deformation parameter $\ep$.
For this geometry it is known\footnote{This will be explicitly derived
  later.} that $\pi_1 = \cdots = \pi_{r+1} \equiv \pi$ for all
components $\Delta_i$, and via an $SL(2,\bZ)$ transformation we can
choose $\pi = \pqvec{1}{0}$ without loss of generality. A junction $J$
is determined by how it ends on the $r+1$ components $\Delta_i$, and
therefore can be represented as a vector in $\bZ^{r+1}$.

The $A_r$ algebra emerges from this data immediately.  The $r+1$
components $\Delta_i$ are arranged on a circle around $z=0$,
as drawn below, and the separation between the $\Delta_i$ is determined by
the deformation parameter $\ep$. Pick an ordered set $Z$ of the $\Delta_i$ by
picking one to be the first element in $Z$ and choosing a direction
around the circle.
Consider two-cycles that occur in the geometry, utilizing an
orthonormal basis $e_i$ for $\bZ^{r+1}$ for convenience and defining
the junctions $J_{ij}\equiv e_i - e_j$ with $i\ne j$. These satisfy $a(J_{ij})=0$
and give two-spheres with two marked points in $X$.  Let us
briefly consider $A_4$
for the sake of visualization. A useful picture is
\begin{equation}
  \begin{tikzpicture}[scale=0.8]
    \fill[color=red] (0:10mm) circle (1mm);
    \fill[color=red] (72:10mm) circle (1mm);
    \fill[color=red] (144:10mm) circle (1mm);
    \fill[color=red] (216:10mm) circle (1mm);
    \fill[color=red] (288:10mm) circle (1mm);
    \draw[thick,->] (68:10mm) arc (68:4:10mm);
    \draw[thick,->] (-4:10mm) arc (-4:-68:10mm);
    \draw[thick,->] (-78:10mm) arc (-78:-140:10mm);
    \draw[thick,->] (-148:10mm) arc (-148:-212:10mm);
    \draw[thick,->] (-220:10mm) arc (-220:-284:10mm);
    \draw[xshift=1cm,thick,dotted,->] (122:12mm) arc (122:-122:12mm);
    \node at (0:3.9cm){$J_{13}=J_{12} + J_{23}$};
    \node at (36-0*72:13.5mm){$J_{12}$};
    \node at (36-1*72:13.5mm){$J_{23}$};
    \node at (36-2*72:12.5mm){$J_{34}$};
    \node at (36-3*72:13.5mm){$J_{45}$};
    \node at (36-4*72:12.5mm){$J_{51}$};
  \end{tikzpicture}
\end{equation}
where the red dots are the components $\Delta_i$ and the arrows depict
some junctions, which are one-dimensional strings in $B$ but are
two-spheres in $X$.  $J_{51} = -J_{12}-J_{23}-J_{34}-J_{45}$ is not an
independent junction, and $J_{13} = J_{12} + J_{23}$ where the
addition is in the second homology. For general $r$, there are
$(r+1)^2 - r-1$ $J_{ij}$ with $i\ne j$, which also happens to be the
number of the roots of $A_r$. Defining $\alpha_i \equiv J_{i,i+1}$, we
see that $\alpha_1$ and $\alpha_2$ can be added to give
$J_{1,3}=\alpha_1 + \alpha_2$, as depicted above. Letting $( \cdot,
\cdot)$ be the negative of the standard inner product on $\bZ^{r+1}$,
$A_{ij} \equiv (\alpha_i, \alpha_j)$ is the negative Cartan matrix of
$A_r$.  This is in accord with standard convention in algebraic
geometry, where simple roots obtained by resolution are $(-2)$-curves.
Though we have arrived at the roots $J_{ij}$ intuitively, one could
have also found them by searching for all $J$ with $(J,J)=-2$ and
vanishing asymptotic charge.  Similarly, solving for $J$ with
$(J,J)=-1$ and $a(J) = \pqvec{1}{0}$ or $a(J)=\pqvec{-1}{0}$ give the
fundamental and antifundamental, respectively. $J$ with $(J,J)=-1$ and
$a(J)=\pqvec{2}{0}$ gives the antisymmetric tensor representation.

We considered a one parameter deformation of the $A_r$ singularity
which completely Higgsed the gauge group. More generically, there are
$r$ inequivalent deformations, which can be written naturally in a
Weyl-invariant form \cite{KatzMorrison}.  The Weyl group $S_{r+1}$ permutes
the $r+1$ defects.

\subsection{Topological Intersections Induce a Symmetric Bilinear Form}
\label{sec:topological intersection}
In section \ref{sec:Ar interlude} the product $(\cdot, \cdot)$ played
an important role in determining roots and weights. In the elliptic
surface defined by a one-dimensional neighborhood of discriminant
components and the elliptic fibration over it, $(\cdot, \cdot)$ happens to be the topological
intersection product. Since all vanishing cycles $\pi_i$ are the same
for $A_r$ it took a simple form; this is generically not the case. In
this section we derive the generic form.

Consider a junction $J$ of the discussed type.
The self intersection $(J,J)$ receives
contributions from vanishing cycles and also from junction points; the
latter involve two prongs becoming one in $B$ away from components of
the discriminant, locally giving a pair of pants in $X$, as depicted
in figure~\ref{fig:fourjunc}. Heuristically, we denote these two types
of contributions as\footnote{While the transpose is explicit here, it will
be implied elsewhere in the text, though hopefully clear based on context.}
\begin{equation}
(J,J) = (J,J)_v + (J,J)_j \equiv J_k\, I_{kl} \, J_l = J^T \, I\, J.
\label{eq:(J,J)}
\end{equation}
For junctions constructed from $N$ components $\Delta_i$, the symmetric matrix $I_{kl}$
gives a symmetric bilinear form in the junction basis.  We write
this matrix in terms of contributions from vanishing cycles and
junction points as $I_{kl}=I^v_{kl} + I^j_{kl}$.
The vanishing cycle contribution is straightforward: in a
sufficiently small neighborhood of a puncture, $J$ intersects itself
only at the puncture. For a particular $l$, if $J_l=1$ the
contribution from this vanishing cycle is $-1$. More generically it is $-J_l^2$.  Therefore
we have $I^v = -{\bf 1}_{N\times N}$.

In order to determine the contribution $(J,J)_j$ from junction points
it will be useful to study an example which easily generalizes.
Consider a junction with four punctures, choose an ordering, and
construct pairs of pants in a small neighborhood in an ascending
fashion, as depicted in figure \ref{fig:fourjunc}. Recall that a
junction with coordinate $J_i \in \bZ$ on $\pi_i$ wraps the one-cycle
$J_i \pi_i$.  At the junction point where
$J_1\pi_1$ and $J_2\pi_2$ join, for example, there are $J_1\pi_1 \cdot
J_2\pi_2$ non-trivial intersections in the torus fiber due to the
joining of three one-cycles.  These points contribute to the
self-intersection $(J,J)_j$. It would seem that this contribution
depends on the choice of two legs $J_1\pi_1$ and $J_2\pi_2$ instead of
the other leg $-J_1\pi_1 -J_2\pi_2$ (negative sign so that all one-cycles are
``incoming''), but this is not so since $J_1\pi_1 \cdot J_2\pi_2 = J_2\pi_2
\cdot (-J_1\pi_1 -J_2\pi_2) = (-J_1\pi_1 -J_2\pi_2) \cdot J_1\pi_1$.  The
contribution from all three junction points to the self-intersection
is therefore given by $J_1\pi_1\cdot J_2\pi_2 + (J_1\pi_1 + J_2\pi_2 )\cdot J_3\pi_3
+ (J_1\pi_1 + J_2\pi_2 + J_3\pi_3) \cdot J_4\pi_4 = \sum_{i<j} J_i\pi_i \cdot J_j\pi_j$.

This argument generalizes directly to $n$-punctures; i.e., to $n$-pronged
string junctions. At the
$j$th junction point, $\sum_{i=1}^j J_i\pi_i$ forms a pair of pants with
$J_j\pi_j$. The contribution to the self-intersection from all of the
points is given by
\begin{equation}
(J,J)_j = \sum_{j=1}^{n-1}(\sum_{i=1}^{j-1}J_i\pi_i) \cdot J_j\pi_j = \sum_{i<j} J_i\pi_i \cdot J_j\pi_j \equiv J_m U_{mn} J_n
\end{equation}
where $U$ is the upper triangular $N\times N$ matrix with entries for
$i<j$ given by $\pi_i \cdot \pi_j$.  Defining a symmetric
matrix $I^j= \frac{1}{2} (U + U^T)$ gives $(J,J)_j = J_m I^{j}_{mn}
J_n$.

Computing the topological intersection matrix $I$ only requires
knowing the ordered set $Z$ of vanishing one-cycles, which determines
the junction basis. For any given geometry, there are many equivalent
choices of $Z$, but the product \eqref{eq:(J,J)} is invariant, as we
will now show.

\subsection{Picard-Lefschetz Theory}
\label{sec:picard lefschetz}

In the last four sections we have asked the reader to suspend disbelief and take
an ordered set of vanishing cycles $Z$ as input for an analysis of two-cycles
in an elliptic surface. In this section we justify this input data and discuss
a method for determining $Z$.

Before discussing its relevance for our work, let us review the
basic ideas of Picard-Lefschetz theory.  Consider a holomorphic map
$f: Y\rightarrow B$ where $Y$ is a $d$ dimensional compact complex
manifold and $B$ is $\bP^1$.  Let the patches of $B$ be the northern
and southern hemispheres $P_+$ and $P_-$, chosen so that all of the
non-degenerate critical points $\{p_1, \dots ,p_k \}$ are in the
interior of $P_+$, and let $z$ be a coordinate on $P_+$. Pick a base
point $P$ and remove the singular points and associated singular
fibers from $Y$ and $B$, defining $\tilde Y = Y\setminus
f^{-1}\{p_1,\dots,p_k\}$ and $ \tilde B\equiv B\setminus \{p_1, \dots,
p_k\}$. Let $E_P$ be the smooth elliptic fiber at $P$.

The fundamental
group $\pi_1(\tilde B,P)$ is non-trivial and induces an action on the
homology of $Y$ according to the Picard-Lefschetz formula. Let us
describe its generators. Choose a disk with center $p_i$ in the
interior of $P_+$, and let the disk have radius $\rho > 0$ small
enough such that any $p_j$ with $i\ne j$ is outside of the disk. Let
$\chi_i(s) = p_i + \rho e^{2\pi i s}$ with $0 \le s \le 1$ be a
circular path around $p_i$, and let $l_i$ be any path from $P$ to $p_i
+ \rho$ in $\tilde B$. Then $w_i = l_i^{-1} \, \circ \, \chi_i \,
\circ \, l_i$ is called the $\emph{elementary path}$ encircling
$p_i$. The fundamental group is generated by the homotopy classes
$[w_1], \dots, [w_k]$ of the elementary paths.  Associated to each
$p_i$ is a vanishing cycle $\pi_i$, which we will soon
determine. Given these vanishing cycles, the Picard-Lefschetz theorem
gives the action of the fundamental group on the homology. For $p \ne
{d-1}$ the fundamental group acts trivially on $H_p(Y)$. For $p=d-1$
the elementary path $w_i$ acts on an element $x$ of $H_p(Y)$ as
\begin{equation}
w_i(x) = x + (-1)^{d(d+1)/2} \langle x, \pi_i \rangle \pi_i.
\label{eq:picard lefschetz formula}
\end{equation}
This result is due to Picard for surfaces $(d=2)$, and to Lefschetz in
the higher dimensional case.  For a recent discussion of this effect
in the language of $(p,q)$ seven-branes and $SL(2,\bZ)$ monodromy, see
\cite{Cvetic:2011gp}.

The relevance to our work is clear. Consider the elliptic fibration
$X_d$ with discriminant $\Delta \sim z^N(\dots)$. Define the patch
$P_+$ to be a disk with center $z=0$, and deform $X_d$ such that the
$N$-fold degenerate critical point at $z=0$ becomes $N$ non-degenerate
critical points at $\{p_1, \cdots, p_N \}$ in $P_+$. This is in the
spirit of the work of Arnol'd \cite{Arnold} and Gusein-Zade
\cite{GuseinZade}. We can deform such that $p_i \ne 0$, so we choose
the base point $P$ to be $z=0$. Schematically, for $N=5$ the setup
appears as
\begin{equation}
\begin{tikzpicture}[scale=1]
    \fill[thick,color=red] (45:10mm) circle (1mm);
    \draw[dotted,thick] (45:3mm) -- (45:9mm);
    \fill[thick,color=red] (80:12mm) circle (1mm);
    \draw[dotted,thick] (80:3mm) -- (80:11mm);
    \fill[thick,color=red] (166:12mm) circle (1mm);
    \draw[dotted,thick] (166:3mm) -- (166:11mm);
    \fill[thick,color=red] (260:8mm) circle (1mm);
    \draw[dotted,thick] (260:3mm) -- (260:7mm);
    \fill[thick,color=red] (320:14mm) circle (1mm);
    \draw[dotted,thick] (320:3mm) -- (320:13mm);

    \draw[dotted] (0,0) circle (18mm);
    \draw[thick] (0,0) circle (3mm);
    \draw[thick] (-1mm,1mm) -- (1mm,-1mm);
    \draw[thick] (1mm,1mm) -- (-1mm,-1mm);

    \node at (2.5cm,1.5cm) {$z$};
    \draw[thick,xshift=2.3cm,yshift=1.3cm] (90:0mm) -- (90:4mm);
    \draw[thick,xshift=2.3cm,yshift=1.3cm] (0:0mm) -- (0:4mm);

 \end{tikzpicture}
\end{equation}
where the red dots are the $p_i$, the check at the origin is the base point $P$, the
solid circle is a small neighborhood of $P$ and the large dotted circle is $P^+$.
One can visualize the elementary paths $w_i$ going around each point
$p_i$ by following the straight dotted line towards $p_i$, encircling it in a small
neighborhood, and then following the dotted line back to the base point.
When determining the vanishing cycles $\pi_i$, we will follow the dotted line  all the way to the critical point, calling it a path of approach.

It is simple to determine an ordered set of vanishing cycles. Consider
an elliptic fibration over $P^+$ given in Weierstrass form as
\begin{equation}
  y^2 = x^3 + f \,x + g
  \label{eq:Weierstrass}
\end{equation}
where $f$ and $g$ are polynomials in $z$. At a generic point in the
base the roots of the right hand side determine three marked points in
the $x$-plane
\begin{equation}
\begin{tikzpicture}[scale=1]
    \fill[xshift=7cm,thick,color=xdotcolor] (180:10mm) circle (1mm);
    \fill[xshift=7cm,thick,color=xdotcolor] (180-120:10mm) circle (1mm);
    \fill[xshift=7cm,thick,color=xdotcolor] (180+120:10mm) circle (1mm);
    \node at (9.2cm,1.3cm) {$x$};
    \draw[xshift=7cm,thick,->] (180:10mm)+(30:1.3mm) -- +(30:16mm);
    \draw[xshift=7cm,thick,->] (180-120:10mm)+(-90:1.3mm) -- +(-90:16mm);
    \draw[xshift=7cm,thick,->] (180+120:10mm)+(150:1.3mm) -- +(150:16mm);
    \node at (6.6cm,0.7cm) {$\pi_\alpha$};
    \node at (8cm,0cm) {$\pi_\beta$};
    \node at (6.6cm,-0.7cm) {$\pi_\gamma$};
    \draw[xshift=9cm,thick,yshift=1.0cm] (90:0mm) -- (90:4mm);
    \draw[xshift=9cm,thick,yshift=1.0cm] (0:0mm) -- (0:4mm);
 \end{tikzpicture}
\label{eqn:pixpiypiz}
\end{equation}
where we have drawn three paths $\pi_\alpha$, $\pi_\beta$, and $\pi_\gamma$
between them such that $\pi_\alpha + \pi_\beta + \pi_\gamma = 0$.
In fact, these paths are one-cycles: at the points where $x^3+fx+g=0$ the double cover
(\ref{eq:Weierstrass}) degenerates, but every other point on the path is a double cover,
so that topologically the path is an $S^1$. More generic paths between the
marked points also give one-cycles, and any two of the $\pi_\alpha$, $\pi_\beta$, and $\pi_\gamma$
can be taken as generators of $H_1(E_P,\bZ)=\bZ^2$.

The elliptic fiber becomes singular when $y=0$, $x^3+fx+g = 0$, and
$\frac{\partial}{\partial x} (x^3+fx+g) = 0$.  The latter two equations are
satisfied if and only if the $x^3+fx+g$  has a double root, and thus at any
$p_i$ two of the green points have collided.  In following the path of
approach from $P$ to $p_i$, those two green points take a particular
path, and the homology cycle of this path is the vanishing cycle
associated the point $p_i$. We will demonstrate this phenomenon for $A_r$ and $D_4$
cases in sections \ref{sec:Ardef} and \ref{sec:monodromy non simply-laced}, respectively. This data, together
with the natural ordering of $p_i$ determined by the order of paths of
approach around a neighborhood of $P$, determines the set $Z$ of ordered vanishing cycles which is
the input for a junction analysis. For a $N$-element $Z$, we define
the monodromy around the entire set $w_Z=w_{\pi_N} \circ \cdots \circ
w_{\pi_1}$. We refer to the Picard-Lefschetz monodromy as the PL-monodromy, to
distinguish it from the outer monodromy, which we have denoted by O-monodromy.

Let us study the effect of swapping the order of two adjacent points;
to do so, it is sufficient to consider an example with three points
$p_i$
\begin{equation}
  \begin{tikzpicture}
    \fill[color=red] (0:10mm) circle (1mm);
    \draw[dotted,thick] (0:3mm) -- (0:9mm);
    \fill[color=red] (120:10mm) circle (1mm);
    \draw[dotted,thick] (120:3mm) -- (120:9mm);
    \fill[color=red] (240:10mm) circle (1mm);
    \draw[dotted,thick] (240:3mm) -- (240:9mm);
    \draw[thick] (-1mm,1mm) -- (1mm,-1mm);
    \draw[thick] (1mm,1mm) -- (-1mm,-1mm);

    \draw[thick,color=ForestGreen,->,dashed] (-45:3mm) arc (-45:315:3mm);

    \node at (0:15mm) {$\pi_1$};
    \node at (120:14mm) {$\pi_2$};
    \node at (240:14mm) {$\pi_3$};

    \node at (2.2cm,1.0cm) {$z$};
    \draw[thick,xshift=2cm,yshift=.8cm] (90:0mm) -- (90:4mm);
    \draw[thick,xshift=2cm,yshift=.8cm] (0:0mm) -- (0:4mm);
    \begin{scope}[shift={(5cm,0cm)}]
    \fill[color=red] (0:10mm) circle (1mm);
    \draw[dotted,thick] (0:3mm) -- (0:9mm);
    \fill[color=red] (120:10mm) circle (1mm);
    \fill[color=red] (240:10mm) circle (1mm);
    \draw[dotted,thick] (240:3mm) -- (240:9mm);
    \draw[thick] (-1mm,1mm) -- (1mm,-1mm);
    \draw[thick] (1mm,1mm) -- (-1mm,-1mm);

    \draw[thick,color=ForestGreen,->,dashed] (-45:3mm) arc (-45:315:3mm);

    \draw[dotted, thick] plot [smooth,tension=.6] coordinates{(-60-0*15:3mm) (-60-1*15:8mm) (-60-2*15:13mm) (-60-3*15:16mm) (-60-4*15:18mm) (-60-5*15:18mm)
    (-60-6*15:17mm) (-60-7*15:16mm)(-60-8*15:15mm) (-60-9*15:14mm) (-60-10*15:13mm)(-60-11*15:11.5mm) (-60-12*15+4:10mm)};

    \node at (2.2cm,1.0cm) {$z$};
    \draw[thick,xshift=2cm,yshift=.8cm] (90:0mm) -- (90:4mm);
    \draw[thick,xshift=2cm,yshift=.8cm] (0:0mm) -- (0:4mm);

    \node at (0:15mm) {$\pi_1$};
    \node at (120:14mm) {$\tilde\pi_2$};
    \node at (240:14mm) {$\pi_3$};

    \end{scope}

    \begin{scope}[shift={(10cm,0cm)}]
    \fill[color=red] (0:10mm) circle (1mm);
    \draw[dotted,thick] (0:3mm) -- (0:9mm);
    \fill[color=red] (120:10mm) circle (1mm);
    \draw[dotted,thick] (120:3mm) -- (120:9mm);
    \fill[color=red] (240:10mm) circle (1mm);
    \draw[thick] (-1mm,1mm) -- (1mm,-1mm);
    \draw[thick] (1mm,1mm) -- (-1mm,-1mm);
    \draw[thick,color=ForestGreen,->,dashed] (-45:3mm) arc (-45:315:3mm);

    \node at (2.2cm,1.0cm) {$z$};
    \draw[thick,xshift=2cm,yshift=.8cm] (90:0mm) -- (90:4mm);
    \draw[thick,xshift=2cm,yshift=.8cm] (0:0mm) -- (0:4mm);

    \draw[dotted, thick] plot [smooth,tension=.6] coordinates{(60+0*15:3mm) (60+1*15:8mm) (60+2*15:13mm) (60+3*15:16mm) (60+4*15:18mm) (60+5*15:18mm)
    (60+6*15:17mm) (60+7*15:16mm)(60+8*15:15mm) (60+9*15:14mm) (60+10*15:13mm)(60+11*15:11.5mm) (60+12*15-4:10mm)};

    \node at (0:15mm) {$\pi_1$};
    \node at (120:14mm) {$\pi_2$};
    \node at (240:14mm) {$\tilde\pi_3$};

    \end{scope}
  \end{tikzpicture}
\end{equation}
where the figure on the left-hand side corresponds to the set
$Z_l=\{\pi_1,\pi_2,\pi_3\}$ by following the direction of the green
arrow encircling the base point.  The PL-monodromy on an arbitrary one-cycle $\gamma$ is $(w_{\pi_3}
\circ w_{\pi_2} \circ w_{\pi_1})(\gamma)$. Suppose one chooses a path which goes
around $p_3$ before $p_2$. There are two distinct choices, given by
the center and right diagrams. Consider the center diagram. Since the
path to $p_3$ has not changed, it must have the same vanishing cycle
$\pi_3$, while the vanishing cycle associated to $p_2$ is now $\tilde
\pi_2 = w_{\pi_3}(\pi_2) = \pi_2 + \pi_2 \cdot \pi_3 \,\,\, \pi_3$ and $Z_c=\{\pi_1, \pi_3, \tilde \pi_2\}$.
A simple computation shows that $w_{\tilde \pi_2}= w_{\pi_{3}} \circ w_{\pi_{2}} \circ w^{-1}_{\pi_{3}}$,
and therefore $w_{Z_c}=w_{\tilde \pi_{2}} \circ w_{\pi_{3}} \circ w_{\pi_1}= w_{Z_l}$. A similar analysis
with $\tilde \pi_3 = w_{\pi_2}^{-1}(\pi_3)$ and $Z_r=\{\pi_1, \tilde \pi_3, \pi_2 \}$ gives $w_{Z_r} = w_{Z_l}$. We see that the monodromy is invariant
under changes of path in which two adjacent points in the path ordering are swapped.

Any change in path can be obtained by successive swaps, and therefore
the PL-monodromy is invariant under such manipulations. Therefore, for
a deformation of any ADE singularity the choice of an ordered set $Z$
is not unique, but instead all such $Z$ fall into an equivalence class
with the same PL-monodromy; they are related by an elementary transformation
of strong bases \cite{GuseinZade}. In particular, given a deformed ADE
singularity the homology cannot depend on the choice of path which
determines $Z$. Since $Z$ determines the junction basis this has
important consequences for junctions: given two orderings $Z$ and
$\tilde Z$ in the same equivalence class, the corresponding
intersection matrices $I$ and $\tilde I$ will differ, as will the
vectors in $\bZ^N$ representing the junctions; however, they will have
the same lattice structure.  For example, we will see that certain
one-parameter deformations of ADE singularities do not give rise to
the canonical orderings $Z$ of Zwiebach and DeWolfe, but
instead determine different sets $Z$. In both cases, though,
junctions $J$ of with $(J,J)=-2$ and $a(J)=0$ give rise to the root
lattice of the corresponding ADE algebra, despite giving
different embeddings of the roots in $\bZ^N$.

Let us
comment on the product $(J,J)$. We would like to show that it is
invariant under a swapping, and therefore under choice of
path. Consider a junction $J=(J_1,J_2)$ with vanishing cycles
$\{\pi_1,\pi_2\}$. From (\ref{eq:(J,J)}), we see that $(J,J) = -J_1^2
-J_2^2 + J_1J_2 \,\, \pi_1 \cdot \pi_2$. Performing the path swap so
that the vanishing cycles are given by $\{\pi_4,\pi_1\}$ where $\pi_4 = w_{\pi_1}^{-1}(\pi_2)$, the junction
$J$ becomes $J=(J_2,J_1 + J_2 \,\, \pi_2 \cdot \pi_1)$, and we compute
\begin{align}
  (J,J) &= -J_2^2 - (J_1 + J_2 \,\, \pi_2 \cdot \pi_1)^2
+ [J_2(\pi_2 - (\pi_2 \cdot \pi_1)\,\, \pi_1) \cdot(J_1 + J_2 \,\, \pi_2 \cdot \pi_1)] \,\, \pi_1 \nonumber \\
&= -J_1^2 -J_2^2 +
J_1J_2 \,\, \pi_1 \cdot \pi_2,
\end{align}
showing that the $(J,J)$ is invariant under the swap. A similar calculation
shows that $(J,J)$ is invariant under the other path swap, where the
paths give the ordering of vanishing cycles $\{ \pi_2, \pi_3\}$ with
$\pi_3 = w_{\pi_2}(\pi_1)$;
the junction $J$ is $J=(J_2-J_1\,\, \pi_1\cdot \pi_2, J_1)$ in that basis.

 Finally, many works on string junctions study
the homological equivalence of junctions under so-called Hanany-Witten
\cite{Hanany:1996ie} moves. The basic idea is simple. Consider a
junction $J$ with a junction point, as discussed above, and suppose
that one of the prongs contributing to the junction point ends on a
marked point with vanishing cycle $\pi_i$. The prong ending on it can
disappear by moving the path $j_1$ in $B$ which determines the
junction such that it crosses the marked point and is now a path
$j_2$. In such a case the asymptotic charge is left invariant because
the contribution from the lost prong is compensated for by the
Picard-Lefschetz action of $\pi_i$ on the one-cycle above $j_2$. The
two-cycles in $X$ determined by $j_1$ and $j_2$ are homologically
equivalent. Without loss of generality, in this paper we study
homological representatives of junctions with the maximal number of
prongs; i.e. where there is no Picard-Lefschetz action on the
one-cycle above the corresponding path in $B$. This is natural for the
deformations we consider and makes Lie algebraic computations
considerably easier. See Figure 3 of \cite{Cvetic:2011gp} for a recent
depiction of the equivalence of junctions under a Hanany-Witten move.

\section{Deformations and ADE Algebras}
In this section we study surface deformations and ADE algebras using
the techniques discussed in section \ref{sec:deformations and
  junctions}. We begin by giving an ordered set $Z$ for each ADE
algebra which  allows for a junction
analysis. We also perform a one-parameter smooth
deformation of a singularity for each ADE algebra and determine an associated ordered
set of vanishing cycles $Z$ which differs from the canonical examples
in \cite{DeWolfe:1998zf} for $D_r$, $E_6$, $E_7$, and $E_8$; nevertheless, the root lattices are
isomorphic. We then discuss the appearance of non-trivial representations, a map from roots
to weights in the Dynkin basis, and the importance and application of Freudenthal's recursion
formula to junctions.
\label{sec:deformations and ADE}
\subsection{The Canonical Basis}
\label{sec:defm and root systems}

In \cite{DeWolfe:1998zf} particular sets $Z$ are used to study junctions
filling out ADE representations. They are given by
\begin{align}
A_r&: \qquad Z= \{\pi_{A_1}, \,\, \dots \,\, , \pi_{A_{r+1}} \} \nonumber \\
D_r&: \qquad Z= \{\pi_{A_1}, \,\, \dots \,\, , \pi_{A_{r}}, \pi_B, \pi_C \} \nonumber \\
E_6&: \qquad Z= \{\pi_{A_1}, \,\, \dots \,\, , \pi_{A_{5}}, \pi_B, \pi_{C_1},\pi_{C_2} \} \nonumber \\
E_7&: \qquad Z= \{\pi_{A_1}, \,\, \dots \,\, , \pi_{A_{6}}, \pi_B, \pi_{C_1},\pi_{C_2} \} \nonumber \\
E_8&: \qquad Z= \{\pi_{A_1}, \,\, \dots \,\, , \pi_{A_{7}}, \pi_B, \pi_{C_1},\pi_{C_2} \}
\label{eqn:junctionbasis}
\end{align}
where the subscripts on $A$ denote that there are multiple components
with vanishing cycle $\pi_A$, and similarly for $C$. Note that the
number of discriminant components $N$, and thus $\pi_i$'s, satisfy
$N=r+1$ for $A_r$ algebras and $N=r+2$ for the rest.  Given this data,
the monodromy can be easily computed using the Picard-Lefschetz
formula (\ref{eq:picard lefschetz formula}). In this basis, the
associated intersection products $I$ can be computed from the general
formula (\ref{eq:(J,J)}) and are presented in Table \ref{table:ADE
  S}. The results agree with \cite{DeWolfe:1998zf}.  We may refer to
these canonical sets $Z$ in shorthand form as $AAAABC$ or $A^4BC$ for
$D_4$, for example.

\subsection{$A_r$ Deformations}
\label{sec:Ardef}
We now turn to the deformation of $A_r$ algebras.
The local equation of
a Weierstrass model with $A_r$ surface singularities can be written as
\begin{equation}
  \label{eq:ArlocalWeierstrass}
  y^2 = x^3 - 3a^2 x + 2a^3 + z^{r+1}  \qquad \text{with} \qquad a\ne 0.
\end{equation}
We do not consider the most general deformation, but instead a one-parameter deformation
which completely Higgses the gauge group. Such a deformation is given by
\begin{equation}
y^2 = x^3 -3a^2x + 2a^3 + z^{r+1} - \ep \qquad \qquad \text{with}\qquad \ep \ll 1.
\label{eq:Ar deformation}
\end{equation}
and the discriminant takes the form
$\Delta = (z^{r+1}-\ep)(2a^3 + z^{r+1} - \ep)$. Since the deformed $A_r$ singularity
arises entirely from the first factor, we study a neighborhood of $z^{r+1} - \ep=0$,
and the discriminant becomes
\begin{equation}
\Delta \sim \prod_{j=1}^{r+1} (z-\ep_j),
\end{equation}
where $\ep_j\equiv |\ep|^{1/(r+1)} e^{2\pi i \, j/(r+1)}$. For the $SU(4)$ case where $N=r+1=4$ the
discriminant components have been deformed away from $z=0$ as
\begin{equation}
\begin{tikzpicture}[scale=1]

    \fill[thick,color=red] (45:12mm) circle (1mm);
    \draw[dotted,thick] (45:3mm) -- (45:11mm);
    \fill[thick,color=red] (135:12mm) circle (1mm);
    \draw[dotted,thick] (135:3mm) -- (135:11mm);
    \fill[thick,color=red] (225:12mm) circle (1mm);
    \draw[dotted,thick] (225:3mm) -- (225:11mm);
    \fill[thick,color=red] (315:12mm) circle (1mm);
    \draw[dotted,thick] (315:3mm) -- (315:11mm);

    \draw[dotted] (0,0) circle (18mm);
    \draw[thick] (0,0) circle (3mm);
    \draw[thick] (-1mm,1mm) -- (1mm,-1mm);
    \draw[thick] (1mm,1mm) -- (-1mm,-1mm);

    \node at (2.2cm,1.7cm) {$z$};
    \draw[thick,xshift=2cm,yshift=1.5cm] (90:0mm) -- (90:4mm);
    \draw[thick,xshift=2cm,yshift=1.5cm] (0:0mm) -- (0:4mm);
 \end{tikzpicture}
\end{equation}
where an analogous result clearly applies to other values of $r$.

Let us determine the vanishing cycles, as described in section
\ref{sec:picard lefschetz}.  Two of the roots of the right hand side
of (\ref{eq:Ar deformation}) collide at $z=\ep_j$, making the elliptic
fiber singular. It is clear that the same two roots must collide for
any $z = \ep_j$, since the phase data does not enter into (\ref{eq:Ar
  deformation}) due to the $r+1$ power. Thus, each of the $r+1$ marked
points have the same vanishing cycle, which we can take to be $\pi_A$
in a particular $SL(2,\bZ)$ frame, matching the known result
(\ref{eqn:junctionbasis}). With this $Z$, one can proceed with a
junction analysis. The results are as discussed in section \ref{sec:Ar
  interlude}.

\subsection{$D_r$ Deformations}
\label{sec:Drdef}
Let us proceed similarly in the $D_r$ case.
We consider the local equation of a Weierstrass model with $D_r$ singularities
along $z=0$:
\begin{equation}
  \label{eq:DrWeierstrass}
  y^2 = x^3 - 3c^2 z^2\, x + 2c^3 \, z^3 + \mu \, z^{r-1}
\end{equation}
where $y$ and $x$ are again fiber coordinates and $c$, $\mu$ are parameters. This is always the case if $r \geq 5$.
Consider a simple one-parameter deformation which
completely Higgses the gauge group:
\begin{equation}
  \label{eq:DrDeform}
y^2 = x^3 - 3c^2 s^2\, x + 2c^3 \, z^3 + \mu \, z^{r-1} - \ep.
\end{equation}
The discriminant is given by $\Delta = (z^{r-1} - \ep)(4c^3z^3 + z^{r-1} - \ep)$. Defining
$a\equiv 4c^3 + z^{r-4}$ and studying $\Delta$ in a neighborhood of $z^{r-1}-\ep=0$, we
have
\begin{equation}
  \Delta \sim (z^3 - \ep/a) \,\prod_{j=1}^{r-1} (z-\ep_j)
\end{equation}
where the $\ep_j$ are again determined by $\ep$. We see that the
discriminant has split into $N=r+2$ components, which are unique for
generic values of $a$. The $D_r$ gauge theory is completely Higgsed.
In section \ref{sec:monodromy non simply-laced} we explicitly
determine the vanishing cycles in the $D_4$ case. The set
$Z_{D_4}=\{\pi_\alpha,\pi_\beta,\pi_\gamma,\pi_\alpha,\pi_\beta,\pi_\gamma\}$
that we find is different from the set
$Z=\{\pi_A,\pi_A,\pi_A,\pi_A,\pi_B, \pi_C\}$ of the canonical basis
(\ref{eqn:junctionbasis}), but are equivalent under an elementary transformation
of strong bases, as discussed in section \ref{sec:picard lefschetz}.  For
$D_5$ we have derived the set is given by $Z_{D_5} = \{\pi_\alpha,
\pi_\alpha,\pi_\beta,\pi_\gamma,\pi_\alpha,\pi_\beta,\pi_\gamma
\}$. Though we have not derived the generic result directly from
geometry, a natural guess given this pattern is $Z_{D_{4+k}} =
\{\pi_{\alpha_1}, \dots, \pi_{\alpha_k},
\pi_\alpha,\pi_\beta,\pi_\gamma,\pi_\alpha,\pi_\beta,\pi_\gamma \}$; this
matches expectations from the canonical basis, and moreover we have
explicitly checked the root junctions for this $Z_{D_{4+k}}$ up through
the $D_8$ case, finding agreement.

\subsection{$E_6$, $E_7$, and $E_8$ Deformations}
\label{sec:E678def}

In this section we give the surface deformations of $E_6$, $E_7$, and $E_8$.

Consider the deformed local Weierstrass model
\begin{equation}
  \label{eq:E6E8}
  y^2 = x^3 + c\, z^k - 3 \ep\, (x+1)
\end{equation}
where the $E_6$ case has $k=4$ and the $E_8$ case has $k=5$. In the
$\ep \rightarrow 0$ limit the deformation is undone and the
Weierstrass model has singularities along $z=0$. The discriminant is
given by $\Delta = (c\, z^k - 3 \ep)^2 - 4 \ep^3$ and we see that the
discriminant splits into two sets of $k$ marked points, the solutions
of $c\, z^k - 3\ep = \pm 2 \ep^{3/2}$. For generic values of the
parameters the discriminant is non-degenerate and the gauge group is
completely Higgsed. Performing an analysis as in section
\ref{sec:picard lefschetz}, we find that that for this deformation the
ordered set of vanishing cycles are $Z_{E_6} = \{ \pi_\alpha,
\pi_\gamma, \pi_\alpha, \pi_\gamma, \pi_\alpha, \pi_\gamma,
\pi_\alpha, \pi_\gamma \}$ and $Z_{E_8} = \{ \pi_\alpha, \pi_\gamma,
\pi_\alpha, \pi_\gamma, \pi_\alpha, \pi_\gamma\,\pi_\alpha,
\pi_\gamma, \pi_\alpha, \pi_\gamma \}$. We see again that simple
deformations of surface singularities do not reproduce the canonical
junction basis (\ref{eqn:junctionbasis}). Taking specific values for
the one-cycles, $\pi_\alpha = \pi_A$ and $\pi_\gamma = -\pi_C$, a junction
analysis using $Z_{E_6}$ and $Z_{E_8}$ as input data finds $72$ and
$240$ junctions $J$ with $(J,J) = -2$ and $a(J)= (0,0)$, matching the
number of roots of $E_6$ and $E_8$ as expected. Furthermore, in the
$E_6$ case there are $27$ junctions $J$ with $(J,J) = -1$ and $a(J) =
(1,0)$, as expected from the canonical basis $AAAAABCC$. The
intersection matrix $I$ for $Z_{E_6}$ is given by
\begin{equation}
I=  \begin{pmatrix}
    $-1$& $1/2$& $0$& $1/2$& $0$& $1/2$& $0$& $1/2$ \\
$1/2$& $-1$& $-1/2$& $0$& $-1/2$& $0$& $-1/2$& $0$ \\
$0$& $-1/2$& $-1$& $1/2$& $0$& $1/2$& $0$& $1/2$ \\
$1/2$& $0$& $1/2$& $-1$& $-1/2$& $0$& $-1/2$& $0$ \\
$0$& $-1/2$& $0$& $-1/2$& $-1$& $1/2$& $0$& $1/2$ \\
$1/2$& $0$& $1/2$& $0$& $1/2$& $-1$& $-1/2$& $0$ \\
$0$& $-1/2$& $0$& $-1/2$& $0$& $-1/2$& $-1$& $1/2$ \\
$1/2$& $0$& $1/2$& $0$& $1/2$& $0$& $1/2$& $-1$ \\
  \end{pmatrix}
\end{equation}
and the intersection matrix for $Z_{E_8}$ can also be computed easily.
We leave an in-depth analysis of these sets $Z_{E_6}$ and $Z_{E_8}$
for future work.

Let us turn to the $E_7$ case. Consider the deformed Weierstrass equation
\begin{equation}
y^2 = x^3+(3c\, z^3 - 3\ep)\, x + 2 \ep
\end{equation}
which has an $E_7$ singularity along $z=0$ in the $\ep \rightarrow 0$
limit. The discriminant is given by $\Delta = (c\, z^3 - \ep)^3 +
\ep^2$ and components $\Delta_i$ break into three sets of $3$ marked
points, the solutions of $cz^3 - \ep = l$ with $l$ a third root of
$-\ep^2$.  The ordered set of vanishing cycles associated with this
deformation is
$Z_{E_7}=\{\pi_\alpha,\pi_\beta,\pi_\gamma,\pi_\alpha,\pi_\beta,\pi_\gamma,\pi_\alpha,\pi_\beta,\pi_\gamma\}$
and performing a junction analysis with the concrete values $\pi_\alpha =
\pi_A$, $\pi_\beta= \pqvec{0}{1}$, $\pi_\gamma = -\pi_C$, one discovers
there are $126$ junctions $J$ with $(J,J)=-2$ and $a(J)=0$. This
matches the number of roots of $E_7$, as expected.  We leave an
in-depth study of this $Z_{E_7}$ for future work.

\subsection{Non-trivial Representations and Freudenthal's Formula}
\label{sec:non-trivial reps}

Having discussed deformations of ADE singularities and the realization
of ADE root systems as $r$-dimensional lattices in $\bZ^N$, we will now
utilize this formalism to describe non-trivial representations.

The representation theory of Lie algebras is rich. Let us briefly
remind the reader of the basic formalism. A non-trivial representation
$\rho$ of a simple Lie algebra $\g$ is determined by a \emph{highest
  weight vector} $\rho$, clearly abusing notation. Any weight in the
\emph{weight lattice} of $\rho$ can be obtained by subtracting some
number of simple roots $\alpha_i$ from $\rho$. For example, the weight
lattice of the adjoint representation $ad(\g)$ is the root lattice,
and all roots can be obtained by subtracting simple roots from the
highest root.  The algorithm of the previous section\footnote{This
  algorithm was advocated for in \cite{DeWolfe:1998zf} and used to
  great effect, when applicable.}  of searching for junctions $J$ with
$(J,J)=-2$ and $a(J) = \pqvec{0}{0}$ recovered the root lattices of
the ADE algebras precisely because they are simply-laced; that is, all
non-trivial roots $J$ of ADE algebras have the same length $(J,J)=-2$
and multiplicity one. A generic representation does not satisfy this
property, and therefore the previous algorithm is not applicable in
general. In fact, even in the case of root lattices, the roots of the
Cartan subalgebra were added by hand, knowing that they are trivial
weights with multiplicity $rk(\g)$.

It is important to note that a set of junctions associated to a
representation $\rho$ of $\g$ are \emph{not} the weights of $\g$,
since the rank of their span is typically greater than $rk(\g)$;
rather, there is a map from junctions to weights. The former can be
seen in a simple example.  Consider $SU(2)$: the $2$ is composed of two
junctions $J_1 = (1,0)$ and $J_2 = (0,1)$, which are those $J$ with
$(J,J)=-1$ and $a(J)=\pqvec{1}{0}$. Clearly $rk(span_\bZ(J_1,J_2))>rk(\g)$, and
therefore it isn't technically correct to call $\{J_1,J_2\}$ the
weight lattice of the $2$. This is an artifact of
embedding in the higher dimensional space $\bZ^N$.  Specifically, any
vector in $span_\bZ(J_1,J_2)$ proportional to $J_1 + J_2$ doesn't
intersect any of the simple roots, and therefore can't be in the
weight lattice. Taking into account this fact, the rank is correct. In general there will be
$N-r$ such relations, since the roots span an $r$-dimensional subspace
of $\bZ^N$; these relations are derived by computing $ker(R)$,
where $R$ is the $N\times r$ matrix whose columns are the simple roots
junctions.

It is simple to give a generic map from junctions to weights; for
simplicity, we will choose to use the \emph{Dynkin basis}, where the
simple roots are vectors given by the rows of the Cartan matrix. A map
$F$ from a set of junctions $J$ to the weight lattice in the Dynkin
basis is a map $F:\bZ^N \rightarrow \bZ^r$, and this can be computed
easily. Since the intersection of root junctions gives the negative
Cartan matrix $-A$, we have $R^T \, I \,R = -A$, where $I$ is the
intersection product (\ref{eq:(J,J)}). Since the columns of $R$ are the root
junctions in $\bZ^N$, we see that $F = -R^T\, I$ maps the roots to the
Cartan matrix, and therefore their Dynkin labels. In the canonical
basis (\ref{eqn:junctionbasis}) the maps $F$ from junctions to Dynkin labels match the
results of \cite{DeWolfe:1998zf}. They are in Table \ref{table:ADE F} for convenience.
Since this map exists we will often abuse language and refer to junctions
and weights interchangeably.

Much of the junction literature thus far has focused on simple
representations.  In a generic representation, the weights $\lambda$
have a variety of lengths and the multiplicities are non-trivial.  The
multiplicity $m_\lambda$ of arbitrary weight $\lambda$ can be
determined by Freudenthal's formula
\begin{equation}
  [(\rho + \xi)^2 -(\lambda + \xi)^2] \,\, m_\lambda = 2 \sum_{\alpha > 0} \sum_{j \ge 1}
  m_{\lambda + j\alpha},
\label{eqn:freudenthal}
\end{equation}
where $\rho$ is the highest weight and $\xi$ is the Weyl vector $\xi
\equiv \frac{1}{2}\sum_{\alpha > 0}\alpha$, given by half the sum of
the positive roots. Since $m_\rho=1$, the multiplicities of all
weights in a representation can be determined recursively.

If deformation theory and string junctions are to reproduce the
generic structure of ADE algebras, junctions must be able to describe
arbitrary representations. For non-trivial representations
with weights of a variety of lengths and multiplicities, this this
amounts to the question of whether Freudenthal's formula holds for
junctions. The formula (\ref{eqn:freudenthal}) requires a
product on the algebra, which for junctions is given by
(\ref{eq:(J,J)}), and the weight junctions are vectors in
$\bZ^N$, not $\bZ^r$, as emphasized previously. It is simple to check
that the highest root junctions reproduce the root lattices in
$\bZ^N$, as they should, but in lieu of a mathematical proof it is
also important to check non-trivial examples.

We have performed many non-trivial checks in carrying out the
computations in this paper, but let us briefly discuss one that may
convince the reader the Freudenthal's formula can be utilized for
junctions. Consider the case of $\g=E_6$. In the Dynkin basis, the
highest weight of the adjoint representation is $(0,0,0,0,0,1)$ and,
as mentioned, Freudenthal's recursion formula reproduces the correct
root lattice from this data. With the simple roots as given in Table \ref{table:simpleroots},
the highest root is given by $J_{hr} = (1, 1, 1, 1, 0, -2, -1,
-1)$. $E_6$ also has a representation\footnote{See, e.g.  table 47 of
  \cite{Slansky:1981yr} for a list of irreducible $E_6$
  representations of dimension $<100,000$.} with highest weight
$(0,0,0,0,0,3)$ in the Dynkin basis, and this representation has
dimension 43,758. If Freudenthal's formula applies to junctions, it
must compute a representation of dimension $43,758$ when applied to a
highest weight junction of $J_{big} \equiv 3J_{hr}$; indeed, it
does\footnote{The interested reader may consult the computation in the
  code referenced in appendix \ref{sec:code}.}. Similar methods will
be utilized elsewhere in the paper, including discussions of the $126$
dimensional representation of $SO(10)$ often considered in models of
particle physics.

\section{Massless Matter in Codimension Two}
\label{sec:codimension two}
Having discussed the basic formalism in section
\ref{sec:deformations and junctions} and the deformation of ADE singularities and
associated representation theoretic data in section \ref{sec:deformations and ADE}, we
will now turn to discuss massless matter representations.

As we have seen, non-trivial representations of a Lie algebra $\g$ can
be identified with two-manifolds which emanate from the deformation of
a \emph{codimension one} ADE singularity.  Vector multiplets in the
adjoint representation arise as two-spheres $J$ with $(J,J)=-2$; as
they do not have boundary, the asymptotic charge is
$a(J)=(0,0)$. Other representations $\rho$ arise from junctions which
wrap a non-vanishing one-cycle $a(J) \ne (0,0)$ a finite distance from
$z=0$, and thus appear to have a boundary $b$ in this
neighborhood. However, it is possible that $b$ ends on other
seven-brane components $\Delta_o$, giving a two-cycle which may become
massless on the codimension two locus $\Delta_z \cap \Delta_o$. Since
this requires that the boundary $b$ ``pinch off'', the asymptotic
charge $a(J)$ puts constraints on the allowed $\Delta_o$.
We would like to again stress a main point: the data of a representation
$\rho$ of $\g$ arises from codimension one data of the elliptic fibration; the
importance of codimension two data is that it determines whether or not
the two-cycles associated to $\rho$ shrink to zero size on that locus, giving
rise to massless matter.

There are many works
\cite{Witten:1996qb,Katz:1996xe,Bershadsky:1996nh,Aspinwall:2000kf,Grassi:2000we,Morrison:2011mb,Grassi:2011hq}
studying the codimension two localization of matter via singularity
resolution. These works have employed a number of approaches, and in
this section we discuss three of them from the point of view of
deformations. Specifically, we discuss codimension two singularities
arising in $K3$-fibrations, where representations $\rho$ of $\g$ arise
from the branching of adjoints of an enhanced codimension two algebra
$\g'$, as utilized in \cite{Katz:1996xe}.  We study the case of
codimension two singularities with simple normal crossing, and argue
that ideas from $K3$-fibrations can be applied in this context as
well. We also discuss outer automorphisms of $\g$ induced by
O-monodromy around codimension two loci; this phenomenon can break
$\g$ to a non-simply laced group $\g'$ and give non-trivial
representations of $\g'$ in codimension two. We will exhibit all of
these ideas in concrete junction examples in section
\ref{sec:examples}.  Due to the many possibilities, we do not attempt
to present an exhaustive list of codimension two enhancements and
their study via deformation. For further examples in the resolution picture, see
\cite{Morrison:2011mb} or \cite{Grassi:2011hq}.

We emphasize that in all three approaches the appearance of massless matter in
codimension two can be understood in terms of junctions in families of elliptic
surfaces. In these cases the Lie algebra product $(\cdot, \cdot)$ is in fact the topological
intersection of two-cycles, represented as junctions.

\vspace{.5cm}
\noindent \emph{\bf Elliptic K3-fibrations and Heterotic Duality}

Consider the case where $X$ is not only elliptically fibered, but also
admits an elliptic $K3$ fibration $X_{d} \xrightarrow{\pi_K} B_{d-2}$
where the K3 fiber is in the stable degeneration limit. In this case
the elliptic $K3$ splits into two rational elliptic surfaces meeting
along a common elliptic curve. We denote such $X$ as $\tilde X$ for
convenience.  Compactifications of F-theory on $\tilde X$ admit
heterotic duals \cite{Vafa:1996xn,Morrison:1996na,Morrison:1996pp},
where the heterotic compactification manifold $Y$ is an elliptically
fibered Calabi-Yau $(d-1)$-fold over $B_{d-2}$.  Indeed, via an
appropriate specialization in the complex structure moduli space of
$\tilde X$, an $E_8$ gauge theory can be engineered at codimension one
in each of the rational elliptic surfaces.

Heterotic duality necessitates the existence of the string junction picture.
This can be seen as follows. The heterotic dual is endowed with a holomorphic
vector bundle $V_1 \oplus V_2$ on $Y$. The bundle $V_1$ with structure group
$H$ breaks one $E_8$ factor to a group $G= [E_8,H]$. If $V_1$ is trivial the entire
$E_8$ gauge theory is intact and therefore the $240$ W-bosons of $E_8$ are massless;
turning on bundle moduli such that the structure group $H=E_8$, the gauge group is
completely broken and all $240$ W-bosons receive a mass. The F-theory dual of this
process is the complete deformation of the $E_8$ singularity in one of the
rational elliptic surfaces, and therefore $240$ finite volume two-cycles must
arise from deformation. We emphasize that these are \emph{not} the resolution two-cycles,
but instead are junctions represented as vectors in $\bZ^{10}$.

Let us give an explicit description of the (Calabi-Yau) geometry  $\tilde X$ before
discussing the appearance of matter in codimension two. It is defined
by the Weierstrass equation $y^2 = x^3 + f \, x + g$ with $f \in
\Gamma(K_{B_{d-1}}^{-4})$ and $g \in \Gamma(K_{B_{d-1}}^{-6})$ as usual,
but since $\tilde X$ is fibered by elliptic $K3$ the base $B_{d-1}$ is itself
$\bP^1$ fibered over $B_{d-2}$, so that
\begin{equation}
  f = \sum_{a=0}^8 z^a s^{8-a}\,\, f_a \qquad \qquad g = \sum_{b=0}^{12} z^a s^{12-b}\,\, g_b
\end{equation}
where $f_a \in H^{0}(B_{d-2}, K_{B_{d-2}}^{-4}\otimes
\cO_{B_{d-2}}(\tilde \eta)^{\otimes(4-a)})$ and $g_b \in
H^{0}(B_{d-2}, K_{B_{d-2}}^{-6}\otimes \cO_{B_{d-2}}(\tilde
\eta)^{\otimes(6-b)})$
are global sections dependent upon the choice\footnote{The interested reader can find more
details on heterotic F-theory duality, using identical notation, in \cite{Cvetic:2011gp}.} of a divisor
class $\tilde \eta$ in $B_{d-2}$.  Via
the appropriate tuning of complex structure moduli in $f_a$ and $g_b$,
an ADE singularity with group $G$ can be engineered along the
component $\Delta_z\equiv\{z=0\}$ with multiplicity $N$. Suppose that
there is another component of the discriminant $\Delta_t$, defined in
terms of a local coordinate by $t=0$ which intersects $\Delta_z$ in
codimension two, and that the ADE singularity enhances to $G'$ at
$\Delta_z \cap \Delta_t$ with multiplicity $N'$.

In this case it is simple to see how matter arises.  Consider $\tilde X$ in a
neighborhood of $t=0$. This neighborhood specifies a family of
elliptic $K3$'s $F_t$, and a generic $F_t$ has $N$ marked points at
$z=0$ and $N'-N$ marked points away from $z=0$. There exists a deformation
of the $N$ marked points in a generic $F_t$ which gives junctions in the
adjoint of $\g$; they become massless upon undoing the deformation.
Similarly, the $N'$ marked points in the codimension two fiber $F_0$
yield an adjoint of $\g'$. In passing from $F_0$ to generic $F_t$
the $dim(\g)-rk(\g)$ junctions between the $N$ marked points remain
massless, giving a massless adjoint of $\g$, while the other
$(dim(\g') - rk(\g'))-(dim(\g)-rk(\g))$ states become massive; the
latter are the weights in $ad(\g')$ which aren't in $ad(\g)$, and hence
can be studied via branching rules $ad(\g') = ad(\g) \oplus_i \rho_i$. Running
the process in reverse, junctions in the representations $\rho_i$ must become
massless in codimension two.

\vspace{.5cm}
\noindent {\bf \emph{Simple Normal Crossing} }

We have seen that under certain circumstances one can understand the
appearance of matter in codimension two in terms of a family of
elliptic $K3$'s. In this section we would like to use a similar idea
in the case of simple normal crossing. Here the discriminant takes the
form $\Delta \sim z^N\, t^{N'}$, where $z,t$ are local coordinates on
$\bC^2$. If one considers the elliptic fibrations over a
one-dimensional neighborhood of $z=0$ in order to study the $N$
components associated to the algebra $\g$, the other $N'$ components
do not give marked points as they did in the
previous section. This is ultimately an artifact of the choice of neighborhood:
if one considers the elliptic fibration over an appropriate slice
\begin{equation}
  \label{eq:differentslice}
  \begin{tikzpicture}
    \node at (0.0cm,1.7cm) {$z=0$};
    \node at (2.2cm,0.0cm) {$t=0$};
    \draw[color=blue, thick] (-15mm,0mm) -- (15mm,0mm);
  \draw[color=red, thick] (0mm,-15mm) -- (0mm,15mm);
  \draw[thick,dotted] (-15mm,7.5mm) -- (7.5mm,-15mm);
 \end{tikzpicture}
\end{equation}
represented here in dots, the $N + N'$ components all appear as marked
points in this slice. This coordinate change allows one to study the
algebra $\g'$ via junctions. The location of the $N'$ marked points in
relation to the $N$ marked points depends on $t$, and the techniques
of the previous section can be applied in the same fashion.

A schematic picture may help the reader to visualize the process. For an
$A_3$ enhancement to $A_4$, the intersection of
the discriminant with the
the dotted slice is
\begin{equation}
  \begin{tikzpicture}

    \fill[xshift=5mm,color=red] (0:2mm) circle (1mm);
    \fill[xshift=5mm,color=red] (90:2mm) circle (1mm);
    \fill[xshift=5mm,color=red] (180:2mm) circle (1mm);
    \fill[xshift=5mm,color=red] (270:2mm) circle (1mm);
    \fill[xshift=-2mm,color=red] (0:-5mm) circle (1mm);
    \draw[dashed] (0,0) circle (15mm);
    \node at (-7mm,4mm) {$I_1$};
    \node at (6mm,6mm) {$4 \times  I_1$};

    \node at (2.5cm,0cm) {$\xrightarrow{\ep \rightarrow 0}$};
    \fill[xshift=55mm,color=red] (0:0mm) circle (1mm);
    \fill[xshift=48mm,color=red] (0:-5mm) circle (1mm);
    \fill[xshift=50mm,color=red] (0:5mm) circle (1mm);
    \draw[xshift=50mm,dashed] (0,0) circle (15mm);
    \node at (43mm,4mm) {$I_1$};
    \node at (55mm,4mm) {$I_4$};
    \draw[xshift=50mm,thick, color=ForestGreen,->] (0:-6mm) -- (0:4mm);

    \node at (7.5cm,0cm) {$\xrightarrow{z=t=0}$};
    \fill[xshift=100mm,color=red] (0:0mm) circle (1mm);
    \draw[xshift=100mm,dashed] (0,0) circle (15mm);
    \node at (100mm,4mm) {$I_5$};

  \end{tikzpicture}
\end{equation}
where on the left we have $\ep\ne 0$, allowing for a junction analysis
of the $A_4$ algebra which has been Higgsed; in the middle we have the
$\ep=0$ limit where the $SU(4)$ gauge symmetry is restored; and on the
right we have the codimension two locus where the $I_1$ and $I_4$
singularity have collided, enhancing to $I_5$.  The green arrow
represents the motion of the $I_1$ and $I_4$ singularity towards one
another as the plane moves closer towards $z=t=0$. As this happens,
junctions stretching form the $I_4$ point to the $I_1$ point shrink to
zero size, giving a massless fundamental and an
antifundamental of $SU(4)$ there.

\vspace{.5cm}
\noindent {\bf \emph{O-Monodromy and Non-Simply-Laced Algebras}}

Consider the locus $\Delta_z$. The generic fiber above this locus is
singular with algebra $\g$. The resolution of singularities gives rise
to rational curves representing the extended Dynkin diagram of $\g$,
and for some elliptic fibrations these rational curves can be mapped
to one another by taking a closed path around a codimension two locus
$t=0$. This ``O-monodromy'' induces an outer automorphism on $\g$.
Since some nodes of the Dynkin diagram, and therefore $M2$-brane
states wrapped on them, cannot be distinguished under the monodromy,
one must take the quotient of $\g$ by the automorphism group, reducing
$\g$ to some non-simply-laced algebra $\g'$. This phenomenon is
well-known in F-theory \cite{Bershadsky:1996nh}.

A natural question is whether a similar phenomenon holds in the deformation
picture. If so, the O-monodromy action must act on two-cycles associated with
deformation; i.e. on string junctions. It is simple to imagine how
this might occur: considering a family of elliptic surfaces parameterized by
codimension two data as discussed above, movement around a codimension two
$t=0$ locus could permute deformed discriminant components and / or the one-cycles
vanishing above them, inducing an action on the junction lattice. See section
\ref{sec:monodromy non simply-laced} for an explicit example of this phenomenon.

\section{Illustrative Examples}
\label{sec:examples}
In this section we study explicit examples which demonstrate the
localization of massless matter representations in codimension two.
We demonstrate the localization of matter in the deformation picture
arising at a codimension two enhancement $A_r \rightarrow A_{r+1}$. We
describe antisymmetrics of $SU(4)$ and sixteen dimensional spinors of
$SO(10)$ via branching rules from higher rank algebras. We perform an
explicit analysis of O-monodromy describing the reduction of $D_4$ to
$G_2$.  We also comment on a number of
representations which are studied in models of particle physics.

To build intuition, we will not only study examples where the results are obtained
directly from deformation, but will also study examples
which use the canonical basis (\ref{eqn:junctionbasis}). The latter sometimes has the advantage that
it is clear how the breaking of groups occurs, or coincides with IIb intuition.

\subsection{Matter from Branching Rules}
In section \ref{sec:codimension two} we described how massless codimension two
matter can be understood by studying junctions in families of elliptic fibrations over a patch
which depend on a parameter $t$. At $t=0$ a fiber collision enhances the algebra from
$\g$ to $\g'$ and the representations of $\g$ at $z=t=0$ can be determined from branching
rules. We demonstrate this explicitly in examples.

\label{sec:branching}

\vspace{.5cm}
\noindent {\bf \emph{Rank Enhancement}}

Before proceeding on to more complicated examples, it is worthwhile to
discuss a simple example. Consider the Weierstrass equation
\begin{equation}
  y^2 = x^3 - 3a^2 \, x + 2a^3 + (z^{r+1}-\ep) \, t.
\end{equation}
In the $\ep \rightarrow 0$ limit there is an $A_r$ singularity along $z=0$ which enhances
to $A_{r+1}$ at $z=t=0$.  The discriminant is given by $\Delta = (z^{r+1}-\ep)\,\,\,t \,\,\,(2a^3 + (z^{r+1}-\ep) t)$; near  $t=0$, $z^{r+1} = \ep$ we have
\begin{equation}
  \Delta \sim \prod_j^{r+1} (z-\ep_j)\, t := \Delta_t \,\, \prod_j \Delta_{z_j}
\end{equation}
and the simple normal crossing of each of the $r+1$
$\Delta_{z_j}$ components with the $\Delta_t$ component is evident.
In the canonical junction basis the $A_r$ singularity is described by $r+1$
$A$-branes and at $t=0$ this stack collides with another $A$-brane,
enhancing the algebra to $A_{r+1}$.
Consider the simple
example $r=1$, where $\g = SU(2)$ and $\g'=SU(3)$.
The roots of $\g'$ are given by
\begin{align}
  v_1 &= (1,0,-1) \qquad
  v_2 = (1,-1,0) \qquad
  v_3 = (0,1,-1) \qquad
  v_4 = (0,0,0) \nonumber \\
  v_5 &= (0,0,0) \,\,\,\,\, \qquad
  v_6 = (0,-1,1) \qquad
  v_7 = (-1,1,0) \qquad
  v_8 = (-1,0,1).
\end{align}
Upon moving from $F_0$ to $F_t$, $N'-N=1$ of the A-branes moves away
from $z=0$; choose it to be the first in the basis for
convenience. Then $v_3$, $v_4$, and $v_6$ still give massless two
cycles in the adjoint of $SU(2)$, and the extra Cartan element $v_5$
of $SU(3)$ is a singlet of $SU(2)$. The pairs $\{v_1,v_8\}$ and
$\{v_2,v_7\}$ fill out a $2$ of $SU(2)$ which is massive away from
$t=0$ due to the separation between the $N$ marked point of the
$SU(2)$ algebra and the extra marked point of the Higgsed $SU(3)$
algebra.  These multiplets become massless upon taking $t \rightarrow
0$, matching the well known result that fundamentals of $SU(2)$ are
localized at codimension two points of $SU(2) \rightarrow SU(3)$
enhancement.  This same type of analysis holds for the generic
$A_r\rightarrow A_{r+1}$ enhancement given above.

Let us consider examples purely using intuition from
junctions.  Consider a geometry which realizes an $A_3$ singularity
along $z=0$, but enhances to $D_4$ at $z=t=0$. In the canonical junction basis the $A_3$ is
represented\footnote{In the weakly coupled type IIb limit this
  geometry describes the intersection of $4$ D7-branes and an
  O7-plane, which can be obtained by unfolding a stack of D7-branes
  with $SO(8)$ gauge symmetry.} by $AAAA$ and $D_4$ is represented by
$AAAABC$, and thus $AAAA$ collides with $BC$ in codimension two. The
adjoint representation of $D_4$ has dimension $28$. The associated
roots are presented in table \ref{table:D4 roots}, and the coloring in
the table shows how the the adjoint of $D_4$ branches into
representations of $A_3$. For example, the highest root of $D_4$ is
$(1,1,0,0,-1,-1)$, and studying the representation of $A_3$ requires
dropping the last two entries of the junction, since they don't end on
the first four entries which give the $AAAA$ of $A_3$ near a common
codimension one locus. Doing so, we see the junction $(1,1,0,0)$ which
has asymptotic charge $(2,0)$; studying this representation using
Freudenthal's formula and the $A_3$ algebra shows that it is the
$\bf{6}$ of $SU(4)$. Another simple example of branching is given in
table \ref{table:E6 roots}.  There we consider a $D_5$ enhancement to
$E_6$ in codimension two, represented in the canonical junction basis
by $AAAAABC \rightarrow AAAAABCC$, i.e. a $C$-brane collides with the
$D_5$ singularity in codimension two. The highest root of $E_6$ is
$(1,1,1,1,0,-2,-1,-1)$; forgetting the last coordinate in order to
study junctions in representations of $D_5$ gives $(1,1,1,1,0,-2,-1)$,
which is the highest weight of a $\bf{16}$ of SO(10).

\subsection{O-monodromy and Reducing $D_4$ to $G_2$}
\label{sec:monodromy non simply-laced}

In this section we study an example where the generic fiber in $\Delta_z$
exhibits a $D_4$ singularity, but monodromy around the codimension two locus $z=t=0$
induces an outer automorphism on the $D_4$ Dynkin diagram which reduces the gauge
group to $G_2$. While the full automorphism group of $D_4$ is $S_3$, it is easy to see
\begin{equation}
\begin{tikzpicture}[scale=.4]
  \draw[thick] (-30: 3mm) -- (-30: 7mm);
  \draw[thick] (210: 3mm) -- (210: 7mm);
  \draw[thick] (90: 3mm) -- (90: 7mm);
  \draw[thick] (-30: 10mm) circle (3mm);
  \draw[thick] (210: 10mm) circle (3mm);
  \draw[thick] (90: 10mm) circle (3mm);
  \draw[thick] (0: 0mm) circle (3mm);
  \draw[thick,color=red,->] (60:10mm) arc (60:0:10mm); 
  \draw[thick,color=red,->] (-60:10mm) arc (-60:-120:10mm);
  \draw[thick,color=red,->] (-180:10mm) arc (-180:-240:10mm);
  \draw[xshift=3cm,->,thick] (0:0mm) -- (0:15mm);
  \draw[xshift=7cm,thick] (0,0) circle (3mm);
  \draw[xshift=7cm,thick] (0: 15mm) circle (3mm);
  \draw[xshift=7cm,thick] (0: 3mm) -- +(9mm,0);
  \draw[xshift=7cm,thick] (45: 3mm) -- +(10.6mm,0);
  \draw[xshift=7cm,thick] (-45: 3mm) -- +(10.6mm,0);
  \draw[xshift=7.85cm,thick] (120:0mm) -- (120:4mm);
  \draw[xshift=7.85cm,thick] (-120:0mm) -- (-120:4mm);
\end{tikzpicture}
\label{eqfig:D4G2Dynkin}
\end{equation}
that the quotient of $D_4$ by the $Z_3$ subgroup of $S_3$ yields $G_2$ at
the level of Dynkin diagrams and Lie algebras. This is realized
geometrically by the smooth resolution of the local Weierstrass
equation
\begin{equation}
  y^2 = x^3 - 3\,c^2\,z^2\,\,x+2c^3z^3+a\,t\,z^3
  \label{eqn:D4G2WeierstrassSingular}
\end{equation}
with its outer monodromy in $H_2(K,\bZ)$, where $K$ is a
general surface intersecting the discriminant locus $z=0$. In $B$
there is a simple normal crossing intersection of $z=0$ and $t=0$;
the general fiber over $z=0$ is $I^*_0$ while over a general point
of $t=0$ it is $I_1$. The intersection of the discriminant with the $Re(t)-Re(z)$ plane
is
\begin{center}
  \begin{tikzpicture}[scale=.7]
    \draw[thick, ->] (0cm,0cm) -- (6cm,0cm);
    \draw[thick, dotted] (0cm,1.5cm) -- (7cm,1.5cm);
    \draw[thick, ->] (0cm,0cm) -- (0cm,3cm);
    \node at (6.2cm,0cm) {$z$};
    \node at (0cm,3.6cm) {$t$};
    \node at (8.5cm,1.5cm) {$t\ne0$, fixed};

    \draw[dotted,thick] (1.2cm,0cm) -- (1.2cm, 3cm);
    \draw[dotted,thick] (2.9cm,0cm) -- (2.9cm, 3cm);
    \draw[dotted,thick] (5.4cm,0cm) -- (5.4cm, 3cm);

    \fill[color=ForestGreen] (0cm,0cm) circle (1mm);

    \fill[color=ForestGreen] (1.2cm,0cm) circle (1mm);
    \fill[color=ForestGreen] (2.9cm,0cm) circle (1mm);
    \fill[color=ForestGreen] (5.4cm,0cm) circle (1mm);

    \fill[color=red] (1.2cm,1.5cm) circle (1mm);
    \fill[color=red] (2.9cm,1.5cm) circle (1mm);
    \fill[color=red] (5.4cm,1.5cm) circle (1mm);

    \fill[color=blue] (2.2cm,1.5cm) circle (1mm);
    \fill[color=blue] (3.8cm,1.5cm) circle (1mm);
    \fill[color=blue] (5.8cm,1.5cm) circle (1mm);

    \draw[color=blue,thick] (.06cm,.06cm) -- (2.2cm,1.51cm);
    \draw[color=blue,thick] (2.2cm,1.5cm) arc (135:45:1.15cm);
    \draw[color=blue,thick] (3.8cm,1.5cm) arc (-135:-45:1.42cm);
    \draw[color=blue,thick] (5.8cm,1.5cm) -- (6.5cm,2.3cm);
  \end{tikzpicture}
\end{center}
where the blue dots come $t \ne 0$ fixed coalesce at $t=0$ while the
red dots remain fixed for all $t$.

The question we address in this section is whether the appearance of
$G_2$ can be seen in the deformation picture. If it is possible, the
monodromy around $t=0$ must induce an action on the six vanishing
cycles associated with $D_4$, and the induced action on the
string junctions representing simple roots must permute the outer legs of
the Dynkin diagram while keeping the central node fixed.  The relevant
local Weierstrass equation, after deformation, is given by
\begin{equation}
  y^2 = x^3 - 3\,c^2\,z^2\,\,x+2c^3z^3+a\,t\,z^3+\ep\, t
  \label{eqn:D4G2Weierstrass}
\end{equation}
where $x$ and $y$ are local coordinates of the ambient space
containing the elliptic fiber, $c$ and $a$ are parameters of the
elliptic fibration, $z$ and $t$ are local coordinates in $B$ and $\ep$
is a deformation parameter. We denote the fiber at a given $z$ and $t$ by $E_{z,t}$.
The discriminant is given by
\begin{equation}
  \Delta = [(a\, t+ 4\, c^3) z^3 + \ep t] \, \left( a\, z^3 + \ep\right)\,\,t.
\end{equation}
For a given $t\ne 0$, there are two sets of three singular points in
the $z$-plane. Three are located at $az^3 + \ep =0$ and the
corresponding critical points in the Weierstrass model are $x=c \,z$,
$y=0$; the other three occur at ${(4\,c^3 + a\,t)}z^3 + {\ep\, t}=0$
with corresponding critical points $x=-c\,z$, and $y=0$.  In the
$z$-plane for a given $ t \neq 0$, let the former be denoted by red
dots and the latter by blue dots.  For appropriate\footnote{A
  Mathematica notebook which demonstrates these phenomena is included
  in the supplementary files.} choices of $a$, $c$  (e.g. for $a,
c$ real, $a < 0\,,$ $c > 0$) and fixed real $t > 0$ small enough,
the marked points appear in the $z$-plane as
\begin{center}
  \begin{tikzpicture}[scale=1]
    \fill[thick,color=red] (180:20mm) circle (1mm);
    \draw[dotted,thick] (180:3mm) -- (180:19mm);
    \fill[thick,color=red] (180-120:20mm) circle (1mm);
    \draw[dotted,thick] (180-120:3mm) -- (180-120:19mm);
    \fill[thick,color=red] (180+120:20mm) circle (1mm);
    \draw[dotted,thick] (180+120:3mm) -- (180+120:19mm);
    \fill[thick,color=blue] (0:10mm) circle (1mm);
    \draw[dotted,thick] (0:3mm) -- (0:9mm);
    \fill[thick,color=blue] (0-120:10mm) circle (1mm);
    \draw[dotted,thick] (0+120:3mm) -- (0+120:9mm);
    \fill[thick,color=blue] (0+120:10mm) circle (1mm);
    \draw[dotted,thick] (-120:3mm) -- (-120:9mm);
    \draw[thick] (0,0) circle (3mm);
    \draw[thick] (-1mm,1mm) -- (1mm,-1mm);
    \draw[thick] (1mm,1mm) -- (-1mm,-1mm);

    \node at (2.2cm,1.7cm) {$z$};
    \draw[thick,xshift=2cm,yshift=1.5cm] (90:0mm) -- (90:4mm);
    \draw[thick,xshift=2cm,yshift=1.5cm] (0:0mm) -- (0:4mm);

    \fill[xshift=7cm,thick,color=xdotcolor] (180:10mm) circle (1mm);
    \fill[xshift=7cm,thick,color=xdotcolor] (180-120:10mm) circle (1mm);
    \fill[xshift=7cm,thick,color=xdotcolor] (180+120:10mm) circle (1mm);
    \node at (9.2cm,1.7cm) {$x$};
    \draw[xshift=7cm,thick,->] (180:10mm)+(30:1.3mm) -- +(30:16mm);
    \draw[xshift=7cm,thick,->] (180-120:10mm)+(-90:1.3mm) -- +(-90:16mm);
    \draw[xshift=7cm,thick,->] (180+120:10mm)+(150:1.3mm) -- +(150:16mm);
    \node at (6.6cm,0.7cm) {$\pi_\alpha$};
    \node at (8cm,0cm) {$\pi_\beta$};
    \node at (6.6cm,-0.7cm) {$\pi_\gamma$};
    \draw[xshift=9cm,thick,yshift=1.5cm] (90:0mm) -- (90:4mm);
    \draw[xshift=9cm,thick,yshift=1.5cm] (90:0mm) -- (90:4mm);
    \draw[xshift=9cm,thick,yshift=1.5cm] (0:0mm) -- (0:4mm);
  \end{tikzpicture}
\end{center}
where we choose the cross point at $z=0$ to be the base point $P$ and
the solid circle near $P$ denotes the base point neighborhood
discussed in section \ref{sec:picard lefschetz}. The right-hand side
gives a depiction of the complex $x$ line at $P$; the green marked
points are solutions of $x^3+ t\,\ep=0$. The smooth fiber $E_{0,t}$ is
given by the equation $y^2=x^3+ t\,\ep$, a branched cover of the
complex $x$ line branched at the three green points (together with
infinity).  Consider a path in $x$ which meets roots only at the
endpoints of the path; the inverse image in $E_{0,t}$ determines a
closed (real) curve. For example the indicated paths give elements
$\pi_{\alpha}$, $\pi_{\beta}$ and $\pi_{\gamma}$ in the first homology
$H_1(E_{0,t}; \bZ)$; they satisfy the equation $\pi_{\alpha}+
\pi_{\beta}+ \pi _{\gamma}=0$ and any two of them form a basis.  We
can fix coordinates in $H_1(E_{0,t}; \bZ)$ so that
$\pi_\alpha=\pqvec{1}{0}$, $\pi_\beta=\pqvec{0}{1}$ and $\pi_\gamma =
\pqvec{-1}{-1}$\,. We also have, with the suitable (complex)
orientation, $\pi_{\alpha}\cdot\pi_{\beta} =
\pi_{\beta}\cdot\pi_{\gamma} = \pi_{\gamma}\cdot\pi_{\alpha} = 1.$

 Now suppose that we vary $z$--- holding $t$ fixed --- along some  path from zero to one of the six singular points in the $z$-plane without crossing any
other of these points. As this happens, the roots of the equation in $x$
\begin{equation}x^3 - 3\,c^2\,z^2\,\,x+2c^3z^3+a\,t\,z^3+\ep\, t =
  0\end{equation} will move from the green marked points until, in the
singular point, two of them merge. This determines a path between
these two points. The corresponding loop in $E_{0,t}$ shrinks to a
point as $z$ moves along this path from the origin to the singular
point, and the homology class is the associated vanishing cycle. For
example, approaching the uppermost red point along the dotted line
from $z=0$, the two roots on the edge labeled $\pi_\alpha$
coalesce. Furthermore, the path by which they join is homotopic
relative endpoints\footnote{i.e. the endpoints do not move through the
  homotopy.}, via paths that also can only touch green points at their
endpoints, to the indicated path in the $x$-plane joining these two
points; therefore at this point the cycle $\pi_\alpha$ vanishes.
Similarly, moving counterclockwise around the diagram beginning with
the upper right red point, one can compute the rest of the vanishing
cycles using straight line paths from the origin to the critical
points. This determines the ordered set of vanishing cycles to be
$Z=\{\pi_\alpha, \pi_\beta, \pi_\gamma, \pi_\alpha, \pi_\beta,
\pi_\gamma\}$.

To determine the vanishing cycle one technically has to specify the
order of the roots coming together, otherwise there is an ambiguity of
sign. However, if a cycle vanishes at a singular point, so does its
negative, and in the Picard-Lefschetz formula (\ref{eq:picard lefschetz formula}) the sign of the
vanishing cycle has no effect on the local monodromy; therefore we
will take the signs as indicated.  Via successive application of the
Picard-Lefschetz formula the monodromy around the entire set $Z$
is then computed to be $w_S(\lambda) \equiv w_{\pi_{\gamma}}\circ
w_{\pi_{\beta}} \circ w_{\pi_{\alpha}} \circ w_{\pi_{\gamma}} \circ
w_{\pi_{\beta}} \circ w_{\pi_{\alpha}} (\lambda) = -\lambda$. This is
precisely the monodromy of a $D_4$ singularity. However, the basis
differs from the AAAABC case\footnote{See the nice work
  \cite{Bonora:2010bu} for an example obtaining $D_4$ from $G_2$ in
  the AAAABC basis. Unlike that work, however, the $\bZ_3$ action we
  study is dictated by geometry and a different interpretation of the
  outer monodromy in terms of a codimension two locus.}.

Let us study string junctions using the ordered set of vanishing
cycles $Z$. In this basis, the intersection product can be easily
computed from equation (\ref{eq:(J,J)}). It is given by
\begin{equation}
I=\begin{pmatrix}
  $-1$&$1/2$&$1/2$&$0$&$-1/2$&$1/2$ \\
$1/2$&$-1$&$-1/2$&$1/2$&$0$&$-1/2$ \\
$1/2$&$-1/2$&$-1$&$1/2$&$-1/2$&$0$ \\
$0$&$1/2$&$1/2$&$-1$&$1/2$&$-1/2$ \\
$-1/2$&$0$&$-1/2$&$1/2$&$-1$&$1/2$ \\
$1/2$&$-1/2$&$0$&$-1/2$&$1/2$&$-1$
\end{pmatrix}.
\end{equation}
There are $24$ junctions $J \in \bZ^6$ with $a(J) = \pqvec{0}{0}$ and
$(J,J) = -2$. Call this set $R$. If $R$ are the roots of $D_4$, then
there must exist $4$-element subsets of $R$ which generate
$12$-element subsets of $R$ as positive linear combinations; that is,
there must be simple roots which generate positive roots. The number
of such sets must equal $|W_{D_4}|=2^3\cdot 4!=192$, the order of the
Weyl group of $D_4$. Direct computation shows that this is the case,
and the results are presented in Table \ref{table:Weyl D4}. Take the first
of these sets to be the simple roots:
\begin{align}
  \alpha_1 &= (0, 0, 0, -1, -1, -1) \qquad \qquad \alpha_2 = (0, 0, -1, 0, 0, 1) \nonumber \\ \alpha_3 &=  (0, -1, 0, 0, 1, 0) \qquad \qquad
  \,\,\,\,\,\,\,\,\,\alpha_4 =  (-1, 0, 1, 1, 0, -1).
\end{align}
In this Weyl Chamber, the highest root is given by the junction $J =
(-1,-1,-1,0,0,0)$. A computation using Freudenthal's recursion
formula (\ref{eqn:freudenthal}) gives the correct level diagram.

Having established that junctions correctly give the $D_4$ algebra via the deformation (\ref{eqn:D4G2Weierstrass})
for a given $t$, let us now study the influence of encircling the locus $t=0$, parameterizing $t$ as $t= |t| e^{i\varphi}$.
Clearly the depictions of the $z$-plane and $x$-plane above are identical at $\varphi$ a multiple of $2\pi$, but varying $\varphi$
continuously between $2\pi n$ and $2\pi (n+1)$ could change the ordered set of vanishing cycles. Varying $\varphi$ between
$0$ and $2 \pi$, we see the action can be represented in the previous schematic diagram as follows:
\begin{center}
  \begin{tikzpicture}[scale=1]
    \fill[thick,color=red] (180:20mm) circle (1mm);
    \draw[dotted,thick] (180:3mm) -- (180:19mm);
    \fill[thick,color=red] (180-120:20mm) circle (1mm);
    \draw[dotted,thick] (180-120:3mm) -- (180-120:19mm);
    \fill[thick,color=red] (180+120:20mm) circle (1mm);
    \draw[dotted,thick] (180+120:3mm) -- (180+120:19mm);
    \fill[thick,color=blue] (0:10mm) circle (1mm);
    \draw[dotted,thick] (0:3mm) -- (0:9mm);
    \fill[thick,color=blue] (0-120:10mm) circle (1mm);
    \draw[dotted,thick] (0+120:3mm) -- (0+120:9mm);
    \fill[thick,color=blue] (0+120:10mm) circle (1mm);
    \draw[dotted,thick] (-120:3mm) -- (-120:9mm);
    \draw[thick] (0,0) circle (3mm);
    \draw[thick] (-1mm,1mm) -- (1mm,-1mm);
    \draw[thick] (1mm,1mm) -- (-1mm,-1mm);

    \draw[thick,->,color=tmotioncolor] (120+20:10mm) arc (120+20:220:10mm);
    \draw[thick,->,color=tmotioncolor] (240+20:10mm) arc (240+20:340:10mm);
    \draw[thick,->,color=tmotioncolor] (20:10mm) arc (20:100:10mm);

    \node at (2.2cm,1.7cm) {$z$};
    \draw[thick,xshift=2cm,yshift=1.5cm] (90:0mm) -- (90:4mm);
    \draw[thick,xshift=2cm,yshift=1.5cm] (0:0mm) -- (0:4mm);

    \fill[xshift=7cm,thick,color=xdotcolor] (180:10mm) circle (1mm);
    \fill[xshift=7cm,thick,color=xdotcolor] (180-120:10mm) circle (1mm);
    \fill[xshift=7cm,thick,color=xdotcolor] (180+120:10mm) circle (1mm);
    \node at (9.2cm,1.7cm) {$x$};
    \draw[xshift=7cm,thick,->] (180:10mm)+(30:1.3mm) -- +(30:16mm);
    \draw[xshift=7cm,thick,->] (180-120:10mm)+(-90:1.3mm) -- +(-90:16mm);
    \draw[xshift=7cm,thick,->] (180+120:10mm)+(150:1.3mm) -- +(150:16mm);
    \draw[xshift=9cm,thick,yshift=1.5cm] (90:0mm) -- (90:4mm);
    \draw[xshift=9cm,thick,yshift=1.5cm] (90:0mm) -- (90:4mm);
    \draw[xshift=9cm,thick,yshift=1.5cm] (0:0mm) -- (0:4mm);

    \draw[xshift=7cm,color=tmotioncolor,thick,->] (120+20+60:10mm) arc (120+20+60:220+60:10mm);
    \draw[xshift=7cm,color=tmotioncolor,thick,->] (240+20+60:10mm) arc (240+20+60:340+60:10mm);
    \draw[xshift=7cm,color=tmotioncolor,thick,->] (20+60:10mm) arc (20+60:100+60:10mm);
  \end{tikzpicture}
\end{center}
As $\varphi$ goes from $0$ to
$2\pi$ the roots of $x^3 + \epsilon \, t = 0$ rotate through
${2\pi\over 3}\,,$ and the induced action on the torus above $P$ ($y^2 = x^3 +
\epsilon\, t$) acts on homology by carrying $\pi_\alpha$ to
$\pi_\beta$, $\pi_\beta$ to $\pi_\gamma\,,$ and $\pi_\gamma$ back to
$\pi_\alpha\,.$ The three red points in the discriminant locus (in the
$z$-plane) stay fixed, but the three blue ones rotate into one another
in the same way, as indicated.
 By continuity, for each value $\arg t =
\varphi$ (starting from zero and increasing), the two roots
which determine $\pi_\alpha$ at $\varphi=0$ must continue to coalesce as $z$ moves from
zero to the upper red point; however, $\varphi=2\pi$ these same two points determine
$\pi_\beta$ and therefore the vanishing cycle of the upper red dot is $\pi_\beta$.
At the other
two other red dots, going counter clockwise, the vanishing cycles are
$\pi_\alpha$ and $\pi_{\gamma}\,$ for $\varphi=2\pi$. On the other hand, the blue
singular points move around (on a slightly off center circle for small
$|t|$) in the same way as the roots in the $x$-plane, as $\varphi$
goes from zero to $2\pi\,.$ Therefore, starting with the first blue
singular point to the left of the upper right red one and moving
counter clockwise, we now get the vanishing cycles $\pi_\gamma$,
$\pi_\beta$, and $\pi_\alpha\,.$ Altogether, beginning again with the
upper right red dot and moving counterclockwise, the ordered set of
vanishing cycles at $\varphi= 2\pi$ is
$Z_{2\pi}=\{\pi_\beta,\pi_\gamma,
\pi_\alpha,\pi_\beta,\pi_\gamma,\pi_\alpha\}$. Similarly, at $\varphi=
4\pi$ we have $Z_{4\pi} = \{\pi_\gamma, \pi_\alpha,
\pi_\beta,\pi_\gamma, \pi_\alpha, \pi_\beta \}$, and $Z_{0} =
Z_{6\pi}\,,$ so that the effect of the monodromy of going around $t =
0$ is to produce this permutation action of a cyclic group $\bZ_3$ on
the ordered set of vanishing cycles.

Given this action on the geometry we can determine the action on
string junctions and therefore on the $D_4$ algebra. We will present
two arguments which show the reduction of $G_2$ to $D_4$. First we
will present a quick and straightforward argument using the same set
of vectors in $\bZ^6$, but in the different bases $Z_\varphi$; in this
case the ordering in $Z$ begins with the upper red dot and moves
counter-clockwise, with the vanishing cycles changing at different
values of $\varphi$, and in this case the a junction $J$ emanating
from the upper red dot would be $J=(1,0,0,0,0,0)$ for all three values
of $\varphi$. Alternatively, we will present a second and more
detailed argument where the first entry in $Z$ is the red dot with
vanishing cycle $\pi_a$, and the rest follow by counterclockwise
ordering. In this viewpoint, the same junction $J$ emanating from the
upper red dot would be $J_0 = (1,0,0,0,0,0)$, $J_{2\pi}=(0,0,0,0,1,0)$
and $J_{4\pi}=(0,0,1,0,0,0)$ at the respective values of $\varphi$. In
this case a matrix $T$ gives the action on junctions induced by
monodromy around $t=0$.

We begin with the first approach. The three bases $Z_{0}, Z_{2\pi},$
and $Z_{4\pi}$ all have the same intersection product $I$, the same
sets $R$, and all give rise to a $D_4$ algebra.  For these reasons,
any set $\{\beta_1, \beta_2, \beta_3, \beta_4\}$ of vectors in $\bZ^6$
that determine simple roots at $\varphi=0$ also determine a set of
simple roots at $\varphi=2\pi, 4\pi$. Since the Cartan matrix is
identical in all three cases, the central node of the Dynkin diagram
is fixed, but we must also determine if there is any action on the
exterior nodes. We would first like to establish that the outer nodes
of the Dynkin diagram rotate into one another under the monodromy
around $t=0$. This can be done with a simple argument about the eight
dimensional representations, which must be permuted by the
automorphism. Junctions in the $8_v$, $8_s$ and $8_c$ representations
have $(J,J)=-1$ and $a(J)=\pi_\alpha$, $a(J) = \pi_\beta$ and $a(J) =
\pi_\gamma$, respectively. For each phase $\varphi=0,2\pi,4\pi$ of
$t\,,$ a junction ending on the upper red dot,
\begin{center}
  \begin{tikzpicture}[scale=1]
    \fill[thick,color=red] (180:20mm) circle (1mm);
    \draw[dotted,thick] (180:3mm) -- (180:19mm);
    \fill[thick,color=red] (180-120:20mm) circle (1mm);
    \draw[->] (10:0mm) (180-120:20mm) -- +(20:15mm);
    \draw[dotted,thick] (180-120:3mm) -- (180-120:19mm);
    \fill[thick,color=red] (180+120:20mm) circle (1mm);
    \draw[dotted,thick] (180+120:3mm) -- (180+120:19mm);
    \fill[thick,color=blue] (0:10mm) circle (1mm);
    \draw[dotted,thick] (0:3mm) -- (0:9mm);
    \fill[thick,color=blue] (0-120:10mm) circle (1mm);
    \draw[dotted,thick] (0+120:3mm) -- (0+120:9mm);
    \fill[thick,color=blue] (0+120:10mm) circle (1mm);
    \draw[dotted,thick] (-120:3mm) -- (-120:9mm);
    \draw[thick] (0,0) circle (3mm);
    \draw[thick] (-1mm,1mm) -- (1mm,-1mm);
    \draw[thick] (1mm,1mm) -- (-1mm,-1mm);

    \node at (2.2cm,1.7cm) {$z$};
    \draw[thick,xshift=2cm,yshift=1.5cm] (90:0mm) -- (90:4mm);
    \draw[thick,xshift=2cm,yshift=1.5cm] (0:0mm) -- (0:4mm);

  \end{tikzpicture}
\end{center}
for example, has
$(J,J)=-1$, but its asymptotic charge depends on $\varphi$. At
$\varphi=0, 2\pi, 4\pi$ this junction is therefore in the $8_v, 8_s, 8_c$ representation,
respectively, giving the necessary triality permutation. But since the Dynkin
labels of the highest weights of these representations are \cite{Slansky:1981yr}
\begin{equation}
  8_v: (1,0,0,0) \qquad 8_s:(0,0,0,1) \qquad 8_c: (0,0,1,0),
\end{equation}
we see that an action permutes these representations if and only if it
permutes the exterior nodes of the $D_4$ Dynkin diagram.  Therefore we
have the $\bZ_3$ automorphism in (\ref{eqfig:D4G2Dynkin}), and the
monodromy reduces the gauge symmetry from $D_4$ to $G_2$.

Though this argument is brief and correct, we will now use the second
approach and see the direct action of monodromy on the junctions.
This requires performing an analysis using $Z=Z_0$ and
using the action on junctions given by the matrix
\begin{equation}
  T =
  \begin{pmatrix}
    0 & 0 & 1 & 0 & 0 & 0 \\
    0 & 0 & 0 & 1 & 0 & 0 \\
    0 & 0 & 0 & 0 & 1 & 0 \\
    0 & 0 & 0 & 0 & 0 & 1 \\
    1 & 0 & 0 & 0 & 0 & 0 \\
    0 & 1 & 0 & 0 & 0 & 0 \\
  \end{pmatrix},
\end{equation}
which satisfies $T^3=1$. In this picture a junction emanating from the
upper red dot is given by $(1,0,0,0,0,0)$, $(0,0,0,0,1,0)$ and
$(0,0,1,0,0,0)$ at $\varphi=0,2\pi,$ and $4\pi$, respectively. Given
the simple roots $\alpha_i$ as above, $T \alpha_i$ and $T^2 \alpha_i$
give the simple roots at the corresponding $\varphi$; they are
presented in Table \ref{tab:d4omonodromy} for convenience. It is clear
that root junctions are \emph{not} invariant, but in fact $T$ maps the
simple roots to a completely different set of simple roots; we have
moved to a different Weyl chamber.  Automorphism requires that
$\alpha_2$, $T \alpha_2$, and $T^2 \alpha_2$ are the central node of
their respective Dynkin diagrams. This is verified easily by computing
the Cartan matrix at each $\varphi$. At any of these $\varphi$ the
highest weights and next highest weights of the $8_v$, $8_s$, and
$8_c$ representations can be determined. Subtracting them gives a
simple root, and it turns out that $\alpha_4$, $T \alpha_4$ and $T^2
\alpha_4$ are the first simple roots subtracted from the $8_v$, $8_s$,
and $8_c$ representations at the respective $\varphi$ values. Similar
statements hold for $\alpha_1$ and $\alpha_3$, proving that the three
external nodes of the Dynkin are permuted by the $\bZ^3$ action $T$. See
Table \ref{tab:d4omonodromy} for more details.

We have identified the appearance of the $D_4$ algebra and the $\bZ_3$ O-monodromy around a codimension two locus
that reduces it to $G_2$. This is evident in the level diagram for the adjoint of $D_4$
\begin{equation}
  \begin{tikzpicture}
    \draw (0:-5cm) circle (1mm);
    \draw (0:-4cm) circle (1mm);
    \fill (0:-3cm)+(0mm,7mm) circle (1mm);
    \fill (0:-3cm) circle (1mm);
    \fill (0:-3cm)+(0mm,-7mm)  circle (1mm);
    \fill (0:-2cm)+(0mm,7mm) circle (1mm);
    \fill (0:-2cm) circle (1mm);
    \fill (0:-2cm)+(0mm,-7mm)  circle (1mm);
    \fill (0:-1cm)+(0mm,7mm) circle (1mm);
    \fill (0:-1cm) circle (1mm);
    \fill (0:-1cm)+(0mm,-7mm)  circle (1mm);
    \draw (0:-1cm)+(0mm,14mm)  circle (1mm);
    \fill (0:0cm)+(0mm,7mm) circle (1mm);
    \fill (0:0cm) circle (1mm);
    \fill (0:0cm)+(0mm,-7mm)  circle (1mm);
    \draw (0:0cm)+(0mm,14mm)  circle (1mm);
    \draw[thick,dotted,color=ForestGreen] (-53mm,3mm) -- (-47mm,3mm) -- (-47mm,-3mm) -- (-53mm,-3mm) -- cycle;
    \draw[thick,dotted,color=blue] (-3mm,17mm) -- (3mm,17mm) -- (3mm,-10mm) -- (-3mm,-10mm) -- cycle;
    \draw[thick,dotted,color=red] (-13mm,17mm) -- (-7mm,17mm) -- (-7mm,-10mm) -- (-13mm,-10mm) -- cycle;
    \draw (0:5cm) circle (1mm);
    \draw (0:4cm) circle (1mm);
    \fill (0:3cm)+(0mm,7mm) circle (1mm);
    \fill (0:3cm) circle (1mm);
    \fill (0:3cm)+(0mm,-7mm)  circle (1mm);
    \fill (0:2cm)+(0mm,7mm) circle (1mm);
    \fill (0:2cm) circle (1mm);
    \fill (0:2cm)+(0mm,-7mm)  circle (1mm);
    \fill (0:1cm)+(0mm,7mm) circle (1mm);
    \fill (0:1cm) circle (1mm);
    \fill (0:1cm)+(0mm,-7mm)  circle (1mm);
    \draw (0:1cm)+(0mm,14mm)  circle (1mm);
  \end{tikzpicture}
\end{equation}
where the highest root is in a green box, the simple roots are in the
red box, and the Cartan elements are in the blue box. In subtracting
simple roots, one moves from left to right down the diagram. Any set
of three black dots represents roots which permute into one another
under the $\bZ^3$ action; unfilled dots are invariant. This structure
is clear in the Dynkin basis, where the highest weight is $(0,1,0,0)$
and the rest of the roots can be obtained by subtracting rows of the
Cartan matrix in the usual way. For example $(0,1,0,0)$ is
invariant under the permutation of the exterior simple roots $\alpha_1
\rightarrow \alpha_3$, $\alpha_3 \rightarrow \alpha_4$, $\alpha_4
\rightarrow \alpha_1$; the roots at level $3$ are $(0,1,0,0) - \alpha_2 - \alpha_i$
for $i=1,3,4$, which also permute. Similar statements hold for junctions, with the
caveat of the change in Weyl chamber, as discussed. Considering $span_\bZ(\alpha_1,\alpha_3,\alpha_4)$,
the $\bZ_3$ action on this space is induced by the matrix
\begin{equation}
  \begin{pmatrix}
    0 & 0 & 1 \\
    1 & 0 & 0 \\
    0 & 1 & 0
  \end{pmatrix}
\end{equation}
the eigenvalues
of which are the three roots of unity, with the invariant subspace given by elements proportional
to $\alpha_1 + \alpha_3 + \alpha_4$. The eigenvalues $e^{2\pi i /3}$ and $e^{4\pi i /3}$ have corresponding
eigenvectors $(e^{4\pi i /3},e^{2\pi i /3},1)$ and $(e^{2\pi i /3},e^{4\pi i /3},1)$, respectively, with transpose
implied when necessary. Similarly, each set of roots given by three black dots have an invariant subspace; together
with unfilled dots, whose corresponding roots are $\bZ_3$-invariant, there is a $14$-dimensional $\bZ_3$-invariant subspace
of the adjoint of $D_4$. This is the adjoint of $G_2$. Each set of three black dots contributes a two-dimensional
non-invariant subspace, for a total of a $14$ dimensional non-invariant subspace, which can be split into $7\oplus 7$
according to the eigenvalues. In all, we have $Adj(D_4) = 28 = Adj(G_2) \oplus 7 \oplus 7 =  14 \oplus 7 \oplus 7$. At
$t=0$ the monodromy ceases to act and $M2$-branes wrapped on the formerly non-invariant cycles give rise to a hypermultiplet
in the $7$ of $G_2$.

\begin{table}[t]
  \centering
  \scalebox{.8}{\begin{tabular}[h]{c|c|c|c|c}
    \hline $\varphi=0$ & $\alpha_1 = (0, 0, 0, -1, -1, -1)$ & $\alpha_2 = (0, 0, -1, 0, 0, 1)$  & $\alpha_3 = (0, -1, 0, 0, 1, 0)$ & $\alpha_4 = (-1, 0, 1, 1, 0, -1)$ \\
    $\varphi=2\pi$ & $T\alpha_1 = (0, -1, -1, -1, 0, 0)$ & $T\alpha_2 = (-1, 0, 0, 1, 0, 0)$ & $T\alpha_3 = (0, 0, 1, 0, 0, -1)$ & $T\alpha_4 = (1, 1, 0, -1, -1, 0)$ \\
    $\varphi=4\pi$ & $T^2\alpha_1 = (-1, -1, 0, 0, 0, -1)$ & $T^2\alpha_2 =(0, 1, 0, 0, -1, 0)$ & $T^2\alpha_3 =(1, 0, 0, -1, 0, 0)$ & $T^2\alpha_4 =(0, -1, -1, 0, 1, 1)$ \\ \hline
  \end{tabular}}

  \vspace{.5cm}
  \scalebox{.85}{
  \begin{tabular}[h]{r|c|c|c}
    & $\varphi=0$ & $\varphi = 2\pi$ & $\varphi=4\pi$ \\ \hline
    Highest weight & $8_v: (-1, -1, 0, 1, 0, -1)$ & $8_s: (0, 1, 0, -1, -1, -1)$ & $8_c: (0, -1, -1, -1, 0, 1)$ \\
    Next highest weight & $8_v: (0, -1, -1, 0, 0, 0)$ & $8_s: (-1, 0, 0, 0, 0, -1)$ & $8_c: (0, 0, 0, -1, -1, 0)$  \\
    Subtracted root & $\alpha_4$ & $T\alpha_4$ &  $T^2\alpha_4$ \\ \hline
    Highest weight & $8_c: (-1, -1, 0, 0, 0, 0)$ & $8_v: (0, 0, 0, 0, -1, -1)$ & $8_s: (0, 0, -1, -1, 0, 0)$ \\
    Next highest weight & $8_c: (-1, 0, 0, 0, -1, 0)$ & $8_v: (0, 0, -1, 0, -1, 0)$ & $8_s: (-1, 0, -1, 0, 0, 0)$ \\
    Subtracted root & $\alpha_3$ & $T\alpha_3$ & $T^2\alpha_3$ \\ \hline
    Highest weight & $8_s: (0, 0, -1, -1, 0, 0)$ & $8_c: (-1, -1, 0, 0, 0, 0)$ & $8_v: (0, 0, 0, 0, -1, -1)$ \\
    Next highest weight & $8_s: (0, 0, -1, 0, 1, 1)$ & $8_c: (-1, 0, 1, 1, 0, 0)$ & $8_v: (1, 1, 0, 0, -1, 0)$ \\
    Subtracted root & $\alpha_1$ & $T\alpha_1$ & $T^2\alpha_1$ \\ \hline
  \end{tabular}}
\caption{\small The top table gives the action of $T$ on each of the simple roots determined
  at $\varphi=0$, determining the simple roots at $\varphi=2\pi, 4\pi$. Note the Weyl chamber has changed. The first column
  of the bottom table gives the highest and next highest weight junction for each of the $8d$
  representations, together with the associated subtracted root. The data in the second and
  third columns are computed by $T$ and $T^2$ action.}
  \label{tab:d4omonodromy}
\end{table}

\subsection{Matter Representations in Particle Physics}
Many representations of ADE groups are considered in particle physics
models, some of which are more natural than others.  As emphasized
throughout, the ADE representations realized via string junctions are
broader than their weakly coupled type II counterparts. In this
section we discuss phenomenologically relevant representations from
the point of view of string junctions in the canonical basis
(\ref{eqn:junctionbasis}). After recovering well-known facts, we
discuss the realization of certain representations outside of the weak
coupling limit, as well potential difficulties in realizing high
dimensional representations in compact geometries. All Weyl chamber
dependent statements in this section depend on the choice of simple
roots given in Table \ref{table:simpleroots}.

Consider $SU(5)$. The $\bf{5}$ is given by those junctions $J$ with
$a(J) = (1,0)$ and $(J,J) = -1$; the ${\bf \ov 5}$ by $J$ with $a(J) =
(-1,0)$ and $(J,J) = -1$. Their highest weights are given by
$J_5=(1,0,0,0,0)$ and $J_{\ov 5} = (0,0,0,0,-1)$.  From
the type IIb point of view, the asymptotic charges signify that these
representations are formed from fundamental strings coming out of and
into stacks of D7-branes, respectively. The $\bf{10}$ is given by
those $J$ with $a(J) = (2,0)$ and $(J,J) = -2$. The asymptotic charge
$(2,0)$ demonstrates the $\bf{10}$ can end on a $BC$ pair. In the type
IIb limit this becomes an O7-plane and the $\bf{10}$ is localized at a
D7-O7 intersection, as is well-known from CFT quantization and other
techniques. From Table \ref{table:D4 roots} it is easy to see that the junction
$(0,0,0,1,1)$ in the $\bf{10}$ can be interpreted as a branching from
the simple root $(0,0,0,1,1,-1,-1)$ of $SO(10)$; this is the junction
realization of brane unfolding.

Consider $SO(10)$. One generation of the quarks and leptons in the
standard model, together with the right-handed neutrinos, embed into a
single $\bf{16}$ of $SO(10)$. It can be realized by the set of
junctions with $(J,J) = -1$ and $a(J)=(1,1)$.  The highest weight is
$J_{16} = (1, 1, 1, 1, 0, -2, -1)$. From the asymptotic charge
$(1,1)$, we see it is a bound state of a D-string and F-string from
the type IIb perspective, and therefore it does not occur at weak
coupling. Consider $E_6$.  The $\bf{27}$ is given by the set of
junctions with $a(J) = (1,0)$ and $(J,J)=-1$. In contrast to the
$\bf{16}$ of $SO(10)$, this representation has the correct asymptotic
charge for it to end on a D7-brane; however, seven-branes with
gauge symmetry $E_6$ cannot be realized at weak coupling.

Many other representations are considered in particle physics,
frequently exotic particles beyond the standard model introduced to
realize a phenomenological mechanism.  For example, in $SO(10)$ GUT
models this includes the $\bf{126}$ of $SO(10)$, introduced in
\cite{Babu:1992ia} in order to account for small neutrino masses. It
is often remarked that such high-dimensional representations are
difficult to realize in compactifications of string theory, M-theory,
or F-theory; they do not exist in weakly coupled theories with
D-branes, and they cannot embed into an adjoint of $E_8$ in the
heterotic string or F-theory.  They also cannot be realized in the
free field heterotic string \cite{Dienes:1996yh,Dienes:1996wx}.

String junctions can realize a broader spectrum of
possibilities.  It can be checked via Freudenthal's formula that
$J_{126} = 2J_{16} = (2, 2, 2, 2, 0, -4, -2)$ is the highest weight of
the $\bf{126}$. However, we emphasize that though this gives necessary
conditions for holomorphic curves to realize the $\bf{126}$, it does
not mean that such curves are realized in compact Calabi-Yau
varieties. As one considers higher dimensional
representations the maximum $(J,J)$ for junctions in those
representations goes up. This puts conditions on self-intersection
numbers of holomorphic curves. For example, the {\bf 43,758} of $E_6$ has
some junctions with $(J,J) = -18$, and it is unclear whether there exist manifolds
with appropriate curves. For example, in elliptic $K3$ the
holomorphic curves $C$ satisfy $C\cdot C \ge -2$ and thus these higher
representations cannot be realized. It would be interesting to study
concretely whether bounds on $C\cdot C$ in more generic manifolds
limit the representation theory. Perhaps there is a no-go in F-theory
on theories with $\bf{126}$ of $SO(10)$ similar to known results in the
heterotic string \cite{Dienes:1996yh,Dienes:1996wx}.

\label{sec:particle reps}

\section{Conclusions and Future Directions}
We have studied the appearance of Lie algebra representations in elliptically
fibered K\" ahler varieties via the deformation of algebraic singularities. A concise list of results
is given in the introduction.

There are many interesting possibilities for future work. Formally,
our work is a mathematical analysis of elliptic fibrations which is
independent of any particular application within string theory. The
connection between the deformation theory of singularities and Lie
theory has been studied extensively in the mathematics literature,
with seminal contributions from Grothendieck, Brieskorn
\cite{Brieskorn}, and Arnol'd \cite{Arnold}.

In this paper we focused on
the realization of ADE representations on the $N$ deformed discriminant
components and the relationship to Picard-Lefschetz theory. We also
studied the occurrence of the non simply laced algebra $\g_2$ from this
point of view.  We hope that the discussions and examples in this paper will
shed light on this important subject.

We have also emphasized throughout that knowing the junction
realization of an ADE representation in terms of codimension one data
does not necessarily mean that the representation can be realized in a
compact elliptic fibration, let alone as massless representation
localized in codimension two. As we have seen, ADE representations of
high dimension have some junctions $J$ with self-intersection
$(J,J)$ a large negative integer; as pointed out in \cite{Mikhailov:1998bx,DeWolfe:1998bi}
in the case of surfaces, holomorphic representatives have $(J,J) \ge -2$, which
limits the allowed ADE flavor representations of BPS states in $d=4$ $\cN=2$ theories on D3 probes.
It would be interesting to study whether similar constraints exist for compactifications
on higher dimensional elliptic fibrations, in particular $d=4$ $\cN=1$ compactifications
of F-theory on elliptically fibered Calabi-Yau fourfolds. At the very least,
the realization of high dimensional ADE representations in these compactifications requires
specializing to high codimension subloci in moduli space; singularities giving
rise to them are not typically  double points \cite{Grassi:2000we,Morrison:2011mb,Grassi:2011hq}.

Finally, we would again like to emphasize the advantages of
deformations in F-theory compactifications: unlike the K\" ahler
moduli of resolutions, the complex structure moduli of deformations
exist in both the defining M-theory compactification and in the
F-theory limit, and it is therefore the more physical description
of gauge theoretic structure in F-theory. For example, in GUTs described by
the breaking of a higher rank group such as $E_8$, the finite volume
two-cycles of the Higgsed $W$-bosons arise from the deformation, not
resolution. Though there has been recent progress in understanding
F-theory compactifications via the Coulomb branch of the defining
M-theory compactification, it would be advantageous to understand the
same physics via deformation; this may also shed light on open
problems in F-theory.

\vspace{.5cm}

\clearpage
\noindent \textbf{Acknowledgments} \\
We thank F. Chen, M. Cveti{\v c}, R. Donagi, L. Everett, P. Fendley,
D. Klevers, V. Kumar, P. Langacker, and H. Yu for useful
conversations.  We are particularly indebted to K. Dienes for
discussion of related issues in the heterotic string; to G. Kane,
R. Lu, and B. Zheng for related discussions on $G_2$ manifolds; and to
D.R. Morrison for extensive conversations and comments. A.G. and
J.H. thank Z. Guralnik and B. Ovrut, M. Cveti{\v c} and I. Garc\'
ia-Etxebarria respectively for previous collaborations on string
junctions. J.H. is supported by the National Science Foundation under
Grant No. PHY11-25915. J.L.S. is supported by DARPA, fund
no. 553700. is the Class of 1939 Professor in the School of Arts and
Sciences of the University of Pennsylvania and gratefully acknowledges
the generosity of the Class of 1939. A.G. and J.H. gratefully
acknowledge the hospitality and support of the Simons Center for
Geometry and Physics.

\appendix

\section{Appendices on ADE Algebras}
In this appendix we present aspects of junction realizations of ADE
algebras which are used throughout the paper. These are by no means complete,
but may serve as a useful reference for the reader.

Specifically, in this appendix we give:
\begin{itemize}
\item  Sets of simple roots for $D_4$, $D_5$, $E_6$, $E_7$, and $E_8$ in the canonical basis (\ref{eqn:junctionbasis}).
\item The positive roots of $E_6$, $E_7$, and $E_8$ in Table \ref{table:E6roots}, Table \ref{table:E7roots} and Table \ref{table:E8roots}.
\item Intersection matrices S for $D_r$, $E_6$, $E_7$, and $E_8$ in the canonical basis (\ref{eqn:junctionbasis}).
\item Maps $F$ from junctions  $\bZ^N$ in the canonical basis (\ref{eqn:junctionbasis}) to weights in  $\bZ^r$ in the Dynkin basis.
\item The $192$ sets of simple roots for $D_4$ using the vanishing cycles $Z$ of section \ref{sec:monodromy non simply-laced}.
\item An illustration of the branching of adjoints of $E_6$ and $SO(8)$ into irreps of $SO(10)$ and $SU(4)$.
\end{itemize}

\clearpage

\begin{table}
\centering
  \begin{tabular}{|c|c|}
   Algebra & Simple Roots \\ \hline
   $D_4$ & $(1, -1, 0, 0, 0, 0)\qquad$$(0, 1, -1, 0, 0, 0)\qquad$$(0, 0, 1, -1, 0, 0)\qquad$$(0, 0, 1, 1, -1, -1)$ \\ \hline
   $D_5$ & $(1, -1, 0, 0, 0, 0, 0)\qquad$  $(0, 1, -1, 0, 0, 0, 0)\qquad$  $(0, 0, 1, -1, 0, 0, 0)$ \\ &  $(0, 0, 0, 1, -1, 0, 0)\qquad$  $(0, 0, 0, 1, 1, -1, -1)$ \\ \hline
   $E_6$ & $(1, -1, 0, 0, 0, 0, 0, 0)\qquad$$(0, 1, -1, 0, 0, 0, 0, 0)\qquad$ $(0, 0, 1, -1, 0, 0, 0, 0)$ \\ & $(0, 0, 0, 1, 1, -1, -1, 0)\qquad$$(0, 0, 0, 0, 0, 0, 1, -1)\qquad$$(0, 0, 0, 1, -1, 0, 0, 0)$ \\ \hline
   $E_7$ & $(1, -1, 0, 0, 0, 0, 0, 0, 0)\qquad$$(0, 1, -1, 0, 0, 0, 0, 0, 0)\qquad$$(0, 0, 1, -1, 0, 0, 0, 0, 0)$ \\ &$(0, 0, 0, 1, -1, 0, 0, 0, 0)\qquad$ $(0, 0, 0, 0, 1, 1, -1, -1, 0)$ \\ & $(0, 0, 0, 0, 0, 0, 0, 1, -1)\qquad$$(0, 0, 0, 0, 1, -1, 0, 0, 0)$ \\ \hline
   $E_8$ & $(1, -1, 0, 0, 0, 0, 0, 0, 0, 0)\qquad$$(0, 1, -1, 0, 0, 0, 0, 0, 0, 0)\qquad$$(0, 0, 1, -1, 0, 0, 0, 0, 0, 0)$ \\ & $(0, 0, 0, 1, -1, 0, 0, 0, 0, 0)\qquad$$(0, 0, 0, 0, 1, -1, 0, 0, 0, 0)\qquad$$(0, 0, 0, 0, 0, 1, 1, -1, -1, 0)$ \\ & $(0, 0, 0, 0, 0, 0, 0, 0, 1, -1)\qquad$$(0, 0, 0, 0, 0, 1, -1, 0, 0, 0)$ \\ \hline
  \end{tabular}
\caption{Junctions which are simple roots for a number of algebras. All examples use the
canonical basis (\ref{eqn:junctionbasis}).}
\label{table:simpleroots}
\end{table}

\begin{table}
\centering
\scriptsize
\begin{tabular}{c}
\ja{(1, 1, 0, 0, -1, -1)}, \nonumber \\
\ja{(1, 0, 1, 0, -1, -1)},  \nonumber \\
\ja{(0, 1, 1, 0, -1, -1)}, \ja{(1, 0, 0, 1, -1, -1)}, \jb{(1, 0, 0, -1, 0, 0)},  \nonumber \\
\ja{(0, 1, 0, 1, -1, -1)}, \jb{(0, 1, 0, -1, 0, 0)}, \jb{(1, 0, -1, 0, 0, 0)},  \nonumber \\
\ja{(0, 0, 1, 1, -1, -1)}, \jb{(0, 1, -1, 0, 0, 0)}, \jb{(0, 0, 1, -1, 0, 0)}, \jb{(1, -1, 0, 0, 0, 0)},  \nonumber \\
\jd{(0, 0, 0, 0, 0, 0)},\jb{(0, 0, 0, 0, 0, 0)},\jb{(0, 0, 0, 0, 0, 0)},\jb{(0, 0, 0, 0, 0, 0)}, \\
\jc{(0, 0, -1, -1, 1, 1)},\jb{(-1, 1, 0, 0, 0, 0)}, \jb{(0, -1, 1, 0, 0, 0)}, \jb{(0, 0, -1, 1, 0, 0)},   \nonumber \\
\jc{(0, -1, 0, -1, 1, 1)},\jb{(-1, 0, 1, 0, 0, 0)}, \jb{(0, -1, 0, 1, 0, 0)},  \nonumber  \\
\jc{(0, -1, -1, 0, 1, 1)},\jc{(-1, 0, 0, -1, 1, 1)} ,\jb{(-1, 0, 0, 1, 0, 0)}  \nonumber \\
\jc{(-1, 0, -1, 0, 1, 1)},  \nonumber \\
\jc{(-1, -1, 0, 0, 1, 1)}.  \nonumber \\
\end{tabular}
\caption{The weight diagram for the roots of $SO(8)$ in the $AAAABC$ basis. Upon Higgsing to $SU(4)$, the roots
which give rise to the \bface{6}, \obface{6}, \bface{15} and \bface{1} of $SU(4)$ are
colored in green, blue, red, and black, respectively.}
\label{table:D4 roots}
\end{table}
\begin{table}
\centering
\tiny

\begin{tabular}{c}
\ja{(1, 1, 1, 1, 0, -2, -1, -1)}, \\
\ja{(1, 1, 1, 0, 1, -2, -1, -1)}, \\
\ja{(1, 1, 0, 1, 1, -2, -1, -1)}, \\
\ja{(1, 0, 1, 1, 1, -2, -1, -1)}, \ja{(1, 1, 0, 0, 0, -1, 0, -1)}, \\
\ja{(0, 1, 1, 1, 1, -2, -1, -1)}, \ja{(1, 0, 1, 0, 0, -1, 0, -1)}, \jb{(1, 1, 0, 0, 0, -1, -1, 0)}, \\
\ja{(0, 1, 1, 0, 0, -1, 0, -1)}, \ja{(1, 0, 0, 1, 0, -1, 0, -1)}, \jb{(1, 0, 1, 0, 0, -1, -1, 0)}, \\
\ja{(0, 1, 0, 1, 0, -1, 0, -1)},  \ja{(1, 0, 0, 0, 1, -1, 0, -1)}, \jb{(0, 1, 1, 0, 0, -1, -1, 0)}, \jb{(1, 0, 0, 1, 0, -1, -1, 0)},\\
\ja{(0, 0, 1, 1, 0, -1, 0, -1)}, \ja{(0, 1, 0, 0, 1, -1, 0, -1)}, \jb{(1, 0, 0, 0, -1, 0, 0, 0)}, \jb{(1, 0, 0, 0, 1, -1, -1, 0)},\jb{(0, 1, 0, 1, 0, -1, -1, 0)},  \\
\ja{(0, 0, 1, 0, 1, -1, 0, -1)}, \jb{(0, 1, 0, 0, -1, 0, 0, 0)}, \jb{(0, 1, 0, 0, 1, -1, -1, 0)}, \jb{(0, 0, 1, 1, 0, -1, -1, 0)}, \jb{(1, 0, 0, -1, 0, 0, 0, 0)}, \\
\ja{(0, 0, 0, 1, 1, -1, 0, -1)}, \jb{(0, 0, 1, 0, -1, 0, 0, 0)}, \jb{(0, 0, 1, 0, 1, -1, -1, 0)},  \jb{(0, 1, 0, -1, 0, 0, 0, 0)}, \jb{(1, 0, -1, 0, 0, 0, 0, 0)}, \\
 \ja{(0, 0, 0, 0, 0, 0, 1, -1)}, \jb{(0, 0, 0, 1, 1, -1, -1, 0)}, \jb{(0, 0, 0, 1, -1, 0, 0, 0)}, \jb{(0, 0, 1, -1, 0, 0, 0, 0)},  \jb{(0, 1, -1, 0, 0, 0, 0, 0)}, \jb{(1, -1, 0, 0, 0, 0, 0, 0)}, \\
\jd{(0, 0, 0, 0, 0, 0, 0, 0)}, \jb{(0, 0, 0, 0, 0, 0, 0, 0)},\jb{(0, 0, 0, 0, 0, 0, 0, 0)},\jb{(0, 0, 0, 0, 0, 0, 0, 0)},\jb{(0, 0, 0, 0, 0, 0, 0, 0)},\jb{(0, 0, 0, 0, 0, 0, 0, 0)}, \\
\jc{(0, 0, 0, 0, 0, 0, -1, 1)},\jb{(-1, 1, 0, 0, 0, 0, 0, 0)}, \jb{(0, -1, 1, 0, 0, 0, 0, 0)}, \jb{(0, 0, -1, 1, 0, 0, 0, 0)}, \jb{(0, 0, 0, -1, -1, 1, 1, 0)},  \jb{(0, 0, 0, -1, 1, 0, 0, 0)}, \\
\jc{(0, 0, 0, -1, -1, 1, 0, 1)},\jb{(-1, 0, 1, 0, 0, 0, 0, 0)}, \jb{(0, -1, 0, 1, 0, 0, 0, 0)}, \jb{(0, 0, -1, 0, -1, 1, 1, 0)}, \jb{(0, 0, -1, 0, 1, 0, 0, 0)},  \\
\jc{(0, 0, -1, 0, -1, 1, 0, 1)},\jb{(-1, 0, 0, 1, 0, 0, 0, 0)}, \jb{(0, -1, 0, 0, -1, 1, 1, 0)}, \jb{(0, -1, 0, 0, 1, 0, 0, 0)},  \jb{(0, 0, -1, -1, 0, 1, 1, 0)}, \\
\jc{(0, -1, 0, 0, -1, 1, 0, 1)},\jc{(0, 0, -1, -1, 0, 1, 0, 1)},\jb{(-1, 0, 0, 0, -1, 1, 1, 0)}, \jb{(-1, 0, 0, 0, 1, 0, 0, 0)},  \jb{(0, -1, 0, -1, 0, 1, 1, 0)},  \\
\jc{(-1, 0, 0, 0, -1, 1, 0, 1)}, \jc{(0, -1, 0, -1, 0, 1, 0, 1)},\jb{(-1, 0, 0, -1, 0, 1, 1, 0)},  \jb{(0, -1, -1, 0, 0, 1, 1, 0)}, \\
\jc{(-1, 0, 0, -1, 0, 1, 0, 1)},\jc{(0, -1, -1, 0, 0, 1, 0, 1)}, \jb{(-1, 0, -1, 0, 0, 1, 1, 0)},  \\
\jc{(-1, 0, -1, 0, 0, 1, 0, 1)},  \jc{(0, -1, -1, -1, -1, 2, 1, 1)},\jb{(-1, -1, 0, 0, 0, 1, 1, 0)}, \\
\jc{(-1, -1, 0, 0, 0, 1, 0, 1)}, \jc{(-1, 0, -1, -1, -1, 2, 1, 1)}, \\
\jc{(-1, -1, 0, -1, -1, 2, 1, 1)}, \\
\jc{(-1, -1, -1, 0, -1, 2, 1, 1)}, \\
\jc{(-1, -1, -1, -1, 0, 2, 1, 1)}, \\
\end{tabular}
\caption{The weight diagram for the roots of $E_6$ in the $AAAAABCC$ basis. Upon Higgsing to $SO(10)$, the roots
which give rise to the \bface{16}, \bface{16'}, \bface{45} and \bface{1} of $SO(10)$ are
colored in green, blue, red, and black, respectively.}
\label{table:E6 roots}
\end{table}

\scriptsize
\begin{table}
\centering
\[ I^{D_r}=\begin{pmatrix}
  -1 &   \dots &   0  &-\frac{1}{2} & \frac{1}{2}  \\
   0 &  \ddots   & 0 &\vdots & \vdots  \\
   0 &   \dots  & -1   &-\frac{1}{2} & \frac{1}{2}  \\
  -\frac{1}{2} &\dots& -\frac{1}{2}  & -1   & 1  \\
 \frac{1}{2} & \dots& \frac{1}{2}  &  1 &   -1  \\
\end{pmatrix}
\hspace{2.2cm}
I^{E_6} = \begin{pmatrix}
  -1 &   0 &   0 &   0 &   0 &-\frac{1}{2} & \frac{1}{2} & \frac{1}{2} \\
   0 &  -1   & 0 &   0  &  0 &-\frac{1}{2} & \frac{1}{2} & \frac{1}{2} \\
   0 &   0  & -1  &  0  &  0 &-\frac{1}{2} & \frac{1}{2} & \frac{1}{2} \\
   0 &   0  &  0  & -1  &  0 &-\frac{1}{2} & \frac{1}{2} & \frac{1}{2} \\
   0 &   0  &  0  &  0  & -1 &-\frac{1}{2} & \frac{1}{2} & \frac{1}{2} \\
-\frac{1}{2} &-\frac{1}{2}& -\frac{1}{2}& -\frac{1}{2}& -\frac{1}{2}  & -1   & 1 &   1 \\
 \frac{1}{2} & \frac{1}{2}&  \frac{1}{2}&  \frac{1}{2} & \frac{1}{2}  &  1  & -1  &  0 \\
 \frac{1}{2} & \frac{1}{2} & \frac{1}{2}&  \frac{1}{2} & \frac{1}{2}  &  1 &   0   &-1 \\
\end{pmatrix}\]

\vspace{.5cm}
\hspace{-.8cm}

\[I^{E_7} = \begin{pmatrix}
  -1 &   0 &   0 &   0 &   0 & 0 &-\frac{1}{2} & \frac{1}{2} & \frac{1}{2} \\
   0 &  -1   & 0 &   0  &  0 & 0&-\frac{1}{2} & \frac{1}{2} & \frac{1}{2} \\
   0 &   0  & -1  &  0  &  0 & 0&-\frac{1}{2} & \frac{1}{2} & \frac{1}{2} \\
   0 &   0  &  0  & -1  &  0 & 0&-\frac{1}{2} & \frac{1}{2} & \frac{1}{2} \\
   0 &   0  &  0  &  0  & -1 & 0 &-\frac{1}{2} & \frac{1}{2} & \frac{1}{2} \\
 0 & 0 & 0 & 0 & 0 & -1 &-\frac{1}{2} & \frac{1}{2} & \frac{1}{2} \\
-\frac{1}{2} &-\frac{1}{2}& -\frac{1}{2}& -\frac{1}{2}& -\frac{1}{2}& -\frac{1}{2}  & -1   & 1 &   1 \\
 \frac{1}{2} & \frac{1}{2}&  \frac{1}{2}&  \frac{1}{2} & \frac{1}{2}& \frac{1}{2}  &  1  & -1  &  0 \\
 \frac{1}{2} & \frac{1}{2} & \frac{1}{2}&  \frac{1}{2} & \frac{1}{2}& \frac{1}{2}  &  1 &   0   &-1 \\
\end{pmatrix}
\vspace{.5cm}
 I^{E_8} = \begin{pmatrix}
  -1 &   0 &   0 &   0 &   0 & 0& 0 &-\frac{1}{2} & \frac{1}{2} & \frac{1}{2} \\
   0 &  -1   & 0 &   0  &  0 & 0& 0&-\frac{1}{2} & \frac{1}{2} & \frac{1}{2} \\
   0 &   0  & -1  &  0  &  0 & 0& 0&-\frac{1}{2} & \frac{1}{2} & \frac{1}{2} \\
   0 &   0  &  0  & -1  &  0 & 0& 0&-\frac{1}{2} & \frac{1}{2} & \frac{1}{2} \\
   0 &   0  &  0  &  0  & -1 & 0& 0 &-\frac{1}{2} & \frac{1}{2} & \frac{1}{2} \\
 0 & 0 & 0 & 0 & 0 & -1 & 0 &-\frac{1}{2} & \frac{1}{2} & \frac{1}{2} \\
 0 & 0 & 0 & 0 & 0 & 0 & -1 &-\frac{1}{2} & \frac{1}{2} & \frac{1}{2} \\
-\frac{1}{2} &-\frac{1}{2}& -\frac{1}{2}& -\frac{1}{2}& -\frac{1}{2}& -\frac{1}{2}& -\frac{1}{2}  & -1   & 1 &   1 \\
 \frac{1}{2} & \frac{1}{2}&  \frac{1}{2}&  \frac{1}{2} & \frac{1}{2}& \frac{1}{2}& \frac{1}{2}  &  1  & -1  &  0 \\
 \frac{1}{2} & \frac{1}{2} & \frac{1}{2}&  \frac{1}{2} & \frac{1}{2}& \frac{1}{2} & \frac{1}{2} &  1 &   0   &-1 \\
\end{pmatrix} \]
\caption{Intersection matrices for $D_r$, $E_6$, $E_7$, and $E_8$ in the canonical basis (\ref{eq:(J,J)}), computed from the generic formula (\ref{eq:(J,J)}). These match \cite{DeWolfe:1998zf}. }
\label{table:ADE S}
\end{table}

\normalsize

\small
\begin{table}
\centering
\[ F^{D_4}=\begin{pmatrix}
1&-1&0&0&0&0&0\\
0&1&-1&0&0&0&0\\
0&0&1&-1&0&0&0\\
0&0&0&1&-1&0&0\\
0&0&0&1&1&1&-1
\end{pmatrix} \qquad \qquad
F^{D_5}=\begin{pmatrix}
  1&-1&0&0&0&0&0\\
0&1&-1&0&0&0&0\\
0&0&1&-1&0&0&0\\
0&0&0&1&-1&0&0\\
0&0&0&1&1&1&-1\\
\end{pmatrix} \]

\vspace{.5cm}
\[F^{E_6} = \begin{pmatrix}
1&-1&0&0&0&0&0&0\\
0&1&-1&0&0&0&0&0\\
0&0&1&-1&0&0&0&0\\
0&0&0&1&1&1&-1&0\\
0&0&0&0&0&0&1&-1\\
0&0&0&1&-1&0&0&0
\end{pmatrix} \qquad \qquad
F^{E_7} = \begin{pmatrix}
1&-1&0&0&0&0&0&0&0\\
0&1&-1&0&0&0&0&0&0\\
0&0&1&-1&0&0&0&0&0\\
0&0&0&1&-1&0&0&0&0\\
0&0&0&0&1&1&1&-1&0\\
0&0&0&0&0&0&0&1&-1\\
0&0&0&0&1&-1&0&0&0
\end{pmatrix} \]

\vspace{.5cm}
 \[F^{E_8} = \begin{pmatrix}
 1&-1&0&0&0&0&0&0&0&0\\
0&1&-1&0&0&0&0&0&0&0\\
0&0&1&-1&0&0&0&0&0&0\\
0&0&0&1&-1&0&0&0&0&0\\
0&0&0&0&1&-1&0&0&0&0\\
0&0&0&0&0&1&1&1&-1&0\\
0&0&0&0&0&0&0&0&1&-1\\
0&0&0&0&0&1&-1&0&0&0
\end{pmatrix} \]
\caption{In the canonical basis (\ref{eqn:junctionbasis}), these are maps $F:\bZ^N \rightarrow \bZ^r$ which map junctions to weights in the Dynkin basis. A
general formula is given in section \ref{sec:non-trivial reps}. These reproduce the maps of \cite{DeWolfe:1998zf}.}
\label{table:ADE F}
\end{table}

\normalsize

\clearpage

\begin{landscape}
\centering
\begin{table}
\scriptsize
\centering
\scalebox{.65}{\begin{tabular}{|c|c|c|}
L& \# & Positive Roots of $E_6$ \\ \hline
0 & 1& (1, 1, 1, 1, 0, -2, -1, -1)\,\,\, \\
1 & 1& (1, 1, 1, 0, 1, -2, -1, -1)\,\,\, \\
2 & 1& (1, 1, 0, 1, 1, -2, -1, -1)\,\,\, \\
3 & 2& (1, 0, 1, 1, 1, -2, -1, -1)\,\,\,(1, 1, 0, 0, 0, -1, 0, -1)\,\,\, \\
4 & 3& (0, 1, 1, 1, 1, -2, -1, -1)\,\,\,(1, 0, 1, 0, 0, -1, 0, -1)\,\,\,(1, 1, 0, 0, 0, -1, -1, 0)\,\,\, \\
5 & 3& (0, 1, 1, 0, 0, -1, 0, -1)\,\,\,(1, 0, 0, 1, 0, -1, 0, -1)\,\,\,(1, 0, 1, 0, 0, -1, -1, 0)\,\,\, \\
6 & 4& (0, 1, 0, 1, 0, -1, 0, -1)\,\,\,(0, 1, 1, 0, 0, -1, -1, 0)\,\,\,(1, 0, 0, 1, 0, -1, -1, 0)\,\,\,(1, 0, 0, 0, 1, -1, 0, -1)\,\,\, \\
7 & 5& (0, 0, 1, 1, 0, -1, 0, -1)\,\,\,(0, 1, 0, 1, 0, -1, -1, 0)\,\,\,(0, 1, 0, 0, 1, -1, 0, -1)\,\,\,(1, 0, 0, 0, -1, 0, 0, 0)\,\,\,(1, 0, 0, 0, 1, -1, -1, 0)\,\,\, \\
8 & 5& (0, 0, 1, 1, 0, -1, -1, 0)\,\,\,(0, 0, 1, 0, 1, -1, 0, -1)\,\,\,(0, 1, 0, 0, -1, 0, 0, 0)\,\,\,(0, 1, 0, 0, 1, -1, -1, 0)\,\,\,(1, 0, 0, -1, 0, 0, 0, 0)\,\,\, \\
9 & 5& (0, 0, 1, 0, -1, 0, 0, 0)\,\,\,(0, 0, 1, 0, 1, -1, -1, 0)\,\,\,(0, 0, 0, 1, 1, -1, 0, -1)\,\,\,(0, 1, 0, -1, 0, 0, 0, 0)\,\,\,(1, 0, -1, 0, 0, 0, 0, 0)\,\,\, \\
10 & 6& (0, 0, 0, 1, -1, 0, 0, 0)\,\,\,(0, 0, 1, -1, 0, 0, 0, 0)\,\,\,(0, 0, 0, 1, 1, -1, -1, 0)\,\,\,(0, 0, 0, 0, 0, 0, 1, -1)\,\,\,(0, 1, -1, 0, 0, 0, 0, 0)\,\,\,(1, -1, 0, 0, 0, 0, 0, 0)\,\,\, \\\end{tabular}}
\caption{\footnotesize Positive roots of $E_6$. Listed are the level $L$, the $\#$ of roots at that level, and the roots themselves in the canonical junction basis (\ref{eqn:junctionbasis}).}
\label{table:E6roots}
\end{table}
\begin{table}
\scriptsize
\centering
\scalebox{.65}{\begin{tabular}{|c|c|c|}
L& \# & Positive Roots $E_7$\\ \hline
0&1&(1, 1, 1, 1, 1, 1, -3, -1, -2)\,\,\, \\
1&1&(1, 1, 1, 1, 1, 1, -3, -2, -1)\,\,\, \\
2&1&(1, 1, 1, 1, 0, 0, -2, -1, -1)\,\,\, \\
3&1&(1, 1, 1, 0, 1, 0, -2, -1, -1)\,\,\, \\
4&2&(1, 1, 0, 1, 1, 0, -2, -1, -1)\,\,\,(1, 1, 1, 0, 0, 1, -2, -1, -1)\,\,\, \\
5&2&(1, 0, 1, 1, 1, 0, -2, -1, -1)\,\,\,(1, 1, 0, 1, 0, 1, -2, -1, -1)\,\,\, \\
6&3&(0, 1, 1, 1, 1, 0, -2, -1, -1)\,\,\,(1, 0, 1, 1, 0, 1, -2, -1, -1)\,\,\,(1, 1, 0, 0, 1, 1, -2, -1, -1)\,\,\, \\
7&3&(0, 1, 1, 1, 0, 1, -2, -1, -1)\,\,\,(1, 0, 1, 0, 1, 1, -2, -1, -1)\,\,\,(1, 1, 0, 0, 0, 0, -1, 0, -1)\,\,\, \\
8&4&(0, 1, 1, 0, 1, 1, -2, -1, -1)\,\,\,(1, 0, 0, 1, 1, 1, -2, -1, -1)\,\,\,(1, 0, 1, 0, 0, 0, -1, 0, -1)\,\,\,(1, 1, 0, 0, 0, 0, -1, -1, 0)\,\,\, \\
9&4&(0, 1, 0, 1, 1, 1, -2, -1, -1)\,\,\,(0, 1, 1, 0, 0, 0, -1, 0, -1)\,\,\,(1, 0, 0, 1, 0, 0, -1, 0, -1)\,\,\,(1, 0, 1, 0, 0, 0, -1, -1, 0)\,\,\, \\
10&5&(0, 0, 1, 1, 1, 1, -2, -1, -1)\,\,\,(0, 1, 0, 1, 0, 0, -1, 0, -1)\,\,\,(0, 1, 1, 0, 0, 0, -1, -1, 0)\,\,\,(1, 0, 0, 0, 1, 0, -1, 0, -1)\,\,\,(1, 0, 0, 1, 0, 0, -1, -1, 0)\,\,\, \\
11&5&(0, 0, 1, 1, 0, 0, -1, 0, -1)\,\,\,(0, 1, 0, 0, 1, 0, -1, 0, -1)\,\,\,(0, 1, 0, 1, 0, 0, -1, -1, 0)\,\,\,(1, 0, 0, 0, 1, 0, -1, -1, 0)\,\,\,(1, 0, 0, 0, 0, 1, -1, 0, -1)\,\,\, \\
12&6&(0, 0, 1, 0, 1, 0, -1, 0, -1)\,\,\,(0, 0, 1, 1, 0, 0, -1, -1, 0)\,\,\,(0, 1, 0, 0, 1, 0, -1, -1, 0)\,\,\,(0, 1, 0, 0, 0, 1, -1, 0, -1)\,\,\,(1, 0, 0, 0, 0, -1, 0, 0, 0)\,\,\,(1, 0, 0, 0, 0, 1, -1, -1, 0)\,\,\, \\
13&6&(0, 0, 0, 1, 1, 0, -1, 0, -1)\,\,\,(0, 0, 1, 0, 1, 0, -1, -1, 0)\,\,\,(0, 0, 1, 0, 0, 1, -1, 0, -1)\,\,\,(0, 1, 0, 0, 0, -1, 0, 0, 0)\,\,\,(0, 1, 0, 0, 0, 1, -1, -1, 0)\,\,\,(1, 0, 0, 0, -1, 0, 0, 0, 0)\,\,\, \\
14&6&(0, 0, 0, 1, 1, 0, -1, -1, 0)\,\,\,(0, 0, 0, 1, 0, 1, -1, 0, -1)\,\,\,(0, 0, 1, 0, 0, -1, 0, 0, 0)\,\,\,(0, 0, 1, 0, 0, 1, -1, -1, 0)\,\,\,(0, 1, 0, 0, -1, 0, 0, 0, 0)\,\,\,(1, 0, 0, -1, 0, 0, 0, 0, 0)\,\,\, \\
15&6&(0, 0, 0, 1, 0, -1, 0, 0, 0)\,\,\,(0, 0, 0, 1, 0, 1, -1, -1, 0)\,\,\,(0, 0, 0, 0, 1, 1, -1, 0, -1)\,\,\,(0, 0, 1, 0, -1, 0, 0, 0, 0)\,\,\,(0, 1, 0, -1, 0, 0, 0, 0, 0)\,\,\,(1, 0, -1, 0, 0, 0, 0, 0, 0)\,\,\, \\
16&7&(0, 0, 0, 0, 1, -1, 0, 0, 0)\,\,\,(0, 0, 0, 1, -1, 0, 0, 0, 0)\,\,\,(0, 0, 0, 0, 1, 1, -1, -1, 0)\,\,\,(0, 0, 0, 0, 0, 0, 0, 1, -1)\,\,\,(0, 0, 1, -1, 0, 0, 0, 0, 0)\,\,\,(0, 1, -1, 0, 0, 0, 0, 0, 0)\,\,\,(1, -1, 0, 0, 0, 0, 0, 0, 0)\,\,\, \\
\end{tabular}}
\caption{\footnotesize Positive roots of $E_7$. Listed are the level $L$, the $\#$ of roots at that level, and the roots themselves in the canonical junction basis (\ref{eqn:junctionbasis}).}
\label{table:E7roots}
\end{table}
\begin{table}
\scriptsize
\centering
\scalebox{.65}{\begin{tabular}{|c|c|c|}
L & \# & Positive Roots of $E_8$ \\ \hline
0&1&(2, 1, 1, 1, 1, 1, 1, -4, -2, -2) \\
1&1&(1, 2, 1, 1, 1, 1, 1, -4, -2, -2) \\
2&1&(1, 1, 2, 1, 1, 1, 1, -4, -2, -2) \\
3&1&(1, 1, 1, 2, 1, 1, 1, -4, -2, -2) \\
4&1&(1, 1, 1, 1, 2, 1, 1, -4, -2, -2) \\
5&1&(1, 1, 1, 1, 1, 2, 1, -4, -2, -2) \\
6&2&(1, 1, 1, 1, 1, 1, 0, -3, -1, -2)(1, 1, 1, 1, 1, 1, 2, -4, -2, -2) \\
7&2&(1, 1, 1, 1, 1, 1, 0, -3, -2, -1)(1, 1, 1, 1, 1, 0, 1, -3, -1, -2) \\
8&2&(1, 1, 1, 1, 1, 0, 1, -3, -2, -1)(1, 1, 1, 1, 0, 1, 1, -3, -1, -2) \\
9&2&(1, 1, 1, 1, 0, 1, 1, -3, -2, -1)(1, 1, 1, 0, 1, 1, 1, -3, -1, -2) \\
10&3&(1, 1, 1, 0, 1, 1, 1, -3, -2, -1)(1, 1, 1, 1, 0, 0, 0, -2, -1, -1)(1, 1, 0, 1, 1, 1, 1, -3, -1, -2) \\
11&3&(1, 1, 0, 1, 1, 1, 1, -3, -2, -1)(1, 1, 1, 0, 1, 0, 0, -2, -1, -1)(1, 0, 1, 1, 1, 1, 1, -3, -1, -2) \\
12&4&(1, 0, 1, 1, 1, 1, 1, -3, -2, -1)(1, 1, 0, 1, 1, 0, 0, -2, -1, -1)(1, 1, 1, 0, 0, 1, 0, -2, -1, -1)(0, 1, 1, 1, 1, 1, 1, -3, -1, -2) \\
13&4&(0, 1, 1, 1, 1, 1, 1, -3, -2, -1)(1, 0, 1, 1, 1, 0, 0, -2, -1, -1)(1, 1, 0, 1, 0, 1, 0, -2, -1, -1)(1, 1, 1, 0, 0, 0, 1, -2, -1, -1) \\
14&4&(0, 1, 1, 1, 1, 0, 0, -2, -1, -1)(1, 0, 1, 1, 0, 1, 0, -2, -1, -1)(1, 1, 0, 0, 1, 1, 0, -2, -1, -1)(1, 1, 0, 1, 0, 0, 1, -2, -1, -1) \\
15&4&(0, 1, 1, 1, 0, 1, 0, -2, -1, -1)(1, 0, 1, 0, 1, 1, 0, -2, -1, -1)(1, 0, 1, 1, 0, 0, 1, -2, -1, -1)(1, 1, 0, 0, 1, 0, 1, -2, -1, -1) \\
16&5&(0, 1, 1, 0, 1, 1, 0, -2, -1, -1)(0, 1, 1, 1, 0, 0, 1, -2, -1, -1)(1, 0, 0, 1, 1, 1, 0, -2, -1, -1)(1, 0, 1, 0, 1, 0, 1, -2, -1, -1)(1, 1, 0, 0, 0, 1, 1, -2, -1, -1) \\
17&5&(0, 1, 0, 1, 1, 1, 0, -2, -1, -1)(0, 1, 1, 0, 1, 0, 1, -2, -1, -1)(1, 0, 0, 1, 1, 0, 1, -2, -1, -1)(1, 0, 1, 0, 0, 1, 1, -2, -1, -1)(1, 1, 0, 0, 0, 0, 0, -1, 0, -1) \\
18&6&(0, 0, 1, 1, 1, 1, 0, -2, -1, -1)(0, 1, 0, 1, 1, 0, 1, -2, -1, -1)(0, 1, 1, 0, 0, 1, 1, -2, -1, -1)(1, 0, 0, 1, 0, 1, 1, -2, -1, -1)(1, 0, 1, 0, 0, 0, 0, -1, 0, -1)(1, 1, 0, 0, 0, 0, 0, -1, -1, 0) \\
19&6&(0, 0, 1, 1, 1, 0, 1, -2, -1, -1)(0, 1, 0, 1, 0, 1, 1, -2, -1, -1)(0, 1, 1, 0, 0, 0, 0, -1, 0, -1)(1, 0, 0, 0, 1, 1, 1, -2, -1, -1)(1, 0, 0, 1, 0, 0, 0, -1, 0, -1)(1, 0, 1, 0, 0, 0, 0, -1, -1, 0) \\
820&6&(0, 0, 1, 1, 0, 1, 1, -2, -1, -1)(0, 1, 0, 0, 1, 1, 1, -2, -1, -1)(0, 1, 0, 1, 0, 0, 0, -1, 0, -1)(0, 1, 1, 0, 0, 0, 0, -1, -1, 0)(1, 0, 0, 0, 1, 0, 0, -1, 0, -1)(1, 0, 0, 1, 0, 0, 0, -1, -1, 0) \\
21&6&(0, 0, 1, 0, 1, 1, 1, -2, -1, -1)(0, 0, 1, 1, 0, 0, 0, -1, 0, -1)(0, 1, 0, 0, 1, 0, 0, -1, 0, -1)(0, 1, 0, 1, 0, 0, 0, -1, -1, 0)(1, 0, 0, 0, 0, 1, 0, -1, 0, -1)(1, 0, 0, 0, 1, 0, 0, -1, -1, 0) \\
22&7&(0, 0, 0, 1, 1, 1, 1, -2, -1, -1)(0, 0, 1, 0, 1, 0, 0, -1, 0, -1)(0, 0, 1, 1, 0, 0, 0, -1, -1, 0)(0, 1, 0, 0, 0, 1, 0, -1, 0, -1)(0, 1, 0, 0, 1, 0, 0, -1, -1, 0)(1, 0, 0, 0, 0, 1, 0, -1, -1, 0)(1, 0, 0, 0, 0, 0, 1, -1, 0, -1) \\
23&7&(0, 0, 0, 1, 1, 0, 0, -1, 0, -1)(0, 0, 1, 0, 0, 1, 0, -1, 0, -1)(0, 0, 1, 0, 1, 0, 0, -1, -1, 0)(0, 1, 0, 0, 0, 1, 0, -1, -1, 0)(0, 1, 0, 0, 0, 0, 1, -1, 0, -1)(1, 0, 0, 0, 0, 0, -1, 0, 0, 0)(1, 0, 0, 0, 0, 0, 1, -1, -1, 0) \\
24&7&(0, 0, 0, 1, 0, 1, 0, -1, 0, -1)(0, 0, 0, 1, 1, 0, 0, -1, -1, 0)(0, 0, 1, 0, 0, 1, 0, -1, -1, 0)(0, 0, 1, 0, 0, 0, 1, -1, 0, -1)(0, 1, 0, 0, 0, 0, -1, 0, 0, 0)(0, 1, 0, 0, 0, 0, 1, -1, -1, 0)(1, 0, 0, 0, 0, -1, 0, 0, 0, 0) \\
25&7&(0, 0, 0, 0, 1, 1, 0, -1, 0, -1)(0, 0, 0, 1, 0, 1, 0, -1, -1, 0)(0, 0, 0, 1, 0, 0, 1, -1, 0, -1)(0, 0, 1, 0, 0, 0, -1, 0, 0, 0)(0, 0, 1, 0, 0, 0, 1, -1, -1, 0)(0, 1, 0, 0, 0, -1, 0, 0, 0, 0)(1, 0, 0, 0, -1, 0, 0, 0, 0, 0) \\
26&7&(0, 0, 0, 0, 1, 1, 0, -1, -1, 0)(0, 0, 0, 0, 1, 0, 1, -1, 0, -1)(0, 0, 0, 1, 0, 0, -1, 0, 0, 0)(0, 0, 0, 1, 0, 0, 1, -1, -1, 0)(0, 0, 1, 0, 0, -1, 0, 0, 0, 0)(0, 1, 0, 0, -1, 0, 0, 0, 0, 0)(1, 0, 0, -1, 0, 0, 0, 0, 0, 0) \\
27&7&(0, 0, 0, 0, 1, 0, -1, 0, 0, 0)(0, 0, 0, 0, 1, 0, 1, -1, -1, 0)(0, 0, 0, 0, 0, 1, 1, -1, 0, -1)(0, 0, 0, 1, 0, -1, 0, 0, 0, 0)(0, 0, 1, 0, -1, 0, 0, 0, 0, 0)(0, 1, 0, -1, 0, 0, 0, 0, 0, 0)(1, 0, -1, 0, 0, 0, 0, 0, 0, 0) \\
28&8&(0, 0, 0, 0, 0, 1, -1, 0, 0, 0)(0, 0, 0, 0, 1, -1, 0, 0, 0, 0)(0, 0, 0, 0, 0, 1, 1, -1, -1, 0)(0, 0, 0, 0, 0, 0, 0, 0, 1, -1)(0, 0, 0, 1, -1, 0, 0, 0, 0, 0)(0, 0, 1, -1, 0, 0, 0, 0, 0, 0)(0, 1, -1, 0, 0, 0, 0, 0, 0, 0)(1, -1, 0, 0, 0, 0, 0, 0, 0, 0)
\end{tabular}}
\caption{\footnotesize Positive roots of $E_8$. Listed are the level $L$, the $\#$ of roots at that level, and the roots themselves in the canonical junction basis (\label{table:E8roots}
\ref{eqn:junctionbasis}).}
\end{table}
\end{landscape}

\clearpage

\begin{landscape}
\begin{table}
\scalebox{.53}{
	\begin{tabular}{|c|c|c|c|c|c|c|c|c|c|c|c|c|c|c|c|c|c|c|c|c|c|c|c|}
$\#$ & Simple Root & Simple Root& Simple Root & Simple Root & $\#$ & Simple Root & Simple Root& Simple Root & Simple Root &$\#$ & Simple Root & Simple Root & Simple Root & Simple Root \\ \hline
$1$&$(0, 0, 0, -1, -1, -1)$&$(0, 0, -1, 0, 0, 1)$&$(0, -1, 0, 0, 1, 0)$&$(-1, 0, 1, 1, 0, -1)$&$65$&$(1, 0, 0, -1, 0, 0)$&$(0, 1, 1, 0, -1, -1)$&$(0, 0, 1, 1, 1, 0)$&$(-1, -1, -1, 0, 0, 0)$&$129$&$(1, 1, 0, 0, 0, 1)$&$(0, -1, -1, -1, 0, 0)$&$(-1, 0, 0, 0, -1, -1)$&$(-1, -1, 0, 1, 1, 0)$\\
$2$&$(0, 0, 0, 1, 1, 1)$&$(0, 0, -1, -1, -1, 0)$&$(0, -1, 0, 0, 1, 0)$&$(-1, 0, 1, 1, 0, -1)$&$66$&$(1, 0, 0, -1, 0, 0)$&$(0, 1, 1, 0, -1, -1)$&$(0, 0, 1, 1, 1, 0)$&$(0, 0, -1, 0, 0, 1)$&$130$&$(1, 1, 0, 0, 0, 1)$&$(0, -1, 0, 0, 1, 0)$&$(-1, 0, 1, 1, 0, -1)$&$(-1, -1, -1, 0, 0, 0)$\\
$3$&$(0, 0, 1, 0, 0, -1)$&$(0, 0, -1, -1, -1, 0)$&$(0, -1, -1, 0, 1, 1)$&$(-1, 0, 0, 1, 0, 0)$&$67$&$(1, 0, 0, -1, 0, 0)$&$(0, 1, 1, 1, 0, 0)$&$(-1, 0, 0, 0, -1, -1)$&$(-1, -1, 0, 1, 1, 0)$&$131$&$(1, 1, 0, 0, 0, 1)$&$(0, 0, -1, -1, -1, 0)$&$(0, -1, -1, 0, 1, 1)$&$(-1, 0, 1, 1, 0, -1)$\\
$4$&$(0, 0, 1, 0, 0, -1)$&$(0, 0, 0, 1, 1, 1)$&$(0, -1, -1, -1, 0, 0)$&$(-1, 0, 0, 0, -1, -1)$&$68$&$(1, 0, 0, -1, 0, 0)$&$(0, 1, 1, 1, 0, 0)$&$(0, -1, -1, 0, 1, 1)$&$(-1, -1, 0, 0, 0, -1)$&$132$&$(1, 1, 0, 0, 0, 1)$&$(0, 0, -1, -1, -1, 0)$&$(0, -1, 0, 0, 1, 0)$&$(-1, 0, 0, 1, 0, 0)$\\
$5$&$(0, 0, 1, 1, 1, 0)$&$(0, 0, -1, 0, 0, 1)$&$(0, -1, -1, -1, 0, 0)$&$(-1, 0, 0, 0, -1, -1)$&$69$&$(1, 0, 0, -1, 0, 0)$&$(0, 1, 1, 1, 0, 0)$&$(0, 0, -1, -1, -1, 0)$&$(-1, -1, 0, 0, 0, -1)$&$133$&$(1, 1, 0, 0, 0, 1)$&$(0, 0, 0, -1, -1, -1)$&$(-1, 0, 1, 1, 0, -1)$&$(-1, -1, -1, 0, 0, 0)$\\
$6$&$(0, 0, 1, 1, 1, 0)$&$(0, 0, 0, -1, -1, -1)$&$(0, -1, -1, 0, 1, 1)$&$(-1, 0, 0, 1, 0, 0)$&$70$&$(1, 0, 0, -1, 0, 0)$&$(0, 1, 1, 1, 0, 0)$&$(0, 0, -1, -1, -1, 0)$&$(0, -1, -1, 0, 1, 1)$&$134$&$(1, 1, 0, 0, 0, 1)$&$(0, 0, 0, -1, -1, -1)$&$(0, -1, -1, 0, 1, 1)$&$(-1, 0, 0, 1, 0, 0)$\\
$7$&$(0, 1, 0, 0, -1, 0)$&$(0, 0, 0, -1, -1, -1)$&$(0, -1, -1, 0, 1, 1)$&$(-1, 0, 1, 1, 0, -1)$&$71$&$(1, 0, 0, -1, 0, 0)$&$(0, 1, 1, 1, 0, 0)$&$(0, 0, -1, 0, 0, 1)$&$(-1, -1, 0, 1, 1, 0)$&$135$&$(1, 1, 0, 0, 0, 1)$&$(0, 0, 0, -1, -1, -1)$&$(0, -1, 0, 0, 1, 0)$&$(-1, -1, -1, 0, 0, 0)$\\
$8$&$(0, 1, 0, 0, -1, 0)$&$(0, 0, 0, 1, 1, 1)$&$(0, -1, -1, -1, 0, 0)$&$(-1, 0, 1, 1, 0, -1)$&$72$&$(1, 0, 0, -1, 0, 0)$&$(0, 1, 1, 1, 0, 0)$&$(0, 0, -1, 0, 0, 1)$&$(-1, 0, 0, 0, -1, -1)$&$136$&$(1, 1, 0, 0, 0, 1)$&$(0, 0, 0, -1, -1, -1)$&$(0, -1, 0, 0, 1, 0)$&$(-1, 0, 1, 1, 0, -1)$\\
$9$&$(0, 1, 0, 0, -1, 0)$&$(0, 0, 1, 0, 0, -1)$&$(0, -1, -1, -1, 0, 0)$&$(-1, -1, 0, 1, 1, 0)$&$73$&$(1, 0, 0, 0, 1, 1)$&$(0, 0, -1, -1, -1, 0)$&$(-1, 0, 0, 1, 0, 0)$&$(-1, -1, 0, 0, 0, -1)$&$137$&$(1, 1, 0, 0, 0, 1)$&$(0, 0, 1, 0, 0, -1)$&$(-1, 0, 0, 0, -1, -1)$&$(-1, -1, 0, 1, 1, 0)$\\
$10$&$(0, 1, 0, 0, -1, 0)$&$(0, 0, 1, 0, 0, -1)$&$(0, 0, 0, 1, 1, 1)$&$(-1, -1, -1, 0, 0, 0)$&$74$&$(1, 0, 0, 0, 1, 1)$&$(0, 0, -1, 0, 0, 1)$&$(0, -1, -1, -1, 0, 0)$&$(-1, 0, 1, 1, 0, -1)$&$138$&$(1, 1, 0, 0, 0, 1)$&$(0, 0, 1, 0, 0, -1)$&$(0, -1, -1, -1, 0, 0)$&$(-1, -1, 0, 1, 1, 0)$\\
$11$&$(0, 1, 0, 0, -1, 0)$&$(0, 0, 1, 1, 1, 0)$&$(0, -1, -1, 0, 1, 1)$&$(-1, -1, 0, 0, 0, -1)$&$75$&$(1, 0, 0, 0, 1, 1)$&$(0, 0, 0, -1, -1, -1)$&$(-1, 0, 1, 1, 0, -1)$&$(-1, -1, -1, 0, 0, 0)$&$139$&$(1, 1, 0, 0, 0, 1)$&$(0, 0, 1, 0, 0, -1)$&$(0, -1, -1, -1, 0, 0)$&$(-1, 0, 0, 0, -1, -1)$\\
$12$&$(0, 1, 0, 0, -1, 0)$&$(0, 0, 1, 1, 1, 0)$&$(0, 0, 0, -1, -1, -1)$&$(-1, -1, -1, 0, 0, 0)$&$76$&$(1, 0, 0, 0, 1, 1)$&$(0, 0, 0, -1, -1, -1)$&$(0, 0, -1, 0, 0, 1)$&$(-1, -1, 0, 1, 1, 0)$&$140$&$(1, 1, 0, 0, 0, 1)$&$(0, 0, 1, 0, 0, -1)$&$(0, -1, -1, 0, 1, 1)$&$(-1, 0, 0, 1, 0, 0)$\\
$13$&$(0, 1, 1, 0, -1, -1)$&$(0, 0, -1, -1, -1, 0)$&$(0, -1, 0, 0, 1, 0)$&$(-1, 0, 0, 1, 0, 0)$&$77$&$(1, 0, 0, 0, 1, 1)$&$(0, 0, 1, 0, 0, -1)$&$(0, -1, -1, -1, 0, 0)$&$(-1, 0, 0, 1, 0, 0)$&$141$&$(1, 1, 0, 0, 0, 1)$&$(0, 0, 1, 0, 0, -1)$&$(0, 0, -1, -1, -1, 0)$&$(-1, 0, 0, 1, 0, 0)$\\
$14$&$(0, 1, 1, 0, -1, -1)$&$(0, 0, -1, 0, 0, 1)$&$(0, -1, -1, -1, 0, 0)$&$(-1, -1, 0, 1, 1, 0)$&$78$&$(1, 0, 0, 0, 1, 1)$&$(0, 0, 1, 0, 0, -1)$&$(0, 0, -1, -1, -1, 0)$&$(-1, -1, 0, 1, 1, 0)$&$142$&$(1, 1, 0, 0, 0, 1)$&$(0, 0, 1, 0, 0, -1)$&$(0, 0, -1, -1, -1, 0)$&$(0, -1, -1, 0, 1, 1)$\\
$15$&$(0, 1, 1, 0, -1, -1)$&$(0, 0, 0, 1, 1, 1)$&$(0, -1, 0, 0, 1, 0)$&$(-1, -1, -1, 0, 0, 0)$&$79$&$(1, 0, 0, 0, 1, 1)$&$(0, 1, 0, 0, -1, 0)$&$(-1, 0, 1, 1, 0, -1)$&$(-1, -1, -1, 0, 0, 0)$&$143$&$(1, 1, 0, 0, 0, 1)$&$(0, 0, 1, 1, 1, 0)$&$(0, -1, -1, -1, 0, 0)$&$(-1, 0, 0, 1, 0, 0)$\\
$16$&$(0, 1, 1, 0, -1, -1)$&$(0, 0, 0, 1, 1, 1)$&$(0, 0, -1, -1, -1, 0)$&$(-1, -1, 0, 0, 0, -1)$&$80$&$(1, 0, 0, 0, 1, 1)$&$(0, 1, 0, 0, -1, 0)$&$(0, -1, -1, -1, 0, 0)$&$(-1, -1, 0, 1, 1, 0)$&$144$&$(1, 1, 0, 0, 0, 1)$&$(0, 0, 1, 1, 1, 0)$&$(0, -1, -1, 0, 1, 1)$&$(-1, 0, 0, 0, -1, -1)$\\
$17$&$(0, 1, 1, 0, -1, -1)$&$(0, 0, 1, 1, 1, 0)$&$(0, -1, -1, -1, 0, 0)$&$(-1, 0, 0, 1, 0, 0)$&$81$&$(1, 0, 0, 0, 1, 1)$&$(0, 1, 0, 0, -1, 0)$&$(0, 0, 0, -1, -1, -1)$&$(-1, -1, -1, 0, 0, 0)$&$145$&$(1, 1, 0, 0, 0, 1)$&$(0, 0, 1, 1, 1, 0)$&$(0, 0, 0, -1, -1, -1)$&$(-1, 0, 0, 1, 0, 0)$\\
$18$&$(0, 1, 1, 0, -1, -1)$&$(0, 0, 1, 1, 1, 0)$&$(0, 0, -1, 0, 0, 1)$&$(-1, -1, 0, 0, 0, -1)$&$82$&$(1, 0, 0, 0, 1, 1)$&$(0, 1, 0, 0, -1, 0)$&$(0, 0, 0, -1, -1, -1)$&$(-1, 0, 1, 1, 0, -1)$&$146$&$(1, 1, 0, 0, 0, 1)$&$(0, 0, 1, 1, 1, 0)$&$(0, 0, 0, -1, -1, -1)$&$(0, -1, -1, 0, 1, 1)$\\
$19$&$(0, 1, 1, 1, 0, 0)$&$(0, 0, -1, -1, -1, 0)$&$(0, -1, -1, 0, 1, 1)$&$(-1, -1, 0, 0, 0, -1)$&$83$&$(1, 0, 0, 0, 1, 1)$&$(0, 1, 0, 0, -1, 0)$&$(0, 0, 1, 0, 0, -1)$&$(-1, -1, 0, 1, 1, 0)$&$147$&$(1, 1, 0, 0, 0, 1)$&$(0, 1, 1, 0, -1, -1)$&$(0, -1, -1, -1, 0, 0)$&$(-1, 0, 0, 1, 0, 0)$\\
$20$&$(0, 1, 1, 1, 0, 0)$&$(0, 0, -1, 0, 0, 1)$&$(0, -1, 0, 0, 1, 0)$&$(-1, 0, 0, 0, -1, -1)$&$84$&$(1, 0, 0, 0, 1, 1)$&$(0, 1, 0, 0, -1, 0)$&$(0, 0, 1, 0, 0, -1)$&$(0, -1, -1, -1, 0, 0)$&$148$&$(1, 1, 0, 0, 0, 1)$&$(0, 1, 1, 0, -1, -1)$&$(0, -1, 0, 0, 1, 0)$&$(-1, 0, 0, 1, 0, 0)$\\
$21$&$(0, 1, 1, 1, 0, 0)$&$(0, 0, 0, -1, -1, -1)$&$(0, -1, 0, 0, 1, 0)$&$(-1, -1, -1, 0, 0, 0)$&$85$&$(1, 0, 0, 0, 1, 1)$&$(0, 1, 1, 0, -1, -1)$&$(-1, 0, 0, 1, 0, 0)$&$(-1, -1, 0, 0, 0, -1)$&$149$&$(1, 1, 0, 0, 0, 1)$&$(0, 1, 1, 0, -1, -1)$&$(0, 0, -1, -1, -1, 0)$&$(-1, -1, 0, 1, 1, 0)$\\
$22$&$(0, 1, 1, 1, 0, 0)$&$(0, 0, 0, -1, -1, -1)$&$(0, 0, -1, 0, 0, 1)$&$(-1, -1, 0, 1, 1, 0)$&$86$&$(1, 0, 0, 0, 1, 1)$&$(0, 1, 1, 0, -1, -1)$&$(0, -1, -1, -1, 0, 0)$&$(-1, -1, 0, 1, 1, 0)$&$150$&$(1, 1, 0, 0, 0, 1)$&$(0, 1, 1, 0, -1, -1)$&$(0, 0, -1, -1, -1, 0)$&$(0, -1, 0, 0, 1, 0)$\\
$23$&$(0, 1, 1, 1, 0, 0)$&$(0, 0, 1, 0, 0, -1)$&$(0, -1, -1, 0, 1, 1)$&$(-1, 0, 0, 0, -1, -1)$&$87$&$(1, 0, 0, 0, 1, 1)$&$(0, 1, 1, 0, -1, -1)$&$(0, 0, -1, -1, -1, 0)$&$(-1, -1, 0, 0, 0, -1)$&$151$&$(1, 1, 0, 0, 0, 1)$&$(0, 1, 1, 0, -1, -1)$&$(0, 0, 1, 1, 1, 0)$&$(-1, -1, -1, 0, 0, 0)$\\
$24$&$(0, 1, 1, 1, 0, 0)$&$(0, 0, 1, 0, 0, -1)$&$(0, 0, -1, -1, -1, 0)$&$(-1, -1, 0, 1, 1, 0)$&$88$&$(1, 0, 0, 0, 1, 1)$&$(0, 1, 1, 0, -1, -1)$&$(0, 0, -1, -1, -1, 0)$&$(-1, 0, 0, 1, 0, 0)$&$152$&$(1, 1, 0, 0, 0, 1)$&$(0, 1, 1, 0, -1, -1)$&$(0, 0, 1, 1, 1, 0)$&$(0, -1, -1, -1, 0, 0)$\\
$25$&$(1, 0, -1, -1, 0, 1)$&$(0, 0, 0, -1, -1, -1)$&$(0, -1, 0, 0, 1, 0)$&$(-1, 0, 0, 1, 0, 0)$&$89$&$(1, 0, 0, 0, 1, 1)$&$(0, 1, 1, 0, -1, -1)$&$(0, 0, -1, 0, 0, 1)$&$(-1, -1, 0, 1, 1, 0)$&$153$&$(1, 1, 0, 0, 0, 1)$&$(1, 0, 0, -1, 0, 0)$&$(0, -1, -1, 0, 1, 1)$&$(-1, 0, 0, 0, -1, -1)$\\
$26$&$(1, 0, -1, -1, 0, 1)$&$(0, 0, 0, 1, 1, 1)$&$(0, -1, 0, 0, 1, 0)$&$(-1, 0, 0, 0, -1, -1)$&$90$&$(1, 0, 0, 0, 1, 1)$&$(0, 1, 1, 0, -1, -1)$&$(0, 0, -1, 0, 0, 1)$&$(0, -1, -1, -1, 0, 0)$&$154$&$(1, 1, 0, 0, 0, 1)$&$(1, 0, 0, -1, 0, 0)$&$(0, -1, -1, 0, 1, 1)$&$(-1, 0, 1, 1, 0, -1)$\\
$27$&$(1, 0, -1, -1, 0, 1)$&$(0, 0, 1, 0, 0, -1)$&$(-1, 0, 0, 0, -1, -1)$&$(-1, -1, 0, 1, 1, 0)$&$91$&$(1, 0, 0, 0, 1, 1)$&$(0, 1, 1, 1, 0, 0)$&$(0, 0, -1, -1, -1, 0)$&$(-1, -1, 0, 1, 1, 0)$&$155$&$(1, 1, 0, 0, 0, 1)$&$(1, 0, 0, -1, 0, 0)$&$(0, 0, -1, -1, -1, 0)$&$(-1, -1, 0, 1, 1, 0)$\\
$28$&$(1, 0, -1, -1, 0, 1)$&$(0, 0, 1, 0, 0, -1)$&$(0, 0, 0, 1, 1, 1)$&$(-1, -1, -1, 0, 0, 0)$&$92$&$(1, 0, 0, 0, 1, 1)$&$(0, 1, 1, 1, 0, 0)$&$(0, 0, -1, 0, 0, 1)$&$(-1, -1, 0, 0, 0, -1)$&$156$&$(1, 1, 0, 0, 0, 1)$&$(1, 0, 0, -1, 0, 0)$&$(0, 0, -1, -1, -1, 0)$&$(-1, 0, 1, 1, 0, -1)$\\
$29$&$(1, 0, -1, -1, 0, 1)$&$(0, 0, 1, 1, 1, 0)$&$(-1, 0, 0, 1, 0, 0)$&$(-1, -1, 0, 0, 0, -1)$&$93$&$(1, 0, 0, 0, 1, 1)$&$(0, 1, 1, 1, 0, 0)$&$(0, 0, 0, -1, -1, -1)$&$(-1, -1, 0, 1, 1, 0)$&$157$&$(1, 1, 0, 0, 0, 1)$&$(1, 0, 0, -1, 0, 0)$&$(0, 0, 1, 1, 1, 0)$&$(-1, -1, -1, 0, 0, 0)$\\
$30$&$(1, 0, -1, -1, 0, 1)$&$(0, 0, 1, 1, 1, 0)$&$(0, 0, 0, -1, -1, -1)$&$(-1, -1, -1, 0, 0, 0)$&$94$&$(1, 0, 0, 0, 1, 1)$&$(0, 1, 1, 1, 0, 0)$&$(0, 0, 0, -1, -1, -1)$&$(0, 0, -1, 0, 0, 1)$&$158$&$(1, 1, 0, 0, 0, 1)$&$(1, 0, 0, -1, 0, 0)$&$(0, 0, 1, 1, 1, 0)$&$(-1, 0, 0, 0, -1, -1)$\\
$31$&$(1, 0, -1, -1, 0, 1)$&$(0, 1, 0, 0, -1, 0)$&$(0, 0, 0, -1, -1, -1)$&$(-1, -1, 0, 1, 1, 0)$&$95$&$(1, 0, 0, 0, 1, 1)$&$(0, 1, 1, 1, 0, 0)$&$(0, 0, 1, 0, 0, -1)$&$(-1, -1, -1, 0, 0, 0)$&$159$&$(1, 1, 0, 0, 0, 1)$&$(1, 0, 0, -1, 0, 0)$&$(0, 1, 1, 0, -1, -1)$&$(-1, -1, -1, 0, 0, 0)$\\
$32$&$(1, 0, -1, -1, 0, 1)$&$(0, 1, 0, 0, -1, 0)$&$(0, 0, 0, 1, 1, 1)$&$(-1, -1, 0, 0, 0, -1)$&$96$&$(1, 0, 0, 0, 1, 1)$&$(0, 1, 1, 1, 0, 0)$&$(0, 0, 1, 0, 0, -1)$&$(0, 0, -1, -1, -1, 0)$&$160$&$(1, 1, 0, 0, 0, 1)$&$(1, 0, 0, -1, 0, 0)$&$(0, 1, 1, 0, -1, -1)$&$(-1, -1, 0, 1, 1, 0)$\\
$33$&$(1, 0, -1, -1, 0, 1)$&$(0, 1, 0, 0, -1, 0)$&$(0, 0, 1, 0, 0, -1)$&$(-1, -1, -1, 0, 0, 0)$&$97$&$(1, 1, 0, -1, -1, 0)$&$(0, -1, -1, 0, 1, 1)$&$(-1, 0, 0, 1, 0, 0)$&$(-1, -1, 0, 0, 0, -1)$&$161$&$(1, 1, 1, 0, 0, 0)$&$(0, -1, -1, -1, 0, 0)$&$(-1, 0, 0, 0, -1, -1)$&$(-1, -1, 0, 1, 1, 0)$\\
$34$&$(1, 0, -1, -1, 0, 1)$&$(0, 1, 0, 0, -1, 0)$&$(0, 0, 1, 0, 0, -1)$&$(0, 0, 0, 1, 1, 1)$&$98$&$(1, 1, 0, -1, -1, 0)$&$(0, -1, 0, 0, 1, 0)$&$(-1, 0, 1, 1, 0, -1)$&$(-1, -1, -1, 0, 0, 0)$&$162$&$(1, 1, 1, 0, 0, 0)$&$(0, -1, -1, 0, 1, 1)$&$(-1, 0, 0, 1, 0, 0)$&$(-1, -1, 0, 0, 0, -1)$\\
$35$&$(1, 0, -1, -1, 0, 1)$&$(0, 1, 0, 0, -1, 0)$&$(0, 0, 1, 1, 1, 0)$&$(-1, -1, -1, 0, 0, 0)$&$99$&$(1, 1, 0, -1, -1, 0)$&$(0, 0, -1, 0, 0, 1)$&$(0, -1, -1, -1, 0, 0)$&$(-1, 0, 1, 1, 0, -1)$&$163$&$(1, 1, 1, 0, 0, 0)$&$(0, 0, -1, -1, -1, 0)$&$(-1, 0, 0, 1, 0, 0)$&$(-1, -1, 0, 0, 0, -1)$\\
$36$&$(1, 0, -1, -1, 0, 1)$&$(0, 1, 0, 0, -1, 0)$&$(0, 0, 1, 1, 1, 0)$&$(0, 0, 0, -1, -1, -1)$&$100$&$(1, 1, 0, -1, -1, 0)$&$(0, 0, -1, 0, 0, 1)$&$(0, -1, 0, 0, 1, 0)$&$(-1, 0, 0, 0, -1, -1)$&$164$&$(1, 1, 1, 0, 0, 0)$&$(0, 0, -1, -1, -1, 0)$&$(0, -1, -1, 0, 1, 1)$&$(-1, -1, 0, 0, 0, -1)$\\
$37$&$(1, 0, -1, -1, 0, 1)$&$(0, 1, 1, 0, -1, -1)$&$(-1, 0, 0, 1, 0, 0)$&$(-1, -1, 0, 0, 0, -1)$&$101$&$(1, 1, 0, -1, -1, 0)$&$(0, 0, 0, 1, 1, 1)$&$(-1, 0, 1, 1, 0, -1)$&$(-1, -1, -1, 0, 0, 0)$&$165$&$(1, 1, 1, 0, 0, 0)$&$(0, 0, -1, -1, -1, 0)$&$(0, -1, -1, 0, 1, 1)$&$(-1, 0, 0, 1, 0, 0)$\\
$38$&$(1, 0, -1, -1, 0, 1)$&$(0, 1, 1, 0, -1, -1)$&$(0, -1, 0, 0, 1, 0)$&$(-1, -1, -1, 0, 0, 0)$&$102$&$(1, 1, 0, -1, -1, 0)$&$(0, 0, 0, 1, 1, 1)$&$(0, -1, -1, -1, 0, 0)$&$(-1, 0, 0, 0, -1, -1)$&$166$&$(1, 1, 1, 0, 0, 0)$&$(0, 0, -1, -1, -1, 0)$&$(0, -1, 0, 0, 1, 0)$&$(-1, 0, 1, 1, 0, -1)$\\
$39$&$(1, 0, -1, -1, 0, 1)$&$(0, 1, 1, 0, -1, -1)$&$(0, 0, 0, 1, 1, 1)$&$(-1, -1, -1, 0, 0, 0)$&$103$&$(1, 1, 0, -1, -1, 0)$&$(0, 0, 0, 1, 1, 1)$&$(0, -1, 0, 0, 1, 0)$&$(-1, -1, -1, 0, 0, 0)$&$167$&$(1, 1, 1, 0, 0, 0)$&$(0, 0, -1, 0, 0, 1)$&$(-1, 0, 0, 0, -1, -1)$&$(-1, -1, 0, 1, 1, 0)$\\
$40$&$(1, 0, -1, -1, 0, 1)$&$(0, 1, 1, 0, -1, -1)$&$(0, 0, 0, 1, 1, 1)$&$(0, -1, 0, 0, 1, 0)$&$104$&$(1, 1, 0, -1, -1, 0)$&$(0, 0, 0, 1, 1, 1)$&$(0, -1, 0, 0, 1, 0)$&$(-1, 0, 1, 1, 0, -1)$&$168$&$(1, 1, 1, 0, 0, 0)$&$(0, 0, -1, 0, 0, 1)$&$(0, -1, -1, -1, 0, 0)$&$(-1, -1, 0, 1, 1, 0)$\\
$41$&$(1, 0, -1, -1, 0, 1)$&$(0, 1, 1, 0, -1, -1)$&$(0, 0, 1, 1, 1, 0)$&$(-1, -1, 0, 0, 0, -1)$&$105$&$(1, 1, 0, -1, -1, 0)$&$(0, 0, 1, 0, 0, -1)$&$(0, -1, -1, -1, 0, 0)$&$(-1, 0, 0, 1, 0, 0)$&$169$&$(1, 1, 1, 0, 0, 0)$&$(0, 0, -1, 0, 0, 1)$&$(0, -1, -1, -1, 0, 0)$&$(-1, 0, 0, 0, -1, -1)$\\
$42$&$(1, 0, -1, -1, 0, 1)$&$(0, 1, 1, 0, -1, -1)$&$(0, 0, 1, 1, 1, 0)$&$(-1, 0, 0, 1, 0, 0)$&$106$&$(1, 1, 0, -1, -1, 0)$&$(0, 0, 1, 0, 0, -1)$&$(0, -1, -1, 0, 1, 1)$&$(-1, 0, 0, 0, -1, -1)$&$170$&$(1, 1, 1, 0, 0, 0)$&$(0, 0, -1, 0, 0, 1)$&$(0, -1, 0, 0, 1, 0)$&$(-1, 0, 1, 1, 0, -1)$\\
$43$&$(1, 0, -1, -1, 0, 1)$&$(0, 1, 1, 1, 0, 0)$&$(-1, 0, 0, 0, -1, -1)$&$(-1, -1, 0, 1, 1, 0)$&$107$&$(1, 1, 0, -1, -1, 0)$&$(0, 0, 1, 0, 0, -1)$&$(0, 0, 0, 1, 1, 1)$&$(-1, 0, 0, 0, -1, -1)$&$171$&$(1, 1, 1, 0, 0, 0)$&$(0, 0, 0, -1, -1, -1)$&$(0, -1, -1, 0, 1, 1)$&$(-1, 0, 1, 1, 0, -1)$\\
$44$&$(1, 0, -1, -1, 0, 1)$&$(0, 1, 1, 1, 0, 0)$&$(0, -1, 0, 0, 1, 0)$&$(-1, -1, -1, 0, 0, 0)$&$108$&$(1, 1, 0, -1, -1, 0)$&$(0, 0, 1, 0, 0, -1)$&$(0, 0, 0, 1, 1, 1)$&$(0, -1, -1, -1, 0, 0)$&$172$&$(1, 1, 1, 0, 0, 0)$&$(0, 0, 0, -1, -1, -1)$&$(0, -1, 0, 0, 1, 0)$&$(-1, 0, 0, 1, 0, 0)$\\
$45$&$(1, 0, -1, -1, 0, 1)$&$(0, 1, 1, 1, 0, 0)$&$(0, 0, 0, -1, -1, -1)$&$(-1, -1, -1, 0, 0, 0)$&$109$&$(1, 1, 0, -1, -1, 0)$&$(0, 0, 1, 1, 1, 0)$&$(-1, 0, 0, 1, 0, 0)$&$(-1, -1, 0, 0, 0, -1)$&$173$&$(1, 1, 1, 0, 0, 0)$&$(0, 0, 0, -1, -1, -1)$&$(0, 0, -1, 0, 0, 1)$&$(-1, 0, 1, 1, 0, -1)$\\
$46$&$(1, 0, -1, -1, 0, 1)$&$(0, 1, 1, 1, 0, 0)$&$(0, 0, 0, -1, -1, -1)$&$(0, -1, 0, 0, 1, 0)$&$110$&$(1, 1, 0, -1, -1, 0)$&$(0, 0, 1, 1, 1, 0)$&$(0, -1, -1, -1, 0, 0)$&$(-1, 0, 0, 0, -1, -1)$&$174$&$(1, 1, 1, 0, 0, 0)$&$(0, 0, 0, -1, -1, -1)$&$(0, 0, -1, 0, 0, 1)$&$(0, -1, 0, 0, 1, 0)$\\
$47$&$(1, 0, -1, -1, 0, 1)$&$(0, 1, 1, 1, 0, 0)$&$(0, 0, 1, 0, 0, -1)$&$(-1, -1, 0, 1, 1, 0)$&$111$&$(1, 1, 0, -1, -1, 0)$&$(0, 0, 1, 1, 1, 0)$&$(0, -1, -1, 0, 1, 1)$&$(-1, -1, 0, 0, 0, -1)$&$175$&$(1, 1, 1, 0, 0, 0)$&$(0, 0, 0, 1, 1, 1)$&$(0, -1, -1, -1, 0, 0)$&$(-1, 0, 1, 1, 0, -1)$\\
$48$&$(1, 0, -1, -1, 0, 1)$&$(0, 1, 1, 1, 0, 0)$&$(0, 0, 1, 0, 0, -1)$&$(-1, 0, 0, 0, -1, -1)$&$112$&$(1, 1, 0, -1, -1, 0)$&$(0, 0, 1, 1, 1, 0)$&$(0, -1, -1, 0, 1, 1)$&$(-1, 0, 0, 1, 0, 0)$&$176$&$(1, 1, 1, 0, 0, 0)$&$(0, 0, 0, 1, 1, 1)$&$(0, -1, 0, 0, 1, 0)$&$(-1, 0, 0, 0, -1, -1)$\\
$49$&$(1, 0, 0, -1, 0, 0)$&$(0, 0, -1, -1, -1, 0)$&$(0, -1, -1, 0, 1, 1)$&$(-1, 0, 1, 1, 0, -1)$&$113$&$(1, 1, 0, -1, -1, 0)$&$(0, 0, 1, 1, 1, 0)$&$(0, 0, -1, 0, 0, 1)$&$(-1, 0, 0, 0, -1, -1)$&$177$&$(1, 1, 1, 0, 0, 0)$&$(0, 0, 0, 1, 1, 1)$&$(0, 0, -1, -1, -1, 0)$&$(-1, 0, 1, 1, 0, -1)$\\
$50$&$(1, 0, 0, -1, 0, 0)$&$(0, 0, -1, 0, 0, 1)$&$(-1, 0, 0, 0, -1, -1)$&$(-1, -1, 0, 1, 1, 0)$&$114$&$(1, 1, 0, -1, -1, 0)$&$(0, 0, 1, 1, 1, 0)$&$(0, 0, -1, 0, 0, 1)$&$(0, -1, -1, -1, 0, 0)$&$178$&$(1, 1, 1, 0, 0, 0)$&$(0, 0, 0, 1, 1, 1)$&$(0, 0, -1, -1, -1, 0)$&$(0, -1, 0, 0, 1, 0)$\\
$51$&$(1, 0, 0, -1, 0, 0)$&$(0, 0, 0, 1, 1, 1)$&$(-1, 0, 1, 1, 0, -1)$&$(-1, -1, -1, 0, 0, 0)$&$115$&$(1, 1, 0, -1, -1, 0)$&$(0, 1, 1, 1, 0, 0)$&$(0, -1, -1, 0, 1, 1)$&$(-1, 0, 0, 0, -1, -1)$&$179$&$(1, 1, 1, 0, 0, 0)$&$(0, 1, 0, 0, -1, 0)$&$(0, -1, -1, -1, 0, 0)$&$(-1, 0, 1, 1, 0, -1)$\\
$52$&$(1, 0, 0, -1, 0, 0)$&$(0, 0, 0, 1, 1, 1)$&$(0, 0, -1, -1, -1, 0)$&$(-1, -1, 0, 0, 0, -1)$&$116$&$(1, 1, 0, -1, -1, 0)$&$(0, 1, 1, 1, 0, 0)$&$(0, -1, 0, 0, 1, 0)$&$(-1, 0, 0, 0, -1, -1)$&$180$&$(1, 1, 1, 0, 0, 0)$&$(0, 1, 0, 0, -1, 0)$&$(0, -1, -1, 0, 1, 1)$&$(-1, 0, 1, 1, 0, -1)$\\
$53$&$(1, 0, 0, -1, 0, 0)$&$(0, 0, 1, 1, 1, 0)$&$(0, -1, -1, 0, 1, 1)$&$(-1, 0, 0, 0, -1, -1)$&$117$&$(1, 1, 0, -1, -1, 0)$&$(0, 1, 1, 1, 0, 0)$&$(0, 0, -1, 0, 0, 1)$&$(-1, -1, 0, 0, 0, -1)$&$181$&$(1, 1, 1, 0, 0, 0)$&$(0, 1, 0, 0, -1, 0)$&$(0, 0, 0, -1, -1, -1)$&$(-1, -1, 0, 1, 1, 0)$\\
$54$&$(1, 0, 0, -1, 0, 0)$&$(0, 0, 1, 1, 1, 0)$&$(0, 0, -1, 0, 0, 1)$&$(-1, -1, 0, 0, 0, -1)$&$118$&$(1, 1, 0, -1, -1, 0)$&$(0, 1, 1, 1, 0, 0)$&$(0, 0, -1, 0, 0, 1)$&$(0, -1, 0, 0, 1, 0)$&$182$&$(1, 1, 1, 0, 0, 0)$&$(0, 1, 0, 0, -1, 0)$&$(0, 0, 0, -1, -1, -1)$&$(0, -1, -1, 0, 1, 1)$\\
$55$&$(1, 0, 0, -1, 0, 0)$&$(0, 1, 0, 0, -1, 0)$&$(-1, 0, 1, 1, 0, -1)$&$(-1, -1, -1, 0, 0, 0)$&$119$&$(1, 1, 0, -1, -1, 0)$&$(0, 1, 1, 1, 0, 0)$&$(0, 0, 1, 0, 0, -1)$&$(-1, -1, -1, 0, 0, 0)$&$183$&$(1, 1, 1, 0, 0, 0)$&$(0, 1, 0, 0, -1, 0)$&$(0, 0, 0, 1, 1, 1)$&$(-1, -1, 0, 0, 0, -1)$\\
$56$&$(1, 0, 0, -1, 0, 0)$&$(0, 1, 0, 0, -1, 0)$&$(0, -1, -1, 0, 1, 1)$&$(-1, -1, 0, 0, 0, -1)$&$120$&$(1, 1, 0, -1, -1, 0)$&$(0, 1, 1, 1, 0, 0)$&$(0, 0, 1, 0, 0, -1)$&$(0, -1, -1, 0, 1, 1)$&$184$&$(1, 1, 1, 0, 0, 0)$&$(0, 1, 0, 0, -1, 0)$&$(0, 0, 0, 1, 1, 1)$&$(0, -1, -1, -1, 0, 0)$\\
$57$&$(1, 0, 0, -1, 0, 0)$&$(0, 1, 0, 0, -1, 0)$&$(0, 0, 0, 1, 1, 1)$&$(-1, -1, -1, 0, 0, 0)$&$121$&$(1, 1, 0, -1, -1, 0)$&$(1, 0, 0, 0, 1, 1)$&$(0, -1, -1, -1, 0, 0)$&$(-1, 0, 0, 1, 0, 0)$&$185$&$(1, 1, 1, 0, 0, 0)$&$(1, 0, -1, -1, 0, 1)$&$(0, -1, 0, 0, 1, 0)$&$(-1, 0, 0, 0, -1, -1)$\\
$58$&$(1, 0, 0, -1, 0, 0)$&$(0, 1, 0, 0, -1, 0)$&$(0, 0, 0, 1, 1, 1)$&$(-1, 0, 1, 1, 0, -1)$&$122$&$(1, 1, 0, -1, -1, 0)$&$(1, 0, 0, 0, 1, 1)$&$(0, -1, -1, -1, 0, 0)$&$(-1, 0, 1, 1, 0, -1)$&$186$&$(1, 1, 1, 0, 0, 0)$&$(1, 0, -1, -1, 0, 1)$&$(0, -1, 0, 0, 1, 0)$&$(-1, 0, 0, 1, 0, 0)$\\
$59$&$(1, 0, 0, -1, 0, 0)$&$(0, 1, 0, 0, -1, 0)$&$(0, 0, 1, 1, 1, 0)$&$(-1, -1, 0, 0, 0, -1)$&$123$&$(1, 1, 0, -1, -1, 0)$&$(1, 0, 0, 0, 1, 1)$&$(0, 0, -1, 0, 0, 1)$&$(-1, -1, 0, 0, 0, -1)$&$187$&$(1, 1, 1, 0, 0, 0)$&$(1, 0, -1, -1, 0, 1)$&$(0, 0, 0, -1, -1, -1)$&$(-1, -1, 0, 1, 1, 0)$\\
$60$&$(1, 0, 0, -1, 0, 0)$&$(0, 1, 0, 0, -1, 0)$&$(0, 0, 1, 1, 1, 0)$&$(0, -1, -1, 0, 1, 1)$&$124$&$(1, 1, 0, -1, -1, 0)$&$(1, 0, 0, 0, 1, 1)$&$(0, 0, -1, 0, 0, 1)$&$(-1, 0, 1, 1, 0, -1)$&$188$&$(1, 1, 1, 0, 0, 0)$&$(1, 0, -1, -1, 0, 1)$&$(0, 0, 0, -1, -1, -1)$&$(-1, 0, 0, 1, 0, 0)$\\
$61$&$(1, 0, 0, -1, 0, 0)$&$(0, 1, 1, 0, -1, -1)$&$(0, 0, -1, -1, -1, 0)$&$(-1, -1, 0, 1, 1, 0)$&$125$&$(1, 1, 0, -1, -1, 0)$&$(1, 0, 0, 0, 1, 1)$&$(0, 0, 1, 0, 0, -1)$&$(-1, -1, -1, 0, 0, 0)$&$189$&$(1, 1, 1, 0, 0, 0)$&$(1, 0, -1, -1, 0, 1)$&$(0, 0, 0, 1, 1, 1)$&$(-1, -1, 0, 0, 0, -1)$\\
$62$&$(1, 0, 0, -1, 0, 0)$&$(0, 1, 1, 0, -1, -1)$&$(0, 0, -1, 0, 0, 1)$&$(-1, -1, 0, 0, 0, -1)$&$126$&$(1, 1, 0, -1, -1, 0)$&$(1, 0, 0, 0, 1, 1)$&$(0, 0, 1, 0, 0, -1)$&$(-1, 0, 0, 1, 0, 0)$&$190$&$(1, 1, 1, 0, 0, 0)$&$(1, 0, -1, -1, 0, 1)$&$(0, 0, 0, 1, 1, 1)$&$(-1, 0, 0, 0, -1, -1)$\\
$63$&$(1, 0, 0, -1, 0, 0)$&$(0, 1, 1, 0, -1, -1)$&$(0, 0, 0, 1, 1, 1)$&$(-1, -1, 0, 0, 0, -1)$&$127$&$(1, 1, 0, -1, -1, 0)$&$(1, 0, 0, 0, 1, 1)$&$(0, 1, 1, 1, 0, 0)$&$(-1, -1, -1, 0, 0, 0)$&$191$&$(1, 1, 1, 0, 0, 0)$&$(1, 0, -1, -1, 0, 1)$&$(0, 1, 0, 0, -1, 0)$&$(-1, -1, 0, 0, 0, -1)$\\
$64$&$(1, 0, 0, -1, 0, 0)$&$(0, 1, 1, 0, -1, -1)$&$(0, 0, 0, 1, 1, 1)$&$(0, 0, -1, -1, -1, 0)$&$128$&$(1, 1, 0, -1, -1, 0)$&$(1, 0, 0, 0, 1, 1)$&$(0, 1, 1, 1, 0, 0)$&$(-1, -1, 0, 0, 0, -1)$&$192$&$(1, 1, 1, 0, 0, 0)$&$(1, 0, -1, -1, 0, 1)$&$(0, 1, 0, 0, -1, 0)$&$(-1, -1, 0, 1, 1, 0)$
\end{tabular}
}
\caption{\footnotesize The possible simple roots of $D_4$, represented in the basis $Z=\{\pi_\alpha,\pi_\beta,\pi_\gamma,\pi_\alpha,\pi_\beta,\pi_\gamma\}$ as discussed in section \ref{sec:monodromy non simply-laced}. There are $192$ possibilities, as there must be since
  the order of $Weyl(D_4) =2^3 \cdot 4!$. We present this data for convenience and to demonstrate the agreement with the Weyl group. }
\label{table:Weyl D4}
\end{table}
\end{landscape}

\section{A Computational Package for Junctions}
\label{sec:code}
Many of the computations in this paper are simple, though tedious. To
facilitate computations, the authors have written a computer code
which performs many of the relevant operations. This code is publicly available
at
\begin{center}
\url{http://github.com/jhhalverson/py-junctions}.
\end {center}
Though it is written in Python, it is best executed through a SAGE
terminal, since it utilizes packages which are automatically included
in SAGE.  All non-trivial computations in this paper were performed
using this package, and there are many example computations
included on the web page. See also the Wiki which demonstrates many of
the relevant methods.

\clearpage
\section{Explicit Junction Depiction\label{app:junction}}
\begin{figure}
  \centering
  \includegraphics[scale=.7]{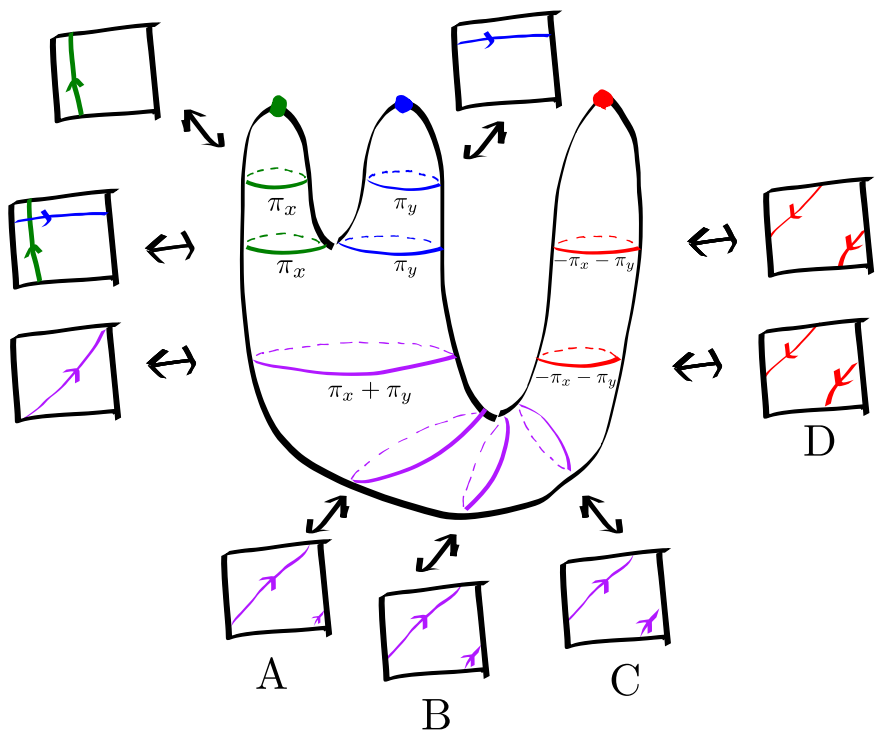}
\caption{An explicit depiction of the topology of a three-pronged junction. See appendix \ref{app:junction} for an explicit description of this two-sphere.}
\label{figure:junction}
\end{figure}

Figure \ref{figure:junction} depicts a two-sphere wrapped by an M2-brane in the deformation
picture of an M-theory compactification; in the F-theory limit this becomes a three-pronged
string junction along a one-manifold in the F-theory base $\tilde B$.
\vspace{.5cm}

Let's discuss the topology of this two-sphere. The green, blue, and
red dots denote the points where the junction ends on
seven-branes. The one-cycles $\pi_x$, $\pi_y$, and $-\pi_x-\pi_y$
vanish over each of these codimension one loci, respectively. The minus sign $-\pi_x-\pi_y$ corresponds to the orientation of this curve as a vanishing cycle for the red singular point, so that we get total asymptotic charge zero. The junction
has one dimension along the elliptic fiber and one along $\tilde B$.
Moving away from the green, blue, and red marked points, the vanishing cycle grow and are
wrapped by one dimension of an M2-brane, depicted by the tori (represented as squares with
periodic boundary conditions) with one-cycles in green, blue, and red.

At the junction point in $\tilde B$, the M2-brane wraps the one-cycle depicted by the
square with both blue and green lines; the self-intersection point is evident. At
this point the junction forms a pair of pants, and moving past the junction point
the M2-brane wraps the purple cycle $\pi_x + \pi_y$.
Continuing down the junction to points A, B, and C the cycle wrapped by the M2-brane
moves towards the red cycle $-\pi_x-\pi_y$; somewhere between points C and D, they merge, or more precisely in terms of the orientations determined as vanishing cycles, bound a common annulus, closing
the two disks (with one and two marked points) into a sphere with three marked points. This sphere is
the two-cycle wrapped by the M2-brane.


\clearpage
\bibliographystyle{utphys}

\begin{thebibliography}{10}


\bibitem{Witten:1995ex}
E.~Witten, ``{String Theory Dynamics in Various Dimensions},'' {\em Nucl.Phys.}
  {\bf B443} (1995) 85--126,
\href{http://www.arXiv.org/abs/hep-th/9503124}{{\tt hep-th/9503124}}.

\bibitem{Vafa:1996xn}
C.~Vafa, ``{Evidence for F-Theory},'' {\em Nucl. Phys.} {\bf B469} (1996)
  403--418,
\href{http://www.arXiv.org/abs/hep-th/9602022}{{\tt hep-th/9602022}}.

\bibitem{Dienes:1996yh}
K.~R. Dienes and J.~March-Russell, ``{Realizing Higher Level Gauge Symmetries
  in String Theory: New Embeddings for String GUTs},'' {\em Nucl.Phys.} {\bf
  B479} (1996) 113--172,
\href{http://www.arXiv.org/abs/hep-th/9604112}{{\tt hep-th/9604112}}.

\bibitem{Dienes:1996wx}
K.~R. Dienes, ``{New Constraints on SO(10) Model Building from String
  Theory},'' {\em Nucl.Phys.} {\bf B488} (1997) 141--158,
\href{http://www.arXiv.org/abs/hep-ph/9606467}{{\tt hep-ph/9606467}}.

\bibitem{Kovalev}
A.~{Kovalev}, ``{Twisted connected sums and special Riemannian holonomy},''
  {\em ArXiv Mathematics e-prints} (Dec., 2000)
  \href{http://www.arXiv.org/abs/arXiv:math/0012189}{{\tt arXiv:math/0012189}}.

\bibitem{KovalevLee}
A.~{Kovalev} and N.-H. {Lee}, ``{K3 surfaces with non-symplectic involution and
  compact irreducible $G_2$-manifolds},'' {\em ArXiv e-prints} (Oct., 2008)
  \href{http://www.arXiv.org/abs/0810.0957}{{\tt 0810.0957}}.

\bibitem{CortiHaskins1}
A.~{Corti}, M.~{Haskins}, J.~{Nordstrom}, and T.~{Pacini}, ``{Asymptotically
  cylindrical Calabi-Yau 3-folds from weak Fano 3-folds},'' {\em ArXiv
  e-prints} (June, 2012) \href{http://www.arXiv.org/abs/1206.2277}{{\tt
  1206.2277}}.

\bibitem{CortiHaskins2}
A.~{Corti}, M.~{Haskins}, J.~{Nordstrom}, and T.~{Pacini}, ``{$G_2$-manifolds
  and associative submanifolds via semi-Fano 3-folds},'' {\em ArXiv e-prints}
  (July, 2012) \href{http://www.arXiv.org/abs/1207.4470}{{\tt 1207.4470}}.

\bibitem{Katz:1996xe}
S.~H. Katz and C.~Vafa, ``{Matter from Geometry},'' {\em Nucl.Phys.} {\bf B497}
  (1997) 146--154,
\href{http://www.arXiv.org/abs/hep-th/9606086}{{\tt hep-th/9606086}}.

\bibitem{Bershadsky:1996nh}
M.~Bershadsky {\em et al.}, ``{Geometric Singularities and Enhanced Gauge
  Symmetries},'' {\em Nucl. Phys.} {\bf B481} (1996) 215--252,
\href{http://www.arXiv.org/abs/hep-th/9605200}{{\tt hep-th/9605200}}.

\bibitem{Grassi:2000we}
A.~Grassi and D.~R. Morrison, ``{Group Representations and the Euler
  Characteristic of Elliptically Fibered Calabi-Yau Threefolds},''
\href{http://www.arXiv.org/abs/math/0005196}{{\tt math/0005196}}.

\bibitem{Aspinwall:2000kf}
P.~S. Aspinwall, S.~H. Katz, and D.~R. Morrison, ``{Lie Groups, Calabi-Yau
  Threefolds, and F Theory},'' {\em Adv.Theor.Math.Phys.} {\bf 4} (2000)
  95--126,
\href{http://www.arXiv.org/abs/hep-th/0002012}{{\tt hep-th/0002012}}.

\bibitem{Morrison:2011mb}
D.~R. Morrison and W.~Taylor, ``{Matter and Singularities},'' {\em JHEP} {\bf
  1201} (2012) 022,
\href{http://www.arXiv.org/abs/1106.3563}{{\tt 1106.3563}}.

\bibitem{Grassi:2011hq}
A.~Grassi and D.~R. Morrison, ``{Anomalies and the Euler Characteristic of
  Elliptic Calabi-Yau Threefolds},''
\href{http://www.arXiv.org/abs/1109.0042}{{\tt 1109.0042}}.

\bibitem{DeWolfe:1998zf}
O.~DeWolfe and B.~Zwiebach, ``{String Junctions for Arbitrary Lie Algebra
  Representations},'' {\em Nucl. Phys.} {\bf B541} (1999) 509--565,
\href{http://www.arXiv.org/abs/hep-th/9804210}{{\tt hep-th/9804210}}.

\bibitem{Gaberdiel:1997ud}
M.~R. Gaberdiel and B.~Zwiebach, ``{Exceptional Groups from Open Strings},''
  {\em Nucl. Phys.} {\bf B518} (1998) 151--172,
\href{http://www.arXiv.org/abs/hep-th/9709013}{{\tt hep-th/9709013}}.

\bibitem{Mikhailov:1998bx}
A.~Mikhailov, N.~Nekrasov, and S.~Sethi, ``{Geometric Realizations of BPS
  States in ${\mathcal{N}}\!=2$ Theories},'' {\em Nucl. Phys.} {\bf B531}
  (1998) 345--362,
\href{http://www.arXiv.org/abs/hep-th/9803142}{{\tt hep-th/9803142}}.

\bibitem{Morrison:1996na}
D.~R. Morrison and C.~Vafa, ``{Compactifications of F-Theory on Calabi--Yau
  Threefolds -- I},'' {\em Nucl. Phys.} {\bf B473} (1996) 74--92,
\href{http://www.arXiv.org/abs/hep-th/9602114}{{\tt hep-th/9602114}}.

\bibitem{Morrison:1996pp}
D.~R. Morrison and C.~Vafa, ``{Compactifications of F-Theory on Calabi--Yau
  Threefolds -- II},'' {\em Nucl. Phys.} {\bf B476} (1996) 437--469,
\href{http://www.arXiv.org/abs/hep-th/9603161}{{\tt hep-th/9603161}}.


\bibitem{GrassiHalversonShaneson:Math}
A.~Grassi, J.~Halverson, and J.~Shaneson, ``{Resolution and Deformation of
  Elliptic Fibrations},''.


\bibitem{KatzMorrison}
S.~{Katz} and D.~R. {Morrison}, ``{Gorenstein Threefold Singularities with
  Small Resolutions via Invariant Theory for Weyl Groups},''.

\bibitem{Arnold}
V.~I. Arnol'd, ``NORMAL FORMS OF FUNCTIONS IN NEIGHBOURHOODS OF DEGENERATE
  CRITICAL POINTS,'' {\em Russian Mathematical Surveys} {\bf 29} (1974), no.~2,
  10.

\bibitem{GuseinZade} Gusein-Zade, S. M., ``{Monodromy groups of
    isolated singularities of hypersurfaces. (Russian)},''
Uspehi Mat. Nauk
(1977).


\bibitem{Hanany:1996ie}
A.~Hanany and E.~Witten, ``{Type IIB Superstrings, BPS Monopoles, and Three-
  Dimensional Gauge Dynamics},'' {\em Nucl. Phys.} {\bf B492} (1997) 152--190,
\href{http://www.arXiv.org/abs/hep-th/9611230}{{\tt hep-th/9611230}}.

\bibitem{Miranda}
R.~Miranda, ``Smooth models for elliptic threefolds,'' in {\em The birational
  geometry of degenerations ({C}ambridge, {M}ass., 1981)}, vol.~29 of {\em
  Progr. Math.}, pp.~85--133.
\newblock Birkh\"auser Boston, Mass., 1983.

\bibitem{Witten:1996qb}
E.~Witten, ``{Phase Transitions in M Theory and F Theory},'' {\em Nucl.Phys.}
  {\bf B471} (1996) 195--216,
\href{http://www.arXiv.org/abs/hep-th/9603150}{{\tt hep-th/9603150}}.


\bibitem{Brieskorn}
E.~Brieskorn, ``Singular elements of semi-simple algebraic groups,''.

\bibitem{DeWolfe:1998bi}
O.~DeWolfe, T.~Hauer, A.~Iqbal, and B.~Zwiebach, ``{Constraints on the BPS
  Spectrum of ${\mathcal{N}}\!=2$, D = 4 Theories with A-D-E Flavor
  Symmetry},'' {\em Nucl. Phys.} {\bf B534} (1998) 261--274,
\href{http://www.arXiv.org/abs/hep-th/9805220}{{\tt hep-th/9805220}}.

\bibitem{Grassi:2000fk}
A.~Grassi, Z.~Guralnik, and B.~A. Ovrut, ``{Five-brane BPS states in heterotic
  M theory},'' {\em JHEP} {\bf 0101} (2001) 037,
\href{http://www.arXiv.org/abs/hep-th/0005121}{{\tt hep-th/0005121}}.

\bibitem{Grassi:2001xu}
A.~Grassi, Z.~Guralnik, and B.~A. Ovrut, ``{Knots, braids and BPS states in M
  theory},'' {\em JHEP} {\bf 0206} (2002) 023,
\href{http://www.arXiv.org/abs/hep-th/0110036}{{\tt hep-th/0110036}}.

\bibitem{Bonora:2010bu}
L.~Bonora and R.~Savelli, ``{Non-Simply-Laced Lie Algebras via F Theory
  Strings},'' {\em JHEP} {\bf 1011} (2010) 025,
\href{http://www.arXiv.org/abs/1007.4668}{{\tt 1007.4668}}.

\bibitem{Donagi:2008ca}
R.~Donagi and M.~Wijnholt, ``{Model Building with F-Theory},''
\href{http://www.arXiv.org/abs/0802.2969}{{\tt 0802.2969}}.

\bibitem{Beasley:2008dc}
C.~Beasley, J.~J. Heckman, and C.~Vafa, ``{Guts and Exceptional Branes in
  F-Theory - I},'' {\em JHEP} {\bf 01} (2009) 058,
\href{http://www.arXiv.org/abs/0802.3391}{{\tt 0802.3391}}.

\bibitem{Andreas:2009uf}
B.~Andreas and G.~Curio, ``{From Local to Global in F-Theory Model Building},''
  {\em J.Geom.Phys.} {\bf 60} (2010) 1089--1102,
\href{http://www.arXiv.org/abs/0902.4143}{{\tt 0902.4143}}.

\bibitem{Marsano:2009ym}
J.~Marsano, N.~Saulina, and S.~Schafe{r-Na}meki, ``{F-Theory Compactifications
  for Supersymmetric Guts},'' {\em JHEP} {\bf 08} (2009) 030,
\href{http://www.arXiv.org/abs/0904.3932}{{\tt 0904.3932}}.

\bibitem{Collinucci:2009uh}
A.~Collinucci, ``{New F-Theory Lifts II: Permutation Orientifolds and Enhanced
  Singularities},'' {\em JHEP} {\bf 04} (2010) 076,
\href{http://www.arXiv.org/abs/0906.0003}{{\tt 0906.0003}}.

\bibitem{Blumenhagen:2009up}
R.~Blumenhagen, T.~W. Grimm, B.~Jurke, and T.~Weigand, ``{F-Theory Uplifts and
  GUTs},'' {\em JHEP} {\bf 09} (2009) 053,
\href{http://www.arXiv.org/abs/0906.0013}{{\tt 0906.0013}}.

\bibitem{Marsano:2009gv}
J.~Marsano, N.~Saulina, and S.~Schafe{r-Na}meki, ``{Monodromies, Fluxes, and
  Compact Three-Generation F-Theory GUTs},'' {\em JHEP} {\bf 08} (2009) 046,
\href{http://www.arXiv.org/abs/0906.4672}{{\tt 0906.4672}}.

\bibitem{Blumenhagen:2009yv}
R.~Blumenhagen, T.~W. Grimm, B.~Jurke, and T.~Weigand, ``{Global F-Theory
  GUTs},'' {\em Nucl. Phys.} {\bf B829} (2010) 325--369,
\href{http://www.arXiv.org/abs/0908.1784}{{\tt 0908.1784}}.

\bibitem{Marsano:2009wr}
J.~Marsano, N.~Saulina, and S.~Schafe{r-Na}meki, ``{Compact F-Theory Guts with
  $U(1)_{PQ}$},'' {\em JHEP} {\bf 04} (2010) 095,
\href{http://www.arXiv.org/abs/0912.0272}{{\tt 0912.0272}}.

\bibitem{Grimm:2009yu}
T.~W. Grimm, S.~Krause, and T.~Weigand, ``{F-Theory GUT Vacua on Compact
  Calabi-Yau Fourfolds},'' {\em JHEP} {\bf 07} (2010) 037,
\href{http://www.arXiv.org/abs/0912.3524}{{\tt 0912.3524}}.

\bibitem{Cvetic:2010rq}
M.~Cveti{\v c}, I.~Garc{\'\i a-}Etxebarria, and J.~Halverson, ``{Global
  F-Theory Models: Instantons and Gauge Dynamics},'' {\em JHEP} {\bf 1101}
  (2011) 073,
\href{http://www.arXiv.org/abs/1003.5337}{{\tt 1003.5337}}.

\bibitem{Chen:2010ts}
C.-M. Chen, J.~Knapp, M.~Kreuzer, and C.~Mayrhofer, ``{Global SO(10) F-Theory
  Guts},'' {\em JHEP} {\bf 10} (2010) 057,
\href{http://www.arXiv.org/abs/1005.5735}{{\tt 1005.5735}}.

\bibitem{Chen:2010tp}
C.-M. Chen and Y.-C. Chung, ``{Flipped $SU(5)$ Guts from $E_{8}$ Singularities
  in F-Theory},'' {\em JHEP} {\bf 03} (2011) 049,
\href{http://www.arXiv.org/abs/1005.5728}{{\tt 1005.5728}}.

\bibitem{Chung:2010bn}
Y.-C. Chung, ``{On Global Flipped $SU(5)$ Guts in F-Theory},'' {\em JHEP} {\bf
  03} (2011) 126,
\href{http://www.arXiv.org/abs/1008.2506}{{\tt 1008.2506}}.

\bibitem{Chen:2010tg}
C.-M. Chen and Y.-C. Chung, ``{On F-Theory $E_{6}$ Guts},'' {\em JHEP} {\bf 03}
  (2011) 129,
\href{http://www.arXiv.org/abs/1010.5536}{{\tt 1010.5536}}.

\bibitem{Knapp:2011wk}
J.~Knapp, M.~Kreuzer, C.~Mayrhofer, and N.-O. Walliser, ``{Toric Construction
  of Global F-Theory Guts},'' {\em JHEP} {\bf 03} (2011) 138,
\href{http://www.arXiv.org/abs/1101.4908}{{\tt 1101.4908}}.

\bibitem{Knapp:2011ip}
J.~Knapp and M.~Kreuzer, ``{Toric Methods in F-Theory Model Building},'' {\em
  Adv.High Energy Phys.} {\bf 2011} (2011) 513436,
\href{http://www.arXiv.org/abs/1103.3358}{{\tt 1103.3358}}.

\bibitem{Marsano:2012yc}
J.~Marsano, H.~Clemens, T.~Pantev, S.~Raby, and H.-H. Tseng, ``{A Global
  $SU(5)$ F-Theory Model with Wilson Line Breaking},''
\href{http://www.arXiv.org/abs/1206.6132}{{\tt 1206.6132}}.

\bibitem{Grimm:2010ez}
T.~W. Grimm and T.~Weigand, ``{On Abelian Gauge Symmetries and Proton Decay in
  Global F- Theory Guts},'' {\em Phys. Rev.} {\bf D82} (2010) 086009,
\href{http://www.arXiv.org/abs/1006.0226}{{\tt 1006.0226}}.

\bibitem{Dolan:2011iu}
M.~J. Dolan, J.~Marsano, N.~Saulina, and S.~Schafe{r-Na}meki, ``{F-Theory Guts
  with U(1) Symmetries: Generalities and Survey},'' {\em Phys.Rev.} {\bf D84}
  (2011) 066008,
\href{http://www.arXiv.org/abs/1102.0290}{{\tt 1102.0290}}.

\bibitem{Marsano:2011nn}
J.~Marsano, N.~Saulina, and S.~Schafe{r-Na}meki, ``{On G-Flux, M5 Instantons,
  and U(1)s in F-Theory},''
\href{http://www.arXiv.org/abs/1107.1718}{{\tt 1107.1718}}.

\bibitem{Grimm:2011tb}
T.~W. Grimm, M.~Kerstan, E.~Palti, and T.~Weigand, ``{Massive Abelian Gauge
  Symmetries and Fluxes in F-Theory},'' {\em JHEP} {\bf 1112} (2011) 004,
\href{http://www.arXiv.org/abs/1107.3842}{{\tt 1107.3842}}.

\bibitem{Morrison:2012ei}
D.~R. Morrison and D.~S. Park, ``{F-Theory and the Mordell-Weil Group of
  Elliptically-Fibered Calabi-Yau Threefolds},''
\href{http://www.arXiv.org/abs/1208.2695}{{\tt 1208.2695}}.

\bibitem{Borchmann:2013jwa}
J.~Borchmann, C.~Mayrhofer, E.~Palti, and T.~Weigand, ``{Elliptic Fibrations
  for $SU(5)$ $\times$ U(1) $\times$ U(1) F-Theory Vacua},''
\href{http://www.arXiv.org/abs/1303.5054}{{\tt 1303.5054}}.

\bibitem{Cvetic:2013nia}
M.~Cveti{\v c}, D.~Klevers, and H.~Piragua, ``{F-Theory Compactifications with
  Multiple U(1)-Factors: Constructing Elliptic Fibrations with Rational
  Sections},''
\href{http://www.arXiv.org/abs/1303.6970}{{\tt 1303.6970}}.

\bibitem{Blumenhagen:2010ja}
R.~Blumenhagen, A.~Collinucci, and B.~Jurke, ``{On Instanton Effects in
  F-Theory},'' {\em JHEP} {\bf 08} (2010) 079,
\href{http://www.arXiv.org/abs/1002.1894}{{\tt 1002.1894}}.

\bibitem{Donagi:2010pd}
R.~Donagi and M.~Wijnholt, ``{MSW Instantons},''
\href{http://www.arXiv.org/abs/1005.5391}{{\tt 1005.5391}}.

\bibitem{Grimm:2011dj}
T.~W. Grimm, M.~Kerstan, E.~Palti, and T.~Weigand, ``{On Fluxed Instantons and
  Moduli Stabilisation in IIB Orientifolds and F-Theory},''
\href{http://www.arXiv.org/abs/1105.3193}{{\tt 1105.3193}}.

\bibitem{Cvetic:2011gp}
M.~Cvetic, I.~Garcia~Etxebarria, and J.~Halverson, ``{Three Looks at Instantons
  in F-theory -- New Insights from Anomaly Inflow, String Junctions and
  Heterotic Duality},'' {\em JHEP} {\bf 1111} (2011) 101,
\href{http://www.arXiv.org/abs/1107.2388}{{\tt 1107.2388}}.

\bibitem{Bianchi:2011qh}
M.~Bianchi, A.~Collinucci, and L.~Martucci, ``{Magnetized E3-Brane Instantons
  in F-Theory},'' {\em JHEP} {\bf 1112} (2011) 045,
\href{http://www.arXiv.org/abs/1107.3732}{{\tt 1107.3732}}.

\bibitem{Kerstan:2012cy}
M.~Kerstan and T.~Weigand, ``{Fluxed M5-Instantons in F-Theory},''
\href{http://www.arXiv.org/abs/1205.4720}{{\tt 1205.4720}}.

\bibitem{Cvetic:2012ts}
M.~Cveti{\v c}, R.~Donagi, J.~Halverson, and J.~Marsano, ``{On Seven-Brane
  Dependent Instanton Prefactors in F-Theory},'' {\em JHEP} {\bf 1211} (2012)
  004,
\href{http://www.arXiv.org/abs/1209.4906}{{\tt 1209.4906}}.

\bibitem{Bianchi:2012kt}
M.~Bianchi, G.~Inverso, and L.~Martucci, ``{Brane Instantons and Fluxes in
  F-Theory},''
\href{http://www.arXiv.org/abs/1212.0024}{{\tt 1212.0024}}.

\bibitem{Marsano:2010ix}
J.~Marsano, N.~Saulina, and S.~Schafe{r-Na}meki, ``{A Note on G-Fluxes for
  F-Theory Model Building},'' {\em JHEP} {\bf 11} (2010) 088,
\href{http://www.arXiv.org/abs/1006.0483}{{\tt 1006.0483}}.

\bibitem{Collinucci:2010gz}
A.~Collinucci and R.~Savelli, ``{On Flux Quantization in F-Theory},'' {\em
  JHEP} {\bf 1202} (2012) 015,
\href{http://www.arXiv.org/abs/1011.6388}{{\tt 1011.6388}}.

\bibitem{Braun:2011zm}
A.~P. Braun, A.~Collinucci, and R.~Valandro, ``{G-Flux in F-Theory and
  Algebraic Cycles},'' {\em Nucl.Phys.} {\bf B856} (2012) 129--179,
\href{http://www.arXiv.org/abs/1107.5337}{{\tt 1107.5337}}.

\bibitem{Marsano:2011hv}
J.~Marsano and S.~Schafe{r-Na}meki, ``{Yukawas, G-Flux, and Spectral Covers from Resolved Calabi- Yau's},'' {\em JHEP} {\bf 11 098} (2011), arXiv:1108.1794.

\bibitem{Krause:2011xj}
S.~Krause, C.~Mayrhofer, and T.~Weigand, ``{$G_4$ Flux, Chiral Matter and
  Singularity Resolution in F-Theory Compactifications},'' {\em Nucl.Phys.}
  {\bf B858} (2012) 1--47,
\href{http://www.arXiv.org/abs/1109.3454}{{\tt 1109.3454}}.

\bibitem{Grimm:2011fx}
T.~W. Grimm and H.~Hayashi, ``{F-Theory Fluxes, Chirality and Chern-Simons
  Theories},''
\href{http://www.arXiv.org/abs/1111.1232}{{\tt 1111.1232}}.

\bibitem{Braun:2012nk}
A.~P. Braun, A.~Collinucci, and R.~Valandro, ``{Algebraic Description of G-Flux
  in F-Theory: New Techniques for F-Theory Phenomenology},''
\href{http://www.arXiv.org/abs/1202.5029}{{\tt 1202.5029}}.

\bibitem{Kuntzler:2012bu}
M.~Kuntzler and S.~ Schafe{r-Na}meki.
``{G-Flux and Spectral Divisors},'' arXiv 1205.5688.

\bibitem{Lawrie:2012gg}
  C.~Lawrie and S.~Schäfer-Nameki,
  JHEP {\bf 1304}, 061 (2013)
  [arXiv:1212.2949 [hep-th]].

\bibitem{Krause:2012he}
S.~Krause, C.~Mayrhofer, and T.~Weigand, ``{Gauge Fluxes in F-Theory and Type
  IIB Orientifolds},'' {\em JHEP} {\bf 1208} (2012) 119,
\href{http://www.arXiv.org/abs/1202.3138}{{\tt 1202.3138}}.

\bibitem{Collinucci:2012as}
A.~Collinucci and R.~Savelli, ``{On Flux Quantization in F-Theory II: Unitary
  and Symplectic Gauge Groups},'' {\em JHEP} {\bf 1208} (2012) 094,
\href{http://www.arXiv.org/abs/1203.4542}{{\tt 1203.4542}}.

\bibitem{Marsano:2012bf}
J.~Marsano, N.~Saulina, and S.~Schafe{r-Na}meki, ``{Global Gluing and
  G-Flux},''
\href{http://www.arXiv.org/abs/1211.1097}{{\tt 1211.1097}}.

\bibitem{Kumar:2009ac}
V.~Kumar, D.~R. Morrison, and W.~Taylor, ``{Mapping 6D ${\mathcal{N}}\!=1$
  Supergravities to F-Theory},'' {\em JHEP} {\bf 02} (2010) 099,
\href{http://www.arXiv.org/abs/0911.3393}{{\tt 0911.3393}}.

\bibitem{Kumar:2010ru}
V.~Kumar, D.~R. Morrison, and W.~Taylor, ``{Global Aspects of the Space of 6D
  ${\mathcal{N}}\!=1$ Supergravities},'' {\em JHEP} {\bf 1011} (2010) 118,
\href{http://www.arXiv.org/abs/1008.1062}{{\tt 1008.1062}}.

\bibitem{Kumar:2010am}
V.~Kumar, D.~S. Park, and W.~Taylor, ``{6D Supergravity without Tensor
  Multiplets},'' {\em JHEP} {\bf 1104} (2011) 080,
\href{http://www.arXiv.org/abs/1011.0726}{{\tt 1011.0726}}.


\bibitem{Morrison:2012np}
D.~R. Morrison and W.~Taylor, ``{Classifying Bases for 6D F-Theory Models},''
  {\em Central Eur.J.Phys.} {\bf 10} (2012) 1072--1088,
\href{http://www.arXiv.org/abs/1201.1943}{{\tt 1201.1943}}.

\bibitem{Esole:2011sm}
M.~Esole and S.-T. Yau, ``{Small Resolutions of $SU(5)$-models in F-Theory},''
\href{http://www.arXiv.org/abs/1107.0733}{{\tt 1107.0733}}.

\bibitem{Cvetic:2012xn}
M.~Cveti{\v c}, T.~W. Grimm, and D.~Klevers, ``{Anomaly Cancellation and
  Abelian Gauge Symmetries in F-Theory},''
\href{http://www.arXiv.org/abs/1210.6034}{{\tt 1210.6034}}.


\bibitem{Sen:1996vd}
A.~Sen, ``{F-Theory and Orientifolds},'' {\em Nucl. Phys.} {\bf B475} (1996)
  562--578,
\href{http://www.arXiv.org/abs/hep-th/9605150}{{\tt hep-th/9605150}}.

\bibitem{Sen:1997gv}
A.~Sen, ``{Orientifold Limit of F-Theory Vacua},'' {\em Phys. Rev.} {\bf D55}
  (1997) 7345--7349,
\href{http://www.arXiv.org/abs/hep-th/9702165}{{\tt hep-th/9702165}}.

\bibitem{Seiberg:1994rs}
N.~Seiberg and E.~Witten, ``{Monopole Condensation, and Confinement in
  ${\mathcal{N}}\!=2$ Supersymmetric Yang-Mills Theory},'' {\em Nucl. Phys.}
  {\bf B426} (1994) 19--52,
\href{http://www.arXiv.org/abs/hep-th/9407087}{{\tt hep-th/9407087}}.

\bibitem{Banks:1996nj}
T.~Banks, M.~R. Douglas, and N.~Seiberg, ``{Probing F-Theory with Branes},''
  {\em Phys. Lett.} {\bf B387} (1996) 278--281,
\href{http://www.arXiv.org/abs/hep-th/9605199}{{\tt hep-th/9605199}}.


\bibitem{Slansky:1981yr}
R.~Slansky, ``{Group Theory for Unified Model Building},'' {\em Phys.Rept.}
  {\bf 79} (1981)
1--128.

\bibitem{Babu:1992ia}
K.~Babu and R.~Mohapatra, ``{Predictive Neutrino Spectrum in Minimal SO(10)
  Grand Unification},'' {\em Phys.Rev.Lett.} {\bf 70} (1993) 2845--2848,
\href{http://www.arXiv.org/abs/hep-ph/9209215}{{\tt hep-ph/9209215}}.

\end{thebibliography}
\providecommand{\href}[2]{#2}\begingroup\raggedright\endgroup

\end{document}